\documentclass[useAMS,usenatbib]{mn2e}
\usepackage{amssymb}
\usepackage{amsmath}
\usepackage{graphicx}
\usepackage{color}
\usepackage{natbib}

\newcommand{\dd}{\mathrm{d}}

\newcommand{\mbh}{\ensuremath{M_\bullet}\,}
\newcommand{\er}{\ensuremath{\lambda\,}}

\newcommand\ion[2]{#1$\,${\scshape{#2}}}

\newcommand{\apjs}{ApJS}
\newcommand{\apj}{ApJ}
\newcommand{\mnras}{MNRAS}

\newcommand{\aj}{AJ}
\newcommand{\aap}{A\&A}
\newcommand{\apjl}{ApJ}
\newcommand{\araa}{ARA\&A}

\title[The cosmic growth of the active black hole population at $1<z<2$]{The cosmic growth of the active black hole population at $1<z<2$ in zCOSMOS, VVDS and SDSS}
\author[Schulze et al.]{A. Schulze$^{1,2}$\thanks{E-mail:
andreas.schulze@ipmu.jp}, A. Bongiorno$^{3}$, I. Gavignaud$^{4}$, M. Schramm$^{1}$, J. Silverman$^{1}$, A. Merloni$^{5}$, \newauthor
G. Zamorani$^{6}$, M. Hirschmann$^7$, V. Mainieri$^{8}$, L. Wisotzki$^{9}$, F. Shankar$^{10}$, F. Fiore$^{3}$, \newauthor A. M. Koekemoer$^{11}$,  G. Temporin$^{12}$
\smallskip \\
$^{1}$ Kavli Institute for the Physics and Mathematics of the Universe (WPI), The University of Tokyo, Kashiwa, Chiba 277-8583, Japan\\
$^{2}$ Kavli Institute for Astronomy and Astrophysics, Peking University, 100871 Beijing, China \\
$^{3}$ INAF-Osservatorio Astronomico di Roma, Via di Frascati 33, 00040, Monteporzio Catone, Rome, Italy \\
$^{4}$  Departamento de Ciencias Fisicas, Universidad Andres Bello, Av. Republica 220, 8370134, Santiago, Chile\\
$^{5}$ Max-Planck-Institut f\"ur extraterrestrische Physik (MPE), Giessenbachstrasse 1, Garching bei M\"unchen, D-85748, Germany\\
$^{6}$ INAF-Osservatorio Astronomico di Bologna, Via Ranzani 1, 40127, Bologna, Italy\\
$^{7}$ UPMC-CNRS, UMR7095, Institut d'Astrophysique de Paris, Boulevard Arago, F-75014 Paris, France\\
$^{8}$ ESO, Karl-Schwarschild-Strasse 2, Garching bei M\"unchen, D-85748, Germany\\
$^{9}$ Leibniz-Institut f\"ur Astrophysik Potsdam (AIP), An der Sternwarte 16, D-14482 Potsdam, Germany\\
$^{10}$ School of Physics \& Astronomy, University of Southampton, Southampton SO17 1BJ, UK\\
$^{11}$ Space Telescope Science Institute, 3700 San Martin Drive, Baltimore, MD 21218, USA\\
$^{12}$ Institute of Astro- and Particle Physics, University of Innsbruck, Technikerstrasse 25, A-6020 Innsbruck, Austria\\
}
\begin{document}

\date{Accepted 2014 November 28.  Received 2014 November 19; in original form 2014 October 6}

\pagerange{\pageref{firstpage}--\pageref{lastpage}} \pubyear{2014}

\maketitle

\label{firstpage}

\begin{abstract}
We present a census of the active black hole population at $1<z<2$, by constructing the bivariate distribution function of black hole mass and Eddington ratio, employing a maximum likelihood fitting technique.  The study of the active black hole mass function (BHMF) and the Eddington ratio distribution function (ERDF) allows us to clearly disentangle the active galactic nuclei (AGN) downsizing phenomenon, present in the AGN luminosity function, into its physical processes of black hole mass downsizing and accretion rate evolution. We are utilizing type-1 AGN samples from three optical surveys (VVDS, zCOSMOS and SDSS), that cover a wide range of 3~dex in luminosity over our redshift interval of interest. 
We investigate the cosmic evolution of the AGN population as a function of AGN luminosity, black hole mass and accretion rate. Compared to $z = 0$, we find a distinct change in the shape of the BHMF and the ERDF, consistent with downsizing in black hole mass. 
The active fraction or duty cycle of type-1 AGN at $z\sim1.5$ is almost flat as a function of black hole mass, while it shows a strong decrease with increasing mass at $z=0$. We are witnessing a phase of intense black hole growth, which is largely driven by the onset of AGN activity in massive black holes towards $z=2$. We finally compare our results to numerical simulations and semi-empirical models and while we find reasonable agreement over certain parameter ranges, we highlight the need to refine these models in order to match our observations.
\end{abstract}

\section{Introduction} \label{sec:intro}
Supermassive black holes (SMBHs) constitute a fundamental component of galaxies, as almost every massive galaxy harbours an SMBH in its centre \citep{Kormendy:1995}.  A complete picture of galaxy evolution requires an understanding of the growth history of SMBHs. This is demanded by observational  evidence for co-evolution between the SMBH and its host galaxy, suggested by the observed tight correlation between SMBH mass and the properties of the galaxy's spheroidal component, e.g. with stellar velocity dispersion \citep[e.g.][]{Ferrarese:2000,Gebhardt:2000, Tremaine:2002, McConnell:2013}, bulge luminosity and bulge mass \citep{Magorrian:1998,Marconi:2003,Haering:2004}. Furthermore, the cosmic black hole (BH) accretion rate density traces well the strong evolution in the cosmic star formation rate density since $z\sim3$ \citep[e.g.][]{Boyle:1998,Marconi:2004,Silverman:2008}, implying a link between star formation and BH accretion in a global sense. Such a connection between star formation and active galactic nuclei (AGN) activity is also directly seen for  luminous AGN \citep[e.g.][]{Netzer:2007,Lutz:2008,Silverman:2009,Rosario:2012}. 

The Soltan argument suggests that most BH growth takes place in luminous AGN phases \citep{Soltan:1982, Salucci:1999,Yu:2002}. Therefore, the demographics of the AGN population enables an assessment of the cosmic SMBH growth history. The main demographic quantity is the AGN luminosity function (AGN LF), which is established over a wide range in redshift, luminosity and wavelength \citep[e.g.][]{Boyle:2000,Wolf:2003,Ueda:2003,Hasinger:2005,LaFranca:2005,Richards:2006a,Bongiorno:2007,Silverman:2008,Croom:2009,Aird:2010,Assef:2011,Fiore:2012,Ueda :2014}. The evolution of the AGN LF allows us to constrain the BH growth history using a continuity equation to evolve the black hole mass function (BHMF) through redshift \citep{Yu:2002,Yu:2004, Marconi:2004, Merloni:2004, Merloni:2008, Yu:2008, Shankar:2009,Draper:2012}. These studies found that the local total BHMF is consistent with being a relic of previous AGN activity. They also link the AGN 'downsizing', observed in the AGN LF \citep{Ueda:2003,Hasinger:2005,LaFranca:2005,Croom:2009}, to the growth of SMBHs, i.e. the most massive BHs grow at  earlier epochs, while lower mass BHs are still actively growing at later times, a behaviour which is also called antihierachical BH growth \citep{Marconi:2004,Merloni:2004}. 

However, having only information on the AGN luminosity does not allow us to distinguish the two quantities that govern SMBH growth i.e. SMBH mass (\mbh) and accretion rate. The latter can be expressed as the Eddington ratio $\lambda=L_\mathrm{Bol}/L_\mathrm{Edd}$, with the Eddington luminosity $L_\mathrm{Edd}=1.3\times10^{38} (\mbh/M_\odot)$~erg s$^{-1}$. This is a critical limitation, since AGN show a wide distribution of accretion rates \citep{Heckman:2004,Babic:2007,Kauffmann:2009,Schulze:2010,Aird:2012,Bongiorno:2012}.
Additional assumptions on the accretion rate distribution or alternatively the mean accretion rate are required  in the above studies to link the bolometric AGN luminosity to BH growth. Thus, while the AGN LF is relatively easy to determine observationally, its information content is limited.  

This situation is similar to galaxy evolution studies, where the galaxy luminosity function (LF) already provides important information, but the study of the stellar mass function in connection to the specific star formation rate (sSFR) largely enhances our understanding \citep[e.g.][]{Elbaz:2007,Peng:2010}.  Galaxies show a downsizing trend, equivalent to BHs, in the sense that for more massive galaxies the bulk of their mass buildup happens at earlier cosmic times compared to less massive galaxies \citep[e.g.][]{Cowie:1996,Bundy:2006,Perez:2008}. While for galaxy evolution, stellar mass and sSFR are key parameters, the same is true for BH growth with BH mass and specific accretion rate (i.e. Eddington ratio).

To break the degeneracy in AGN luminosity and  disentangle the AGN LF into their underlying distribution functions, additional information on the SMBH mass has to be added. The BHMF and Eddington ratio distribution function (ERDF) provide additional observational constraints on models of galaxy evolution, BH growth and BH - galaxy co-evolution using empirical approaches \citep{Shankar:2013}, semi-analytic models \citep{Fanidakis:2012, Hirschmann:2012} or numerical simulations \citep{DiMatteo:2008,Hirschmann:2014,Sijacki:2014}. The ERDF also allows further constraints on BH accretion models and the triggering history of BHs \citep{Yu:2005, Hopkins:2009}. 

Additional constraints on  BH - galaxy co-evolution are obtained from the cosmic evolution of the relation between \mbh and bulge properties \citep{Croton:2006,Booth:2011,Dubois:2011,Angles:2013}. However, its observational determination is affected by sample selection effects \citep{Lauer:2007,Schulze:2011}. To model and account for selection effects that will bias observations, precise knowledge of the underlying distribution functions, including the BHMF, is required \citep{Schulze:2011, Schulze:2014}.

The observational determination of BHMF and ERDF requires a well-defined AGN sample, as it is the case for the LF determination, but in addition requires  a BH mass estimate for every object in the sample.  The most direct BH mass estimates for large statistical AGN samples use the 'virial' method to estimate \mbh from a single-epoch spectrum of a type-1 AGN with broad emission lines in their optical spectra. We here restrict our investigation to these broad line AGN and emphasize that our definition of an active BH only refers to this unobscured AGN population (we will try to correct for obscured AGN in section~\ref{sec:obscured}).

With calibrations available for the broad H$\alpha$, H$\beta$, \ion{Mg}{ii} and \ion{C}{iv} line \citep[e.g.][]{Vestergaard:2002,McLure:2004,Greene:2005,Vestergaard:2006,Trakhtenbrot:2012,Park:2013}, this enables the estimate of \mbh for large AGN samples out to high-$z$ \citep[e.g.][]{McLure:2004,Netzer:2007b,Shen:2008,Trump:2009,Shen:2011}. However, these 'virial'  BH masses are still subject to significant uncertainties and possible systematics \citep[e.g.][]{Vestergaard:2006,Shen:2013,Denney:2013}. While these are particularly important for individual objects, they are less critical for demographic studies of the AGN population, as will be carried out here.

The establishment of this 'virial' method for BH mass estimation allowed the empirical determination of the active BHMF. There have been two main approaches used for the determination of the BHMF and ERDF. The first is directly borrowed from the determination of the AGN LF, using the classical $1/V_\mathrm{max}$ estimator \citep{Schmidt:1968} with the same volume weights as for the AGN LF \citep{Wang:2006,Greene:2007,Vestergaard:2008,Vestergaard:2009,Schulze:2010,Shen:2012,Nobuta:2012}. We refer to this approach as luminosity-weighted $V_\mathrm{max}$ method.
While it provides a non-parametric and model-independent estimate of the distribution function, in general it suffers from severe incompleteness due to the improper volume weights which do not correct for active BHs below the flux limit of the survey \citep[][see also Appendix~\ref{sec:vmax}]{Kelly:2009,Schulze:2010}. Furthermore, it does not account for the uncertainty in the virial \mbh estimates. Therefore, in general it does provide a biased census of the active SMBH population.

For a proper treatment of the survey selection function, the BHMF and ERDF need to be determined jointly by fitting an analytic model for the bivariate distribution function of \mbh and $\lambda$ to the observations. This forward modelling of the distribution functions can be achieved via a maximum likelihood method, as developed by \citet{Schulze:2010}, or within a Bayesian framework  \citep{Kelly:2009}.

\citet{Schulze:2010} used the maximum likelihood method to determine the BHMF and ERDF at $z<0.3$ from the Hamburg/ESO Survey \citep{Wisotzki:2000}, establishing the local zero-point of these distribution functions. Their results support the notion of downsizing in BH mass. However, a full exploration of the growth history of BHs  requires the determination of BHMF and ERDF at higher redshifts. 
The SDSS QSO sample has been used to study the evolution of BH activity by determining the BHMF for $0.4<z<5$  \citep{Kelly:2010,Shen:2012,Kelly:2013}, employing a Bayesian framework \citep{Kelly:2009}. These authors also found evidence for BH mass downsizing in the BHMF and for $z>2$ also in the ERDF \citep{Kelly:2013}. However, since the SDSS only covers the bright end of the AGN LF, their results are largely limited to the high-mass end of the BHMF and to high Eddington ratios and are affected by large incompleteness at lower values, which introduces large uncertainties in this regime. To resolve these limitations and probe the BHMF and ERDF to lower masses and $\lambda$, deep AGN surveys have to be used. \citet{Nobuta:2012} presented the BHMF and ERDF  at $z\sim1.4$ based on an X-ray-selected sample from the Subaru \textit{XMM-Newton} Deep Survey \citep[SXDS;][]{Ueda:2008}, utilizing the maximum likelihood method of \citet{Schulze:2010}. Since deep surveys with spectroscopic follow-up are currently limited to small areas, they did not probe the bright end of the LF. Therefore, for a wide luminosity coverage and thus a wide coverage of BH mass and Eddington ratio, it is important to combine deep, small area surveys with shallower, large-area surveys.

In this paper, we follow this strategy. We are using two well-defined AGN samples from the deep, small-area surveys of the VVDS and zCOSMOS, and combine our results with the large area, shallow SDSS to constrain the BHMF and ERDF within the redshift range  $1.1<z<2.1$, covering a wide range in \mbh and $\lambda$. Our redshift range is chosen such that it covers the broad \ion{Mg}{ii} line in the spectra for all three samples.

In Section~\ref{sec:samples}, we present the individual samples we are using.  The BH masses and Eddington ratios of the three samples are presented in Section~\ref{sec:masses}.  In Section~\ref{sec:df}, we present our results for the bivariate distribution function. We discuss these results in Section~\ref{sec:discussion} and conclude in Section~\ref{sec:conclusions}. 
The appendix gives more details on the selection functions of the employed samples and present results on the BHMF, ERDF and AGN LF based on the $V_\mathrm{max}$ method.

Throughout this paper, we assume a Hubble constant of $H_0 = 70$ km s$^{-1}$ Mpc$^{-1}$ and cosmological density parameters $\Omega_\mathrm{m} = 0.3$ and $\Omega_\Lambda = 0.7$.

\section{The sample}  \label{sec:samples}
\subsection{The VVDS}
The VVDS is a pure magnitude-limited spectroscopic survey \citep{LeFevre:2005,LeFevre:2013,Garilli:2008}. The target selection is purely based on the $I$-band magnitude without any further selection criteria. It consists of a wide survey \citep{Garilli:2008}, spread over three fields, with an effective area of $4.5$~deg$^2$ for $17.5<I_\mathrm{AB}<22.5$ and a deep survey \citep{LeFevre:2005}, spread over two fields, with an effective area of $0.62$~deg$^2$ for $17.5<I_\mathrm{AB}<24.0$. The $I$-band imaging has been designed to be complete to much fainter magnitudes than these limits, with limiting magnitudes of $I_\mathrm{AB} = 24.8$ and $I_\mathrm{AB} = 25.3$ at $5\sigma$ in the wide and the deep fields, respectively \citep{McCracken:2003,LeFevre:2004}.
Targets have been nearly randomly selected for spectroscopy with the VIMOS multi-object spectrograph on the Very Large Telescope (VLT). The survey used the LRRED grism covering the spectral range $5500-9500$~\AA{} at a resolution of $R\sim230$, corresponding to a spectral resolution element of $\sim33$~\AA. Type-1 AGN have been manually identified based on the presence of broad emission lines (FWHM$>1000$~km s$^{-1}$) in their spectra. 
These simple selection criteria provide a well-defined AGN sample, free of a potential bias by pre-selection via colour and/or morphology.
Details on the type-1 AGN sample are given in \citet{Gavignaud:2006} and \citet{Gavignaud:2008}, while the type-1 AGN luminosity function, based on the 'epoch 1' sample is presented in \citet{Bongiorno:2007}.

We here use the AGN sample from VVDS 'epoch 2', presented in \citet{Gavignaud:2008}. It consists of 298 AGN, 235 with a secure redshift and 63 with a degenerate redshift, because only a single broad emission line without any other feature is present in the spectrum. BH masses based on \ion{Mg}{ii} have been measured for 86 out of 95 within $1.0<z<1.9$, for which  the data quality was sufficient to estimate \mbh, and are presented in \citet{Gavignaud:2008}. These 86 AGN form our VVDS-sample. We additionally included the redshift degenerate objects that could lie in our redshift range (61 AGN), measured their \mbh, assuming the broad line is \ion{Mg}{ii}, and statistically account for their contribution to the full sample (see Appendix~\ref{sec:selfunc}).

\subsection{zCOSMOS}
Like the VVDS, zCOSMOS is a magnitude-limited spectroscopic survey \citep{Lilly:2007} using VIMOS. It consists of zCOSMOS-bright \citep{Lilly:2009}, targeting $\sim 20000$ galaxies over the full COSMOS field and zCOSMOS-deep, targeting $\sim10000$ galaxies over the central 1 deg$^2$.

We here use the data set of the zCOSMOS-bright 20k spectroscopic catalogue. We do not include the type-1 AGN detected in the deep sample \citep{Bongiorno:2012,Rosario:2013}, since they do not constitute a well-defined sample, mainly due to the BzK colour pre-selection, targeting $z>1.5$ galaxies. Targets have been chosen from \textit{Hubble Space Telescope} ($HST$) $F814W$ imaging  \citep{Koekemoer:2007} with $15<I_\mathrm{AB}<22.5$ across the 1.7 deg$^2$ COSMOS field. The target selection is based on a combination of nearly random selection and compulsory targets. The latter were observed with a higher chance and were largely based on X-ray detection \citep{Hasinger:2007,Brusa:2010}, which strongly favours AGN. For the statistical analysis, this has to be taken into account by proper weighting (see Appendix~\ref{sec:selfunc}). Spectra have been obtained with the VIMOS multi-object spectrograph within the range $5500-9500$~\AA{} at a resolution of $R\sim580$.

Type-1 AGN are again identified by the presence of broad emission lines with FWHM$>1000$ km s$^{-1}$ in the spectra. Results on the zCOSMOS AGN sample from the 10k catalogue are presented in \citet{Merloni:2010}. 
BH masses from broad \ion{Mg}{ii} can be estimated within the redshift range $1.1<z<2.1$. zCOSMOS-bright contains 160 broad line AGN within this redshift range, for 145 of them we could measure a reliable \mbh. These AGN form our zCOSMOS-sample.

We note that the COSMOS field has a rich multi-wavelength coverage from X-rays to radio. This gives us, for example, a more precise handle on the bolometric luminosities and the host galaxy contribution of the AGN sample.

\begin{figure}
\centering 
\resizebox{\hsize}{!}{\includegraphics[clip]{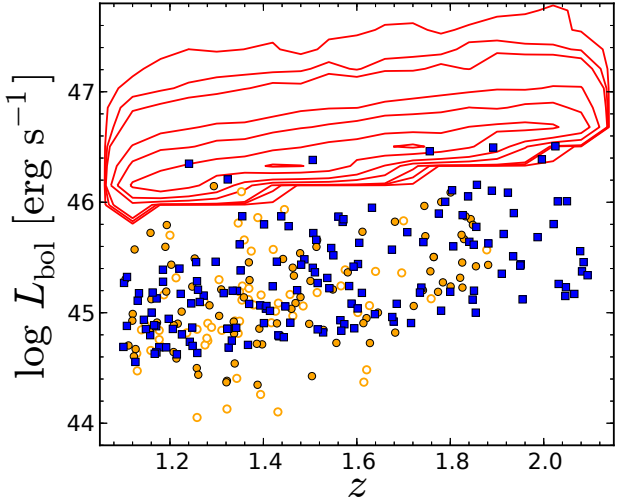}}
\caption{Redshift  versus bolometric AGN luminosity for the three AGN samples within the redshift range considered in this work, $1.1<z<2.1$. The orange circles, blue squares and the red contours are for the VVDS, zCOSMOS and SDSS sample, respectively (the VVDS sample is restricted to $1.1<z<1.9$ and redshift degenerate objects are shown as open circles).}
\label{fig:zdist}
\end{figure}

\subsection{SDSS}
For the SDSS sample, we largely follow the work of \citet{Shen:2012}. We start from the SDSS DR7 QSO sample \citep{Schneider:2010}, consisting of 105\,783 QSOs. We restrict this sample to the subset that has been selected by a uniform target selection algorithm \citep{Richards:2002} and thus form a well-defined broad line AGN sample \citep{Richards:2006a,Shen:2012}. The sample is magnitude limited to $15.0<i<19.1$ below $z=2.9$ and the target selection is based on multicolour information. The effective area of the survey is 6248 deg$^2$ \citep{Shen:2012}.

We further restrict the uniform QSO sample to the redshift range $1.1<z<2.1$, containing 28\,322 QSOs. BH masses, based on the broad \ion{Mg}{ii} line have been estimated by \citet{Shen:2011}. We reject \mbh estimates with measurement error larger than $0.5$~dex, leading to a sample of 27\,257 QSOs with available \mbh, forming our SDSS-sample.

\subsection*{}
In Fig.~\ref{fig:zdist} we show the redshift distribution against luminosity for the three samples within the redshift range we study in this work ($1.1<z<2.1$). 

The study of the demographics of the AGN population from these well-defined samples requires knowledge on the respective selection functions. In Appendix~\ref{sec:selfunc}, we provide details on the applied selection functions for the three surveys.

\section{Black hole masses and Eddington ratios}  \label{sec:masses}

\subsection{Black hole masses}
BH mass estimates from single epoch broad line AGN spectra can be obtained via the 'virial method'. This method employs the virial relation $\mbh= f R_{\mathrm{BLR}} \Delta V^2 / G$, where the broad line cloud velocity $\Delta V$ is derived from the broad emission line width, $f$ is a scale factor that accounts for the geometry and kinematics of the BLR (usually calibrated by scaling to the local \mbh$-\sigma_\ast$ relationship \citep[e.g.][]{Onken:2004,Woo:2010,Woo:2013}, and $R_\mathrm{BLR}$ is the broad line region size. While the latter can be directly measured via reverberation mapping \citep[e.g.][]{Blandford:1982,Peterson:2004}, for single epoch spectra, commonly the thereby-established scaling relation to the continuum luminosity \citep{Kaspi:2000, Bentz:2009, Bentz:2013} is used.

We (re-)computed BH masses consistently for all three samples using the formula from \citet{Shen:2011}:
\begin{equation}
\mbh= 10^{6.74} \left( \frac{L_{3000}}{10^{44}\,\mathrm{erg\,s}^{-1}}\right)^{0.62} \left( \frac{\mathrm{FWHM}}{1000\,\mathrm{km\,s}^{-1} }\right)^2 M_\odot
\end{equation} 

The \ion{Mg}{ii} broad line widths and the continuum luminosities at 3000~\AA{} have been measured for the SDSS sample by \citet{Shen:2011}. For the VVDS sample, they are provided by \citet{Gavignaud:2008}, but only for AGN with secure redshift. For the redshift degenerate AGN in the VVDS sample, we measured the \ion{Mg}{ii} full width at half maximum (FWHM) and $L_{3000}$ using exactly the same procedure as for the rest of the VVDS sample, assuming the broad line in the spectrum is indeed \ion{Mg}{ii}.

For the zCOSMOS sample, we used the VIMOS spectra to fit the broad lines. A few of the objects also have SDSS spectra and we find consistent results for these. The fitting procedure is similar to \citet{Gavignaud:2008} and outlined in \citet{Schramm:2013}.

For both, the VVDS and the zCOSMOS sample, the  \ion{Mg}{ii} line region is modelled by a pseudo-continuum, consisting of a power law and the broadened \ion{Fe}{ii} template from \citet{Vestergaard:2001} and one or two Gaussian components for the broad line (always two Gaussians for the VVDS AGN). The \ion{Mg}{ii} broad line FWHM was measured from the best-fitting line profile and the monochromatic continuum luminosity at 3000~\AA{} $L_{3000}$ is taken from the power-law fit. Details on the BH mass measurement for COSMOS AGN will be given in Schramm et al. \citep[in prep.; see also][]{Merloni:2010}.

\begin{figure}
\centering 
\resizebox{\hsize}{!}{\includegraphics[clip]{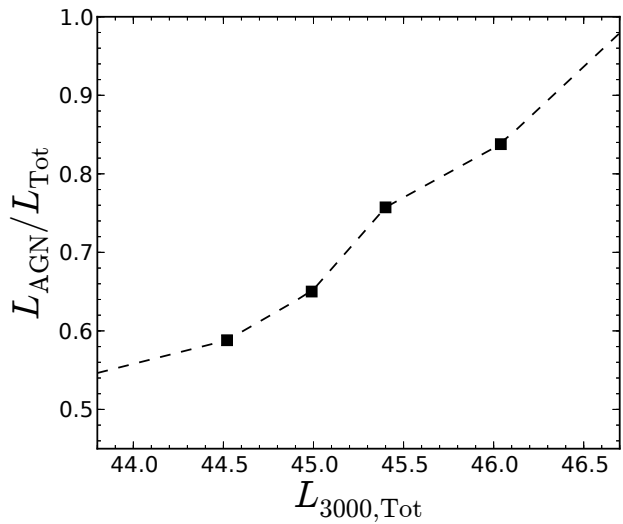}}
\caption{Average host galaxy contamination correction for the continuum luminosity at 3000~\AA{}, derived from the SED decomposition of COSMOS type-1 AGN. The black squares show the average correction for the sample in four bins of total $L_{3000}$, and the dashed line represents the interpolation for the average correction. We applied this correction to $L_{3000}$ for the VVDS and zCOSMOS samples.}
\label{fig:hostcorr}
\end{figure}

For the SDSS AGN sample, the host galaxy contamination to $L_{3000}$ is negligible at these redshifts \citep{Shen:2011}.  However, this is not the case for the deeper VVDS and zCOSMOS samples, thus $L_{3000}$  needs to be corrected for this contamination. For these samples, we applied an average host galaxy correction to $L_{3000}$. 
We used the spectral energy distribution (SED) decomposition of 428 type-1 AGN in the COSMOS field \citep{Bongiorno:2012} to compute the average host correction in several bins of total $L_{3000}$ and interpolated between the bins. The resulting average host correction as a function of $L_{3000}$ is shown in Fig.~\ref{fig:hostcorr} and was applied to the continuum luminosity at 3000~\AA{} for the VVDS and zCOSMOS AGN sample. While we may over- or underestimate the host contribution for individual objects, this will statistically account for the systematic effect of the host contribution. While the zCOSMOS AGN in our sample also have individual correction values, these are consistent with this average host correction and we decided to use the latter also for zCOSMOS, to ensure consistency between the two samples.

\begin{figure}
\centering 
\resizebox{\hsize}{!}{\includegraphics[clip]{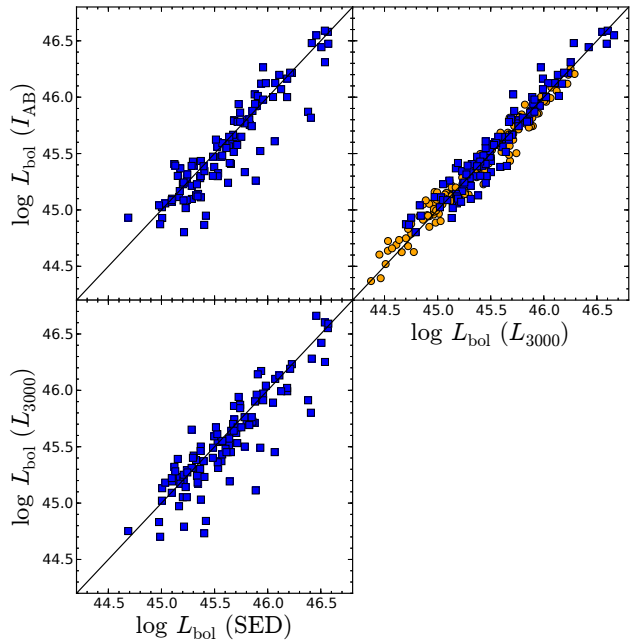}}
\caption{Comparison of bolometric luminosity estimates. The left-hand panels compare bolometric luminosity computed from $I_\mathrm{AB}$ (upper panel) and from $L_{3000}$ (lower panel) to the bolometric luminosity computed from the rest-frame SED  for the 107 AGN common to our zCOSMOS sample and the \textit{XMM}-COSMOS sample studied by \citet{Lusso:2012}. The right-hand panel compares the bolometric luminosities obtained from $I_\mathrm{AB}$ and from $L_{3000}$ with each other. In this panel, we show in addition to the zCOSMOS sample (blue squares) the VVDS sample (orange circles). The solid line shows the one-to-one relation.}
\label{fig:lbolcomp}
\end{figure}

\subsection{Bolometric Luminosities}     \label{sec:lbol}
For the SDSS sample we compute the bolometric luminosity from the $i$-band magnitude, following \citet{Shen:2012}. Since the SDSS selection function is defined by its $i$-band flux limit, this enables a direct mapping of the selection function from magnitude and redshift dependence to \mbh, $\lambda$ and $z$ dependence. This largely simplifies the determination of the BHMF and ERDF. Alternatively, either a distribution of $L_\mathrm{bol}$ at a given magnitude and redshift has to be assumed to determine $\Omega_\mathrm{e}(L_\mathrm{bol},z)$, or an individual selection function for every object has to be defined.

We apply the $K$-correction from \citet{Richards:2006a} to compute the absolute $i$-band magnitude normalized at $z = 2$ from the $i$-band flux and translate this to the continuum luminosity at 2500~\AA{} following \citet{Richards:2006a}. We compute the bolometric luminosity from this $L_{2500}$ estimate, assuming a constant correction factor $f_\mathrm{BC}(2500~\mathrm{\AA{}})=5$ \citep{Richards:2006b,Shen:2012}. The bolometric luminosity computed this way is on average fully consistent with the bolometric luminosity obtained directly from $L_{3000}$, by applying a constant bolometric correction factor of 5.15 to $L_{3000}$ \citep{Richards:2006b,Runnoe:2012}.

For the VVDS and the zCOSMOS sample, we  follow the same strategy of using the apparent magnitude to compute $L_{2500}$, and derive the bolometric luminosity via $L_\mathrm{bol}=5 L_{2500}$. This ensures consistency between the samples and minimizes possible systematics between the samples when combining the data sets. To compute $L_{2500}$, we adopt the $K$-correction from \citet{Richards:2006a} to compute $M_i(z=2)$, but transformed this $K$-correction to the CFHT/CFH12K $I$-band (VVDS) and the \textit{HST}/ACS $F814W$ filter (zCOSMOS), respectively, using the QSO template from \citet{VandenBerk:2001}. For these two filters, this transformation is small over our redshift range. We statistically correct $L_{2500}$ for host galaxy contribution, using the average host correction presented above (see Fig.~\ref{fig:hostcorr}). 
Alternatively, we also compute $L_\mathrm{bol}$ from the $L_{3000}$, as measured from the spectral fitting, applying a constant bolometric correction factor $f_\mathrm{BC}(3000~\mathrm{\AA{}})=5.15$  \citep{Richards:2006b}. We show in the upper-right panel of Fig.~\ref{fig:lbolcomp} that both estimates of $L_\mathrm{bol}$ give consistent results.

In both cases, we assume a constant bolometric correction factor. It has been suggested that the bolometric correction might depend on luminosity \citep{Marconi:2004,Hopkins:2007,Trakhtenbrot:2012}. While the bolometric correction in X-rays shows a clear dependence on luminosity or Eddington ratio \citep[e.g.][]{Marconi:2004,Vasudevan:2007,Jin:2012,Lusso:2012}, this dependence is weaker for the optical bolometric correction \citep[e.g.][]{Marconi:2004,Lusso:2012,Runnoe:2012}, with large uncertainties in the low-luminosity regime. Therefore, a constant bolometric correction is a good approximation over the luminosity regime we are probing.

A large fraction of our zCOSMOS sample is included in the X-ray-selected $XMM$-COSMOS catalogue \citep{Brusa:2010}. \citet{Lusso:2012} measured the bolometric luminosity directly for $XMM$-COSMOS AGN by integration of the rest-frame SED from X-rays to 1$\mu$m. In the left-hand panel of Fig.~\ref{fig:lbolcomp} we compare our bolometric luminosities, using both the $I_\mathrm{AB}$ and $L_{3000}$ prior to the host galaxy correction, with their results for the 107 AGN in common. We find good agreement of these $L_\mathrm{bol}$ estimates to  the direct integration results, which still include host galaxy light, presented in \citet{Lusso:2012}.

\begin{figure}
\centering 
\resizebox{\hsize}{!}{\includegraphics[clip]{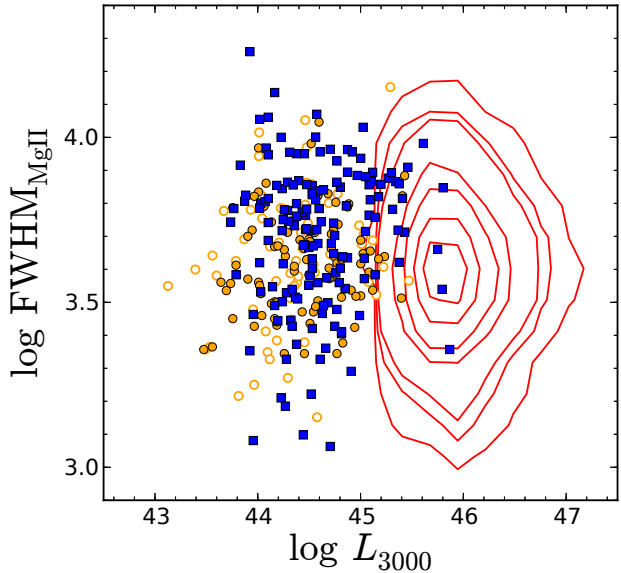}}
\caption{Distribution of the observable quantities 3000~\AA{} continuum luminosity and \ion{Mg}{ii} FWHM for the three samples. The orange circles show the VVDS sample, where filled circles are secure redshift objects and open circles are redshift degenerate objects. The blue squares represent the zCOSMOS sample and the red contours show the SDSS sample.}
\label{fig:lwplane}
\end{figure}

\begin{figure*}
\centering
        \includegraphics[width=16cm,clip]{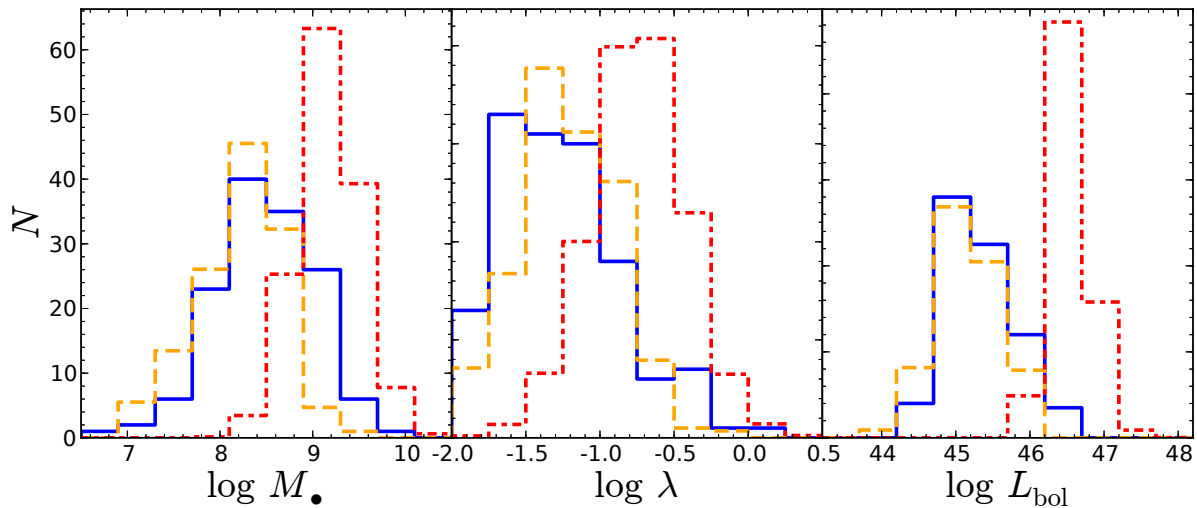}
\caption{Histograms of BH masses (left-hand panel), Eddington ratios (middle panel) and bolometric luminosities (right-hand panel) for the VVDS (orange dashed line), zCOSMOS (blue solid line) and SDSS (red dash-dotted line). The VVDS sample includes redshift regenerate objects with their respective weight. The SDSS sample is scaled down by a factor of 200.}
\label{fig:mer_hist}
\end{figure*}

\subsection{The mass-luminosity plane}   \label{sec:mbh_er_results}
In Fig.~\ref{fig:lwplane} we first show the bivariate distribution of the three AGN samples in the  $L_{3000}$-FWHM plane. This gives the distribution of the underlying observables for \mbh, $\lambda$ and $L_\mathrm{bol}$. We are using the term \textit{'distribution'}, to refer to the observed location of the sample properties in these kind of diagrams, while we use the term \textit{'distribution function' } to explicitly refer to the underlying demographic quantity, i.e. accounting for the selection effects that affect the observed distributions.

We here and in most other figures use the following colour and symbol convention: (1) the VVDS sample is shown by orange circles. Filled circles are objects with secure redshift and open circles indicate degenerate redshift objects. (2) The zCOSMOS sample is shown by blue squares. (3) The SDSS sample is represented by red contours. 

We see the distinct luminosity regimes occupied by the bright SDSS sample and the deep VVDS and zCOSMOS samples. For the VVDS sample, there is little overlap in luminosity with the SDSS sample, while the zCOSMOS sample extends a bit more into the high-luminosity regime. This is also illustrated in the bolometric luminosity histogram, shown in the right-hand panel of Fig.~\ref{fig:mer_hist}.  It demonstrates the need to combine deep and bright surveys for a wide luminosity coverage. On the other hand, the FWHM distributions for the three samples are largely consistent.\footnote{The zCOSMOS sample tends to show a slightly higher mean FWHM compared to SDSS and VVDS (with $\langle \log \mathrm{FWHM} \rangle = 3.70$ vs. $3.61$). We will investigate if this offset is intrinsic or caused by systematics in the spectral analysis in future work.}

In Fig.~\ref{fig:mer_hist} we compare the histograms of \mbh, $\lambda$ and $L_\mathrm{bol}$ for the VVDS, zCOSMOS and SDSS samples. 
The VVDS and zCOSMOS samples have almost consistent BH mass distributions, covering the range $10^7 \lesssim \mbh \lesssim 10^{9.4}$ and Eddington ratio distributions. Both samples extend about 1~dex lower in \mbh, compared to the SDSS sample, which approximately covers the range $10^8 \lesssim \mbh \lesssim 10^{10}$ and also probe lower Eddington ratios than the SDSS sample, down to 0.01 of the Eddington rate. Compared to the SDSS sample, they show a wider dispersion in their observed mass, luminosity and Eddington ratio distributions, consistent with previous studies on deep AGN samples \citep{Gavignaud:2008,Trump:2009,Merloni:2010}. We emphasize that these observed distributions do not constitute the underlying distribution functions, but are affected by the specific survey selection criteria.

\begin{figure*}
\centering
        \includegraphics[width=16cm,clip]{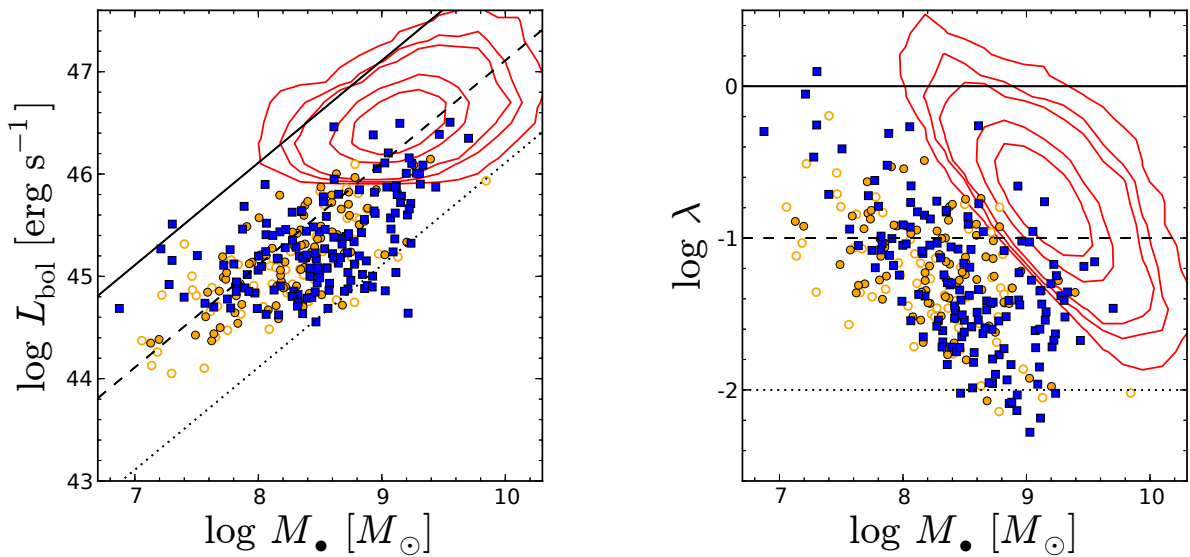}
\caption{Observed bivariate distribution for the three samples in the BH mass$-$luminosity plane (left-hand panel) and BH mass$-$Eddington ratio plane (right-hand panel). The red contours show the apparent distribution for the SDSS sample, the blue squares are for the zCOSMOS sample and the orange circles show the VVDS sample. For the latter, open circles indicate redshift degenerate objects. The black solid, dashed and dotted lines indicate Eddington ratios of 1, 0.1 and 0.01, respectively.}
\label{fig:mleplane}
\end{figure*}

This aids the interpretation of the observed mass$-$luminosity and mass$-$Eddington ratio  bivariate distribution, shown in Fig.~\ref{fig:mleplane}.
The red contours show the apparent bivariate distribution in $\mbh-L_\mathrm{bol}$ or $\mbh-\lambda$ of the SDSS sample. For the studied redshift range, the SDSS flux limit corresponds to an absolute luminosity limit of $\sim46.3$ erg s$^{-1}$, i.e. SDSS is not sensitive to AGN below this luminosity. This becomes evident when adding the VVDS and zCOSMOS samples, which smoothly fill in the lower luminosity parameter range missed by the SDSS. At the lowest masses and Eddington ratios the flux limits of VVDS and zCOSMOS also lead to an apparent absence of objects even in these deep samples. 

On the other hand, the lack of AGN at the highest masses and highest Eddington ratios even in the SDSS (upper-right corner in the right -hand panel) is physical and a direct consequence of the decrease of the space density both in the BHMF and the ERDF in this regime, as further shown below. The same holds for VVDS and zCOSMOS, while their much smaller volumes already lack moderately high \mbh and high $\lambda$ AGN, which necessitates the addition of the SDSS data.  
Combined with the flux limit, this causes the apparent anti-correlation between \mbh and $\lambda$ in the individual surveys, i.e. they cannot detect low mass, low Eddington ratio AGN below the flux limit, while high mass, high Eddington ratio AGN do not exist in these volumes.

For the combined sample, the observed distribution is approximately confined within the Eddington limit and $\lambda=0.01$, consistent with previous results on type-1 AGN \citep[e.g.][]{Kollmeier:2006,Trump:2009,Schulze:2010,Shen:2011}. The Eddington limit serves as an upper boundary for the accretion rate of a BH. The low observed number of $\lambda<0.01$ broad line AGN could be a physical effect, caused by the change in the accretion mode to a radiatively inefficient accretion flow, which leads to the disappearance of the BLR \citep[e.g.][]{Nicastro:2000,Yuan:2007,Trump:2011}. This change in accretion mode is thought to happen around $10^{-2}$-$10^{-3}$. Alternatively, the apparent limit could be driven by observational limitations to detect the corresponding very broad and low-luminosity lines beneath the spectral noise and the broad \ion{Fe}{ii} emission. 

The study of the \textit{active} BHMF, requires a clear definition of an active BH. In general, AGN show a broad range of levels of activity \citep[e.g.][]{Heckman:2004,Hickox:2009,Aird:2012,Bongiorno:2012}, down to accretion rates well below 0.01\citep[e.g.][]{Soria:2006,Ho:2008,Gallo:2010,Miller:2012}. A different definition of an active BH will naturally result in a different active BHMF \citep{Goulding:2010}.

In this study, we define an active BH as a type-1 AGN with an Eddington ratio $\lambda\geq0.01$, analogous to \citet{Schulze:2010}. The lower Eddington ratio limit is mainly a practical choice, corresponding to our observed range. While we observe AGN below this limit in our sample, their low number and restricted BH mass range does not allow us to put strong constraints on the ERDF below $\lambda=0.01$. This differs somewhat from the more standard definition via an AGN luminosity limit but we think is more physically motivated, in particular when studying the bivariate distribution of \mbh and $\lambda$. 
As shown in the  right-hand panel of Fig.~\ref{fig:mleplane}, for this definition of an active BH the SDSS sample suffers from severe incompleteness, while the deeper VVDS and zCOSMOS samples are much less affected. However, even these deep surveys become increasingly incomplete towards the lowest \mbh and lowest Eddington ratios.

Therefore, we restrict the three samples to $\mbh>10^7 \ M_\odot$ and $\lambda>0.01$ for the determination of the BHMF and ERDF,
leaving 147,137 and 27\,238 AGN in the samples from the VVDS, zCOSMOS and SDSS, respectively.

We further note that besides low accretion AGN our census does not include obscured (type-2) AGN. This is simply due to the fact that the virial method to estimate \mbh for large statistical samples is limited to unobscured (broad line) AGN. In section~\ref{sec:discussion}, we will make an attempt to account for the obscured population in the analysis.

\section{The active BHMF and ERDF}  \label{sec:df}

Early studies on the active BHMF focused on the $1/V_\mathrm{max}$ method, directly applying the volume weights from the AGN LF \citep{Greene:2007,Vestergaard:2008,Vestergaard:2009}. However, given our definition of an active BH, this approach introduces unaccounted sample incompleteness at the low-mass end of the BHMF, since in this mass range not the full Eddington ratio range is sampled, due to the  flux limit of the survey and the missing objects are not statistically accounted for \citep{Kelly:2009,Schulze:2010,Shen:2012}. Furthermore, this approach does not correct for the uncertainty in the virial BH mass estimates \citep{Kelly:2009,Shen:2012}. These limitations can be resolved by employing e.g. the maximum likelihood approach described in the next section.

On the other hand, if interpreted with the necessary caution, the determination of the binned BHMF and ERDF via the $1/V_\mathrm{max}$ method can serve as a useful tool, to guide the more refined parametric analysis presented in the following. We present and discuss our results for the BHMF and ERDF as well as for the AGN LF, using the  $1/V_\mathrm{max}$ method, in Appendix~\ref{sec:vmax}.

\subsection{The maximum likelihood method} \label{sec:ml}
A proper determination of the \textit{intrinsic} BHMF and the ERDF requires a joint modelling of these two distribution functions to take the selection function fully into account. In \citet{Schulze:2010}, we developed a maximum likelihood approach to  estimate the \textit{intrinsic} BHMF and ERDF simultaneously. We here present a modified framework, based on our previous work, but generalized to more flexible models for the bivariate distribution function, more diverse selection functions and adapted to the combination of several different surveys.
A further improvement, compared to \citet{Schulze:2010}  is the inclusion of a correction for virial BH mass uncertainties.

We aim at estimating the bivariate distribution function of BH mass end Eddington ratio $\Psi(\mbh,\lambda,z)$, i.e. $\Psi(\mbh,\lambda,z) \dd \log \mbh \dd \log \lambda$ gives the space density of active BHs with BH mass between $\log M_\bullet$ and $\log M_\bullet+\dd\log M_\bullet$ and Eddington ratio between $\log \lambda$ and $\log \lambda+\dd\log \lambda$ at the redshift $z$. The BHMF, ERDF and AGN LF are all different projections of this bivariate distribution function.

The bivariate distribution function can be also expressed as a function of \mbh and $L_\mathrm{bol}$, $\Psi(\mbh,L_\mathrm{bol},z)$. Marginalizing this over \mbh gives the bolometric AGN LF, i.e.
\begin{equation}
\Phi( L_\mathrm{bol},z)= \int\, \Psi(\mbh,L_\mathrm{bol},z) \dd\log M_\bullet \  .  \label{eq:lfbol}
\end{equation}

The survey selection modifies the \textit{intrinsic} distribution function $\Psi(\mbh,\lambda,z)$ to an \textit{observed} distribution
\begin{equation} 
\Psi_\mathrm{o}(\mbh,\lambda,z)= \Omega(\mbh,\lambda,z) \Psi(\mbh,\lambda,z) \ ,
\end{equation} 
where $\Omega(\mbh,\lambda,z)$ is the selection function of the survey. 

The maximum likelihood technique aims at minimizing the likelihood function $S=-2\ln \mathcal{L}$, where $\mathcal{L}= \prod_{i=1}^N p_i$ is the product of the individual likelihoods for the observed objects \citep{Marshall:1983}. 
This multivariate probability distribution $p(\mbh,\lambda,z)$, i.e. the probability  to observe an AGN with BH mass \mbh, Eddington ratio $\lambda$ and redshift $z$, is given by the normalized observed bivariate distribution:
\begin{equation} 
p_i(M_\bullet,\lambda,z) = \frac{1}{N_i}\Omega_i(M_\bullet,\lambda,z)\, \Psi(M_\bullet,\lambda,z)\, \frac{\dd V}{\dd z} \ ,  \label{eq:pi}
\end{equation} 
where $\Omega_i(M_\bullet,\lambda,z)$ is the selection function for the $i$ th object and 
\begin{equation} 
N_i= \iiint \Omega_i(M_\bullet,\lambda,z) \Psi(M_\bullet,\lambda,z) \frac{\dd V}{\dd z} \dd \log M_\bullet\, \dd \log \lambda\, \dd z   \label{eq:ni}
\end{equation} 
is the normalization for the $i$ th object . If the selection function is the same for all objects, also $N_i$ will be the same for them and corresponds to the expected number of objects from the model. However, the method is also applicable for a selection function which varies from object to object. In this case, $N_i$ corresponds to the total number of objects, assuming all objects would have the selection function $\Omega_i(M_\bullet,\lambda,z)$.

Using the probability distribution given by Equation~\ref{eq:pi}, we minimize the function:
\begin{equation}
S=-2  \sum_{i=1}^N \left[ \ln \Psi(\mbh_{i},\lambda_{i},z _{i}) - \ln N_{i} \right] \ ,
\end{equation} 
where  $N$ is the number of objects in the respective survey. This maximum likelihood method represents a forward modelling approach. We have to assume a specific parametric model for the bivariate distribution function and we fit the probability distribution predicted by this model in the $M_\bullet-\lambda$ plane to the observations.

A drawback in this approach is that the normalization $\Psi_\ast$ of the bivariate distribution function cannot be determined directly in the fit. We therefore determined it in a second step by integrating over the best-fitting model and scaling the predicted number of objects to the observed number,
 \begin{equation}
 \Psi_\ast=\frac{N_\mathrm{obs}}{N_\mathrm{mod}} 
  \ . \label{eq:psiast}
 \end{equation} 

We will first determine the bivariate distribution function for each survey individually, but our main analysis is based on the combined data set from the three surveys. We combine multiple surveys with their respective effective area and survey depth using the approach proposed by \cite{Avni:1980}.
The overlap in area and luminosity between SDSS, zCOSMOS and VVDS is marginal and there are no objects in common to several samples. Therefore, these three surveys can be treated as being independent. 

The combined selection function follows as $\Omega_\mathrm{e}(\mbh,L_\mathrm{bol},z) = \sum_{j=1}^{N_j} \Omega^j_\mathrm{e}(\mbh,L_\mathrm{bol},z)$, where $N_j$ is the number of combined surveys and is adopted for the combined dataset. Thus, it is assumed that each object with its given luminosity, redshift and BH mass could potentially have been detected in each of the surveys,  if it otherwise satisfies the respective selection function of that survey.

Since the selection function, specifically the spectroscopic success rate (see Appendix~\ref{sec:selfunc}), is initially defined as a function of magnitude $m$ and redshift $z$, we need to map $\Omega(m,z) \rightarrow \Omega(\mbh,\lambda,z)=\Omega(\mbh,L_\mathrm{bol},z)$. We here follow the approach by \citet{Shen:2012} and \citet{Kelly:2013}, by computing $L_\mathrm{bol}$ from the optical magnitude, as discussed in Section~\ref{sec:lbol}. This ensures a direct and consistent mapping from magnitude to bolometric luminosity and therefore from $\Omega(m,z)$ to $\Omega(\mbh,L_\mathrm{bol},z)$.

For the determination of the underlying distribution function it has to be noted that virial BH masses are thought to have an error of $\sim0.2-0.4$~dex \citep{Vestergaard:2006,Park:2012}, as derived from the scatter around reverberation mapping masses. We here account for this uncertainty, an improvement compared to \citet{Schulze:2010}\footnote{The results for the local BHMF and ERDF presented in \citet{Schulze:2010} only marginally change if we correct for the uncertainty in virial BH masses, assuming $\sigma_\mathrm{VM}=0.2$. The main effect is a slight decrease at the high accretion end of the ERDF \citep[see][]{Schulze:2011b}.}, by convolving  $\Psi(M_\bullet,\lambda,z)$ with the measurement uncertainty before comparing the model distribution to the data:
\begin{equation}
\Psi_\mathrm{e}(M_{\bullet,\mathrm{e}},\er_\mathrm{e},z) = 
\int g(M_{\bullet,\mathrm{e}},\er_ \mathrm{e}\mid M_\bullet,\er)\, \Psi(M_\bullet,\er,z) \, \dd \log M_\bullet \, \dd \log \er \ . \label{eq:pobs}
\end{equation}
We assume a log-normal scatter distribution in $M_\bullet$ and $L_\mathrm{bol}$,
\begin{equation}
g(M_{\bullet,\mathrm{e}},\er_ \mathrm{e}\mid M_\bullet,\er)=\frac{1}{2\pi \sigma_\mathrm{VM} \sigma_\mathrm{bol}} \exp\left\lbrace -\frac{(\mu_\mathrm{e}-\mu)^2}{2 \sigma_\mathrm{VM}^2} -\frac{(l_\mathrm{e}-l)^2}{2 \sigma_\mathrm{bol}^2} \right\rbrace
 \, , \label{eq:gobs}
\end{equation}
where $\mu=\log M_\bullet$ is the BH mass and $l=\log L_\mathrm{bol}$ is the bolometric luminosity.
For the uncertainty in the bolometric correction $\sigma_\mathrm{bol}$ we assumed a scatter of 0.05~dex \citep{Marconi:2004}, and for the uncertainty in the virial mass $\sigma_\mathrm{VM}=0.2$~dex, consistent with previous work \citep{Kelly:2010,Kelly:2013}. The commonly assumed measurement uncertainty associated with an individual viral mass estimate is $\sim0.3-0.4$~dex. However, for our purposes we are only interested in the effect of uncertainties on the apparent distribution of \mbh and $\lambda$ of our statistical sample, i.e. by which amount these distributions are actually broadened by using viral masses instead of true masses. Several studies suggest this value is actually smaller, $\sim0.2$~dex \citep{Kollmeier:2006,Fine:2008,Steinhardt:2010, Kelly:2010}. For a detailed discussion of this difference, its possible origin and consequences see e.g. \citet{Shen:2013}. Adopting a large value for $\sigma_\mathrm{VM}$ corresponds to a significant correction to the distribution functions, in particular when this value becomes comparable to the actual width of either the observed \mbh or $\lambda$ distribution. In this sense, our choice of $\sigma_\mathrm{VM}=0.2$~dex for the scatter in the virial method can be considered as a conservative value, since it corresponds to the smallest correction suggested by our current knowledge of the virial mass uncertainties.

\begin{figure*}
\centering
        \includegraphics[width=16cm,clip]{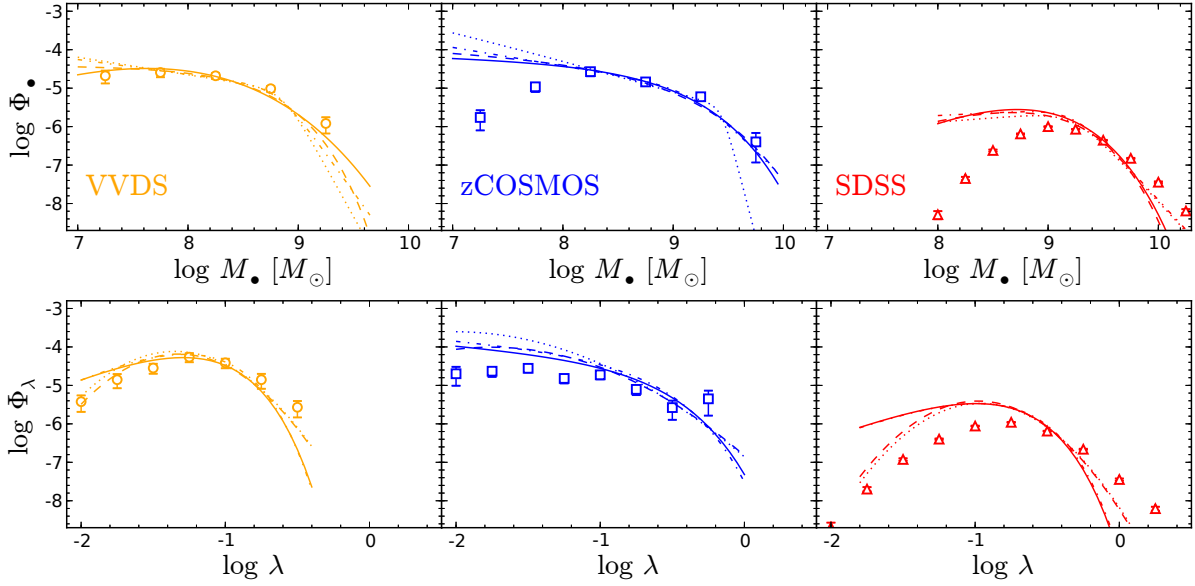}
\caption{Best-fitting maximum likelihood results for the BHMF (upper panels) and ERDF (lower panels) using different parametric models. We show the different distribution functions for our three surveys independently, in the left-hand panels for the VVDS, in the middle panels for zCOSMOS and in the right-hand panels for SDSS. We used four different models for the bivariate distribution function: a modified Schechter function for the BHMF with either a  Schechter function (solid line) or a log-normal function (dashed line) for the ERDF, or a double power law for the BHMF, again either with a  Schechter function (dash-dotted line) or a log-normal function (dotted line) for the ERDF. We also show the luminosity weighted $V_\mathrm{max}$ results with the open circles (VVDS), squares (zCOSMOS) and triangles (SDSS). Note that the  maximum likelihood results are not a fit to these binned data, but correct for incompleteness in the binned estimate as well as for uncertainty in the virial mass estimates. The former causes the deviation between our best fits and the bins at low \mbh / low $\lambda$, while the latter is responsible for the difference at high \mbh / high $\lambda$.}
\label{fig:pamodels}
\end{figure*}

\subsection{Parametric model}  \label{sec:model}
For the maximum likelihood fitting, we have to assume a specific parametric model for the bivariate distribution function $\Psi(M_\bullet,\er,z) $. We assume a BHMF independent of Eddington ratio, while we allow for a mass dependence in the ERDF. This is motivated by the results of \citet{Kelly:2013}, who find a mass dependence in the ERDF when using virial BH mass estimates based on \ion{Mg}{ii}, as we do in this study.
\begin{equation}
\Psi(M_\bullet,\er,z) = \rho_\er(\er,M_\bullet,z)\,\rho_\bullet(M_\bullet,z)\rho_z(z) \ , \label{eq:psi_me}
\end{equation}
where $\rho_\er(\er,M_\bullet,z)$ is an ERDF term, $\rho_\bullet(M_\bullet,z)$ is the BHMF term and $\rho(z)$ is a redshift evolution term. The BHMF is given by integration of the bivariate distribution function over $\lambda$,
\begin{equation}
\Phi_\bullet(M_\bullet,z) = \int \Psi(M_\bullet,\er,z)\, \dd \log \er  \ .
\end{equation}
Equivalently, the ERDF is given by integration of the bivariate distribution function over $M_\bullet$,
\begin{equation}
\Phi_\er(\er,z) = \int \Psi(M_\bullet,\er,z)\, \dd \log M_\bullet \ . 
\end{equation}

We will start with the simplified assumption of a redshift-independent BHMF and ERDF within the given redshift bin.  In particular for a narrow redshift bin this is a valid assumption, while we would expect evolution over the full range $1.1<z<2.1$. We will investigate redshift evolution in the bivariate distribution function within the full redshift range we are probing in more detail further below.

For the BHMF we use two different parametric models, that have been shown to be adequate to describe the low-$z$ BHMF \citep{Schulze:2010}. First, we use a modified Schechter function \citep[e.g.][]{Aller:2002,Schulze:2010}, given by
\begin{equation}
\rho_\bullet(M_\bullet) = \frac{\Psi^*}{\log_{10}e} \left( \frac{M_\bullet}{M_\bullet^*} \right)^{\alpha+1} \exp \left( - \left[ \frac{M_\bullet}{M_\bullet^*}\right]^{\beta}  \right)  \  .  \label{eq:modschecht}
\end{equation}
Secondly, we also utilize a double power law fit for the BHMF, given by
\begin{equation}
\rho_\bullet(M_\bullet) = \frac{\Psi^*/\log_{10}e}{(M_\bullet / M_\bullet^*)^{-(\alpha+1)} + (M_\bullet / M_\bullet^*)^{-(\beta+1)} } \ . \label{eq:dpl}
\end{equation}
For the ERDF we use a Schechter function \citep{Schechter:1976} as our reference model, where we allow for a mass dependence in the break Eddington ratio:
\begin{equation}
\rho_\lambda(\lambda,M_\bullet) = \frac{1}{\log_{10}e} \left( \frac{\lambda}{\lambda_*(M_\bullet)} \right)^{\alpha_\lambda+1} \exp \left( -\frac{\lambda}{\lambda_*(M_\bullet)} \right)  \ . \label{eq:schecht}
\end{equation}
We parametrize the mass dependence in $\lambda_*$ by
\begin{equation}
\log \lambda_*(M_\bullet) = \log \lambda_{*,0}+k_{\lambda} (\log M_\bullet-\log M_{\bullet,c}) \  , \label{eq:lambdaast}
\end{equation}
where we fixed $\log M_{\bullet,c}=8.0$. We also tested a second-order polynomial for the mass dependence in $\lambda_*$, but found the best-fitting second-order term parameter to be consistent with zero.

Additionally, we tested a log-normal distribution, restricted at our lower Eddington ratio limit $\log \lambda=-2$:
\begin{equation}
\rho_\lambda(\lambda,M_\bullet) = \frac{1}{\log_{10}e \sqrt{2\pi} \sigma_\lambda} \exp \left( -\frac{(\log\lambda-\log\lambda_*(M_\bullet))^2}{2 \sigma_\lambda^2} \right)   \ ,\label{eq:lognorm}
\end{equation}
where we also use equation~(\ref{eq:lambdaast}) for the mass dependence in the peak value $\lambda_*(M_\bullet)$.

We parametrize the redshift evolution in the bivariate distribution function of BHMF and ERDF by a density evolution term, where the normalization evolves in a log-linear manner:
\begin{equation}
\rho_z(z) =10^{\gamma (z-z_c)}  \  . \label{eq:pde}
\end{equation}
We fixed $z_c=1.6$ to the central value of our redshift range.

The absolute normalization of the BHMF and ERDF is obtained by integrating $\Psi(M_\bullet,\er,z)$ over  $\lambda$ or \mbh respectively. We chose integration intervals of $-2<\log \lambda <1$ and $7<\log M_\bullet<11$.
Our reference model has seven free parameters determined in the fitting ($M_\ast$, $\alpha$, $\beta$, $\lambda_{\ast,0}$, $\alpha_\lambda$, $k_{\lambda}$,  $\gamma$) plus the normalization $\Psi_\ast$ determined by Equation~(\ref{eq:psiast}).

\begin{figure*}
\centering
        \includegraphics[width=16cm,clip]{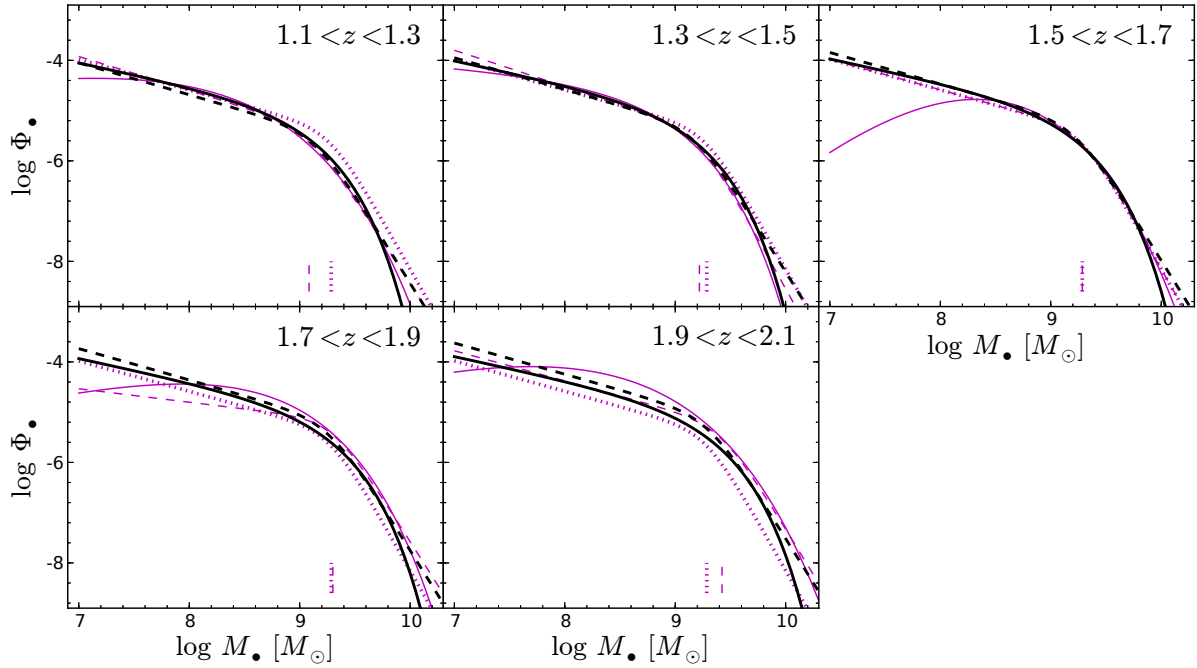}
\caption{Active BHMF for the combined data set from VVDS, zCOSMOS and SDSS in narrow redshift bins over the range $1.1<z<2.1$, based on our maximum likelihood fitting approach. The magenta lines show the best-fitting results within the narrow redshift bin only, while the black lines give the best-fitting model to the full range $1.1<z<2.1$ evaluated at the central redshift of the individual $z$-bin. The solid lines are for a model parametrization with a modified Schechter function BHMF and the dashed lines for a double power-law BHMF. The magenta dotted line indicates the best-fitting double power-law BHMF model for the redshift bin $1.5<z<1.7$, shown for reference. The vertical magenta dotted line and magenta dashed line indicate the position of the break of the BHMF at $1.5<z<1.7$ and in the respective $z$ bin.}
\label{fig:bhmf_zbin}
\end{figure*}

\begin{figure*}
\centering
        \includegraphics[width=16cm,clip]{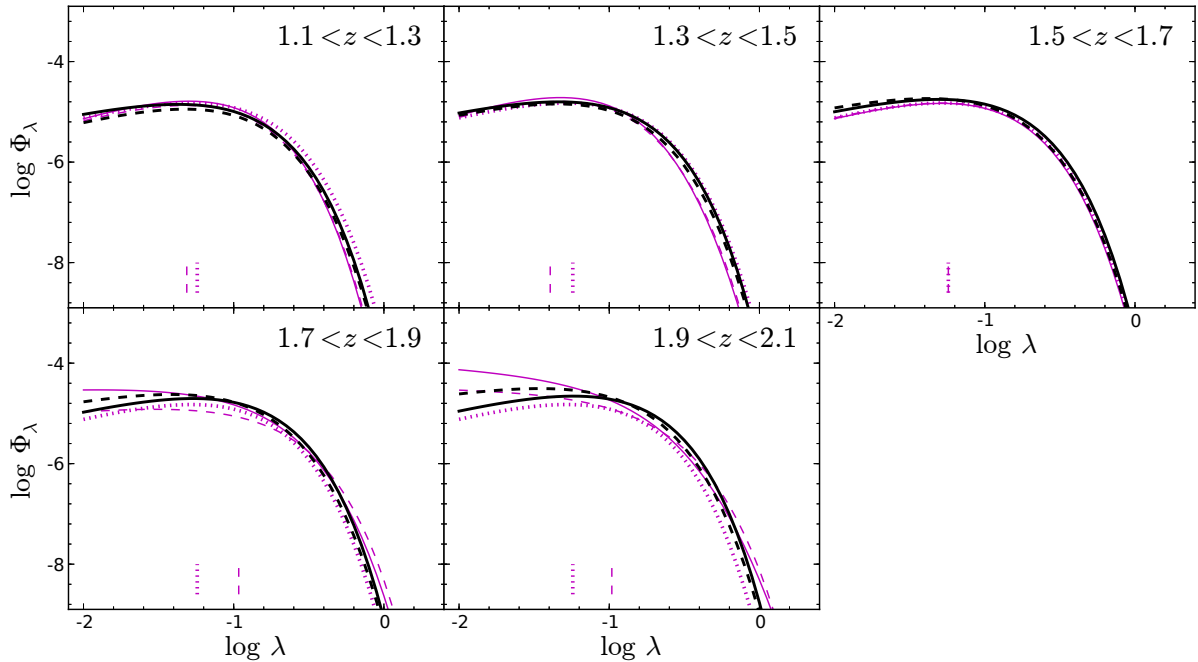}
\caption{ERDF for the combined data set from VVDS, zCOSMOS and SDSS in narrow redshift bins over the range $1.1<z<2.1$, based on our maximum likelihood fitting approach. The different lines are the same as in Fig.~\ref{fig:bhmf_zbin}.}
\label{fig:erdf_zbin}
\end{figure*}

\begin{figure*}
\centering
        \includegraphics[width=16cm,clip]{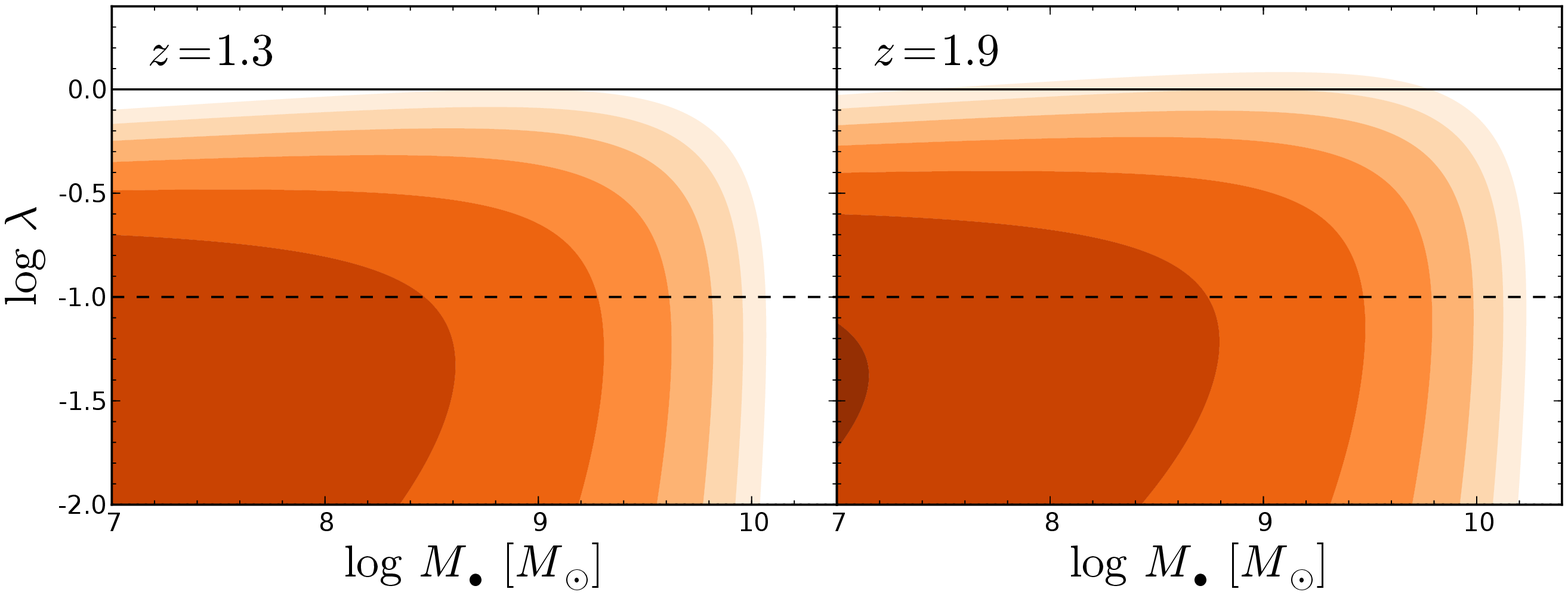}
\caption{Bivariate distribution function of BH mass and Eddington ratio $\Psi(\mbh,\lambda)$ for our best-fitting modified Schechter function BHMF model at two redshifts. The contours show lines of constant space density, from $10^{-10}$ to $10^{-5}$, separated by a factor of 10 each.}
\label{fig:pmodel}
\end{figure*}

\subsection{Results} \label{sec:results}
We start by fitting the bivariate distribution function of BHMF and ERDF to the three surveys independently, before combining the data sets. We fit the data with the maximum likelihood approach outlined in section~\ref{sec:ml}, using the parametric models presented in section~\ref{sec:model} over the full redshift range $1.1<z<2.1$. At this stage, we ignore any possible redshift evolution in $\Psi(\mbh,\er,z)$, apart from the density evolution term (equation~\ref{eq:pde}). We will postpone a more detailed investigation of  evolution within our redshift range to the analysis of the combined sample. We will comment on the results for the SDSS in narrower redshift bins in section~\ref{sec:comp}, when we directly compare our results to the work by \citet{Kelly:2013}.

The results for the BHMF and ERDF for four different parametric models for each survey are shown in Fig.~\ref{fig:pamodels}. We use a modified Schechter function or a double power law for the BHMF and a Schechter function or a (truncated) log-normal distribution for the ERDF. We also show the binned luminosity-weighted $V_\mathrm{max}$ results for reference as open symbols. We emphasize that the maximum likelihood results are not a fit to the binned distribution functions, but account for their inherent limitations, as discussed above. The binned distribution functions are only shown for reference, keeping their limitations in mind.

As expected, for the SDSS sample we find a large completeness correction compared to the luminosity-weighted $1/V_\mathrm{max}$ result, due to the bright flux limit of the sample. Contrary, for both VVDS and zCOSMOS, the completeness corrections are comparatively small. Since the VVDS sample consists  of the wide $I<22.5$~mag and the deep $I<24$~mag fields, its completeness correction is the smallest, also compared to zCOSMOS, which only has a  $I<22.5$~mag field. Note that at the high \mbh / high-$\lambda$ end, the best fit falls off steeper than the binned estimate, since the latter does not account for the virial SMBH mass uncertainties.

The four parametric model combinations give largely consistent results. Differences occur mainly in regions with poor statistics and/or where large completeness corrections have to be applied. This verifies that our results are robust to the specific assumed parametric model as long as the completeness corrections are only moderate. This applies in particular over the range for which the luminosity-weighted $V_\mathrm{max}$ results are consistent with the maximum likelihood model. The SDSS data set shows a larger dependence on the parametric model in the range with significant completeness corrections, i.e. $M_\bullet<10^9\,M_\odot$ and $\lambda<0.1$, demonstrating the limitation of the SDSS alone to securely constrain the low-mass end of the BHMF and the low Eddington ratio end of the ERDF at these redshifts. 

In the following, we will restrict our analysis to the Schechter function ERDF and  will not further consider the log-normal ERDF model. The Schechter function model is more flexible, since it allows for a turnover at low $\lambda$, but does not enforce it, in contrast to the log-normal model. Furthermore, it ensures an exponential decrease of the space density towards the Eddington limit, while in the log-normal model this decrease is in general flatter than exponential. We will utilize two parametric models: always a Schechter function for the ERDF and either a modified Schechter function or a double power law for the BHMF.

These results on the individual surveys demonstrate their respective strengths and weaknesses. The SDSS is powerful in constraining the high mass end of the BHMF and the high-$\lambda$ end of the ERDF, while the uncertainties are large at lower \mbh and $\lambda$, due to the associated large completeness corrections required. On the other hand, VVDS and zCOSMOS can reliably constrain the low \mbh and $\lambda$ regime BHMF and ERDF, while they fail at the high \mbh and high-$\lambda$ end, due to the poor statistics in this regime caused by the relative small area of these surveys. Therefore, in the next step we combine all three surveys to constrain the BHMF and ERDF over a wide range in \mbh and $\lambda$. For the combined fit, we restrict the SDSS sample to the BH mass range $\log \mbh>8.5$, to reduce the number of objects associated with the largest completeness corrections.

To test for redshift evolution in the bivariate distribution function $\Psi(\mbh,\lambda)$ within our redshift range, we first determined the best-fitting maximum likelihood model in narrow bins of redshift. The results for our two reference parametric models are shown in Fig.~\ref{fig:bhmf_zbin} and \ref{fig:erdf_zbin} by the magenta solid line (modified Schechter function BHMF) and the magenta dashed line (double power-law BHMF), respectively. Both parametric models agree reasonably well for the ERDF and for the BHMF at $\log \mbh>8$. At lower masses, the two models show a stronger deviation, indicating the larger uncertainty of the BHMF in this regime in the narrow $z$ bins, due to the relative small number of objects at $\log \mbh<8$ in zCOSMOS and VVDS. At this point, we are mainly interested in any detectable trends of evolution between the $z$ bins. With the magenta dotted line, we indicate the distribution function for the $1.5<z<1.7$ bin for reference in all panels. In addition, we mark the position of the break in the distribution functions at $1.5<z<1.7$  and the respective redshift bin by the vertical dotted and dashed lines in all panels.
We identify two main trends of evolution between the narrow $z$ bins. First, the break of the BHMF seems to shift to higher \mbh with increasing redshift, and secondly, the break of the ERDF seems to move to higher $\lambda$ with increasing redshift. To allow our parametric model to accommodate such trends, we implement a log-linear redshift dependence for the break of the BHMF and for the break in the ERDF:
\begin{equation}
\log M_*(z) = \log M_{*,0}+c_{\bullet} (z-z_c) \  , \label{eq:mbh_zevo}
\end{equation}

\begin{equation}
\log \lambda_*(M_\bullet,z) = \log \lambda_{*,0}+k_{\lambda} (\log M_\bullet-\log M_{\bullet,c}) +c_{\lambda} (z-z_c) \  , \label{eq:lambd_zevo}
\end{equation}
Furthermore, we allow for redshift evolution in the faint end slope of the ERDF:
\begin{equation}
\log \alpha_{\lambda}(z) = \alpha_{\lambda,0}+c_{\alpha_\lambda} (z-z_c) \  . \label{eq:alphalam_zevo}
\end{equation}

\begin{table*}
\caption{Best-fitting model parameters for the bivariate distribution function of BH mass and Eddington ratio, i.e. the BHMF and ERDF.}
\label{tab:dfpara}
\centering
\begin{tabular}{lcccccccccccc}
\hline \hline \noalign{\smallskip}
Sample & Model & $\log (\Phi^\ast)$  & $M_\bullet^\ast$ & $\alpha$ & $\beta$ & $\lambda_{\ast,0}$ & $\alpha_\lambda$ & $k_\lambda$  & $\gamma$ & $c_\bullet$ & $c_\lambda$ & $c_{\alpha_\lambda}$ \\ 
\noalign{\smallskip} \hline \noalign{\smallskip}
SDSS: $z=0.6$ & mod.S+S & -4.68 &  7.69 & -1.30 &  0.50 & -1.01 & -1.42 & 0.146 &  0.00 & $-$ & $-$ & $-$ \\ \noalign{\smallskip} 
SDSS: $z=0.8$ & mod.S+S & -4.88 &  8.06 & -1.19 &  0.57 & -1.02 & -1.09 & 0.074 &  0.00 & $-$ & $-$ & $-$ \\ \noalign{\smallskip} 
SDSS: $z=1.0$ & mod.S+S & -4.53 &  8.16 & -1.21 &  0.59 & -1.21 & -0.87 & 0.155 &  0.00 & $-$ & $-$ & $-$ \\ 
\noalign{\smallskip} \hline \noalign{\smallskip}
SDSS & mod.S+S & -8.73 &  5.83 &  1.50 &  0.33 & -1.17 & -0.23 & 0.068 &  0.03 & 0.275 & 0.086 & 0.230 \\ \noalign{\smallskip} 
zCOSMOS & mod.S+S & -4.73 &  7.28 & -0.67 &  0.37 & -0.84 & -1.44 & -0.107 &  0.12 & 0.292 & 0.148 & 0.048 \\ \noalign{\smallskip} 
VVDS & mod.S+S & -4.38 &  7.41 & -0.46 &  0.47 & -1.55 &  0.72 & 0.032 & -0.22 & -0.014 & -0.325 & 0.521 \\ 
\noalign{\smallskip}  \hline \noalign{\smallskip}
Combined & mod.S+S & -5.32 &  9.09 & -1.50 &  0.96 & -1.19 & -0.29 & 0.099 &  0.05 & 0.270 & 0.100 & 0.094 \\ \noalign{\smallskip} 
Combined & DPL+S & -5.64 &  9.23 & -1.68 & -4.56 & -1.24 & -0.36 & 0.148 &  0.31 & 0.198 & 0.114 & -0.423 \\ 
\noalign{\smallskip} \hline
\end{tabular}
\end{table*}

We fixed $z_c=1.6$ for all three equations. With these three additional free parameters, we fit our 10 parameter models for the bivariate distribution function of \mbh and $\lambda$ to the data. The best-fitting results for the modified Schechter function BHMF (black solid line) and double power-law BHMF (black dashed line) models are shown in Figs~\ref{fig:bhmf_zbin} and \ref{fig:erdf_zbin} and the best-fitting parameters are given in Table~\ref{tab:dfpara}. We find good agreement between the two parametric models. 
We will use the modified Schechter function model as our reference model for further discussion, without implying any clear preference for either of these parametrizations. We comment if any differences arise from the two different model implementations.
In Fig.~\ref{fig:pmodel} we directly show the bivariate distribution function of BH mass and Eddington ratio $\Psi(\mbh,\lambda)$ for our best-fitting modified Schechter function model at two redshifts.

\begin{figure}
\centering 
\resizebox{\hsize}{!}{ \includegraphics[clip]{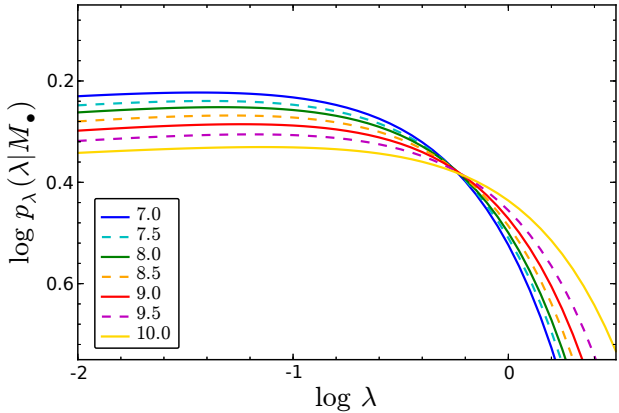}}
\caption{Dependence of the ERDF on BH mass. The different lines show the conditional ERDF at different BH masses, from $10^7$ to $10^{10} M_\odot$.}
\label{fig:erdf_m}
\end{figure}

\begin{figure*}
\centering
        \includegraphics[width=16cm,clip]{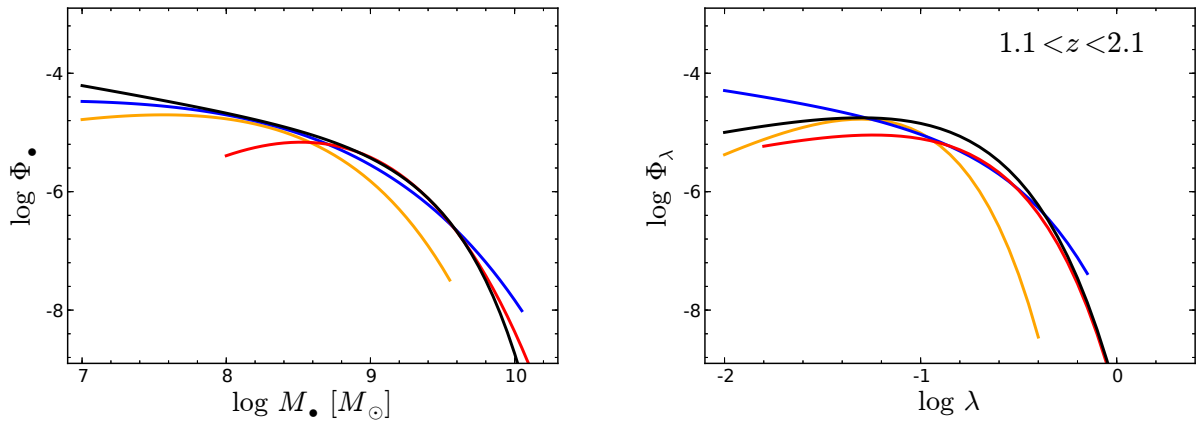}
\caption{Comparison of our best-fitting BHMF (left-hand panel) and ERDF (right-hand panel) between the combined data set (black solid line) and the individual surveys of VVDS (orange dashed line), zCOSMOS (blue solid line) and SDSS (red dash-dotted line). the normalization of the BHMF is based on the ERDF with $\log \lambda>-1.5$ and the normalization of the ERDF uses $\log \mbh>8$.}
\label{fig:df_comps3}
\end{figure*}

\begin{figure*}
\centering
        \includegraphics[width=16cm,clip]{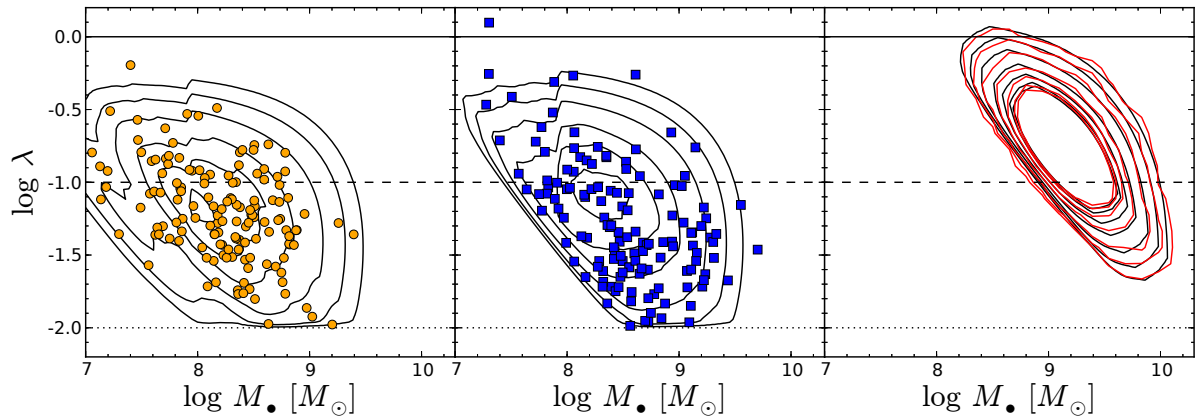}
\caption{Comparison of the distribution of BH masses and Eddington ratios in the $M_\bullet-\lambda$ plane between the observational data and the prediction derived from the best-fitting bivariate distribution function $\Psi(\mbh,\lambda)$. The latter is always shown by the black contours. Left-hand panel: results for the VVDS (orange circles), Middle panel: results for zCOSMOS (blue squares), Right-hand panel: results for SDSS (red contours). The horizontal solid, dashed and dotted lines indicate Eddington ratios of 1, 0.1 and 0.01, respectively.}
\label{fig:pobs}
\end{figure*}

Our model parametrization allows for a \mbh dependence of the ERDF. Our best-fitting results for both models supports such a dependence with $k_\lambda\approx0.10-0.15$. In Fig.~\ref{fig:erdf_m} we show the conditional probability distribution of $\lambda$ at a given \mbh at several masses. At higher masses the conditional Eddington ratio distribution is shifted towards a higher break value. This is also directly visible in the bivariate distribution function shown in Fig.~\ref{fig:pmodel}.
A similar behaviour has been found in \citet{Kelly:2013} only using the SDSS data, however only for BH mass estimates based on \ion{Mg}{ii}, while they see no significant mass dependence using H$\beta$ or \ion{C}{iv}. At this point it is therefore unclear if this represents a real trend or is due to systematic effects in the viral relations.

We also determined the bivariate distribution function with this parametrization for the three individual samples. The best-fitting models are given in Table~\ref{tab:dfpara}. In Fig.~\ref{fig:df_comps3} we compare our best-fitting model to the combined data set to these results for  the three individual surveys. Since the BHMF and the ERDF are only a marginalization over the bivariate distribution function $\Psi(\mbh,\lambda)$, their respective normalization depends on the shape of the other distribution function, i.e. the normalization of the BHMF is determined by the integration over the ERDF and vice versa. This should be always kept in mind when comparing different BHMFs or ERDFs independently, since a difference in the ERDF between two fits will manifest in a different normalization of the BHMF. To reduce the differences caused by this normalization condition, we normalize the distribution function over a range that is well restricted for the three surveys, i.e. we use $\log \lambda>-1.5$ as lower limit for the normalization of the BHMF and $\log \mbh>8$ for the normalization of the ERDF. 
The BHMF for the VVDS and zCOSMOS are in good agreement for $\mbh<10^9 M_\odot$. On the other hand, the ERDFs for the two deep surveys differ. This might be caused by low number statistics and/or cosmic variance between the small area fields. Combining the two data sets, as performed here, will alleviate both issues. Furthermore, we again see the large uncertainties at the high mass end and high Eddington ratio regime from the deep, small-area surveys. This regime is mainly constrained by the data from the SDSS QSO sample.
Overall, the combined best-fitting bivariate distribution function $\Psi(\mbh,\lambda)$ provides a good fit to the three individual data sets and combines their individual strengths.

This can be furthermore demonstrated by a direct comparison to the data. Our maximum likelihood method fits a model probability distribution in the $\mbh-\lambda$ plane to the observed data in that plane. In Fig.~\ref{fig:pobs} we compare the observed distribution in the $\mbh-\lambda$ plane for the three surveys, VVDS (left-hand panel, orange circles), zCOSMOS (middle panel, blue squares) and SDSS (right-hand panel, red contours), with this distribution predicted from our best-fitting bivariate distribution function (black contours). We find a  good agreement between observations and the best-fitting model. Our bivariate distribution function provides an excellent representation of the observed mass and luminosity/ Eddington ratio distributions.

\begin{figure}
\centering 
\resizebox{\hsize}{!}{ \includegraphics[clip]{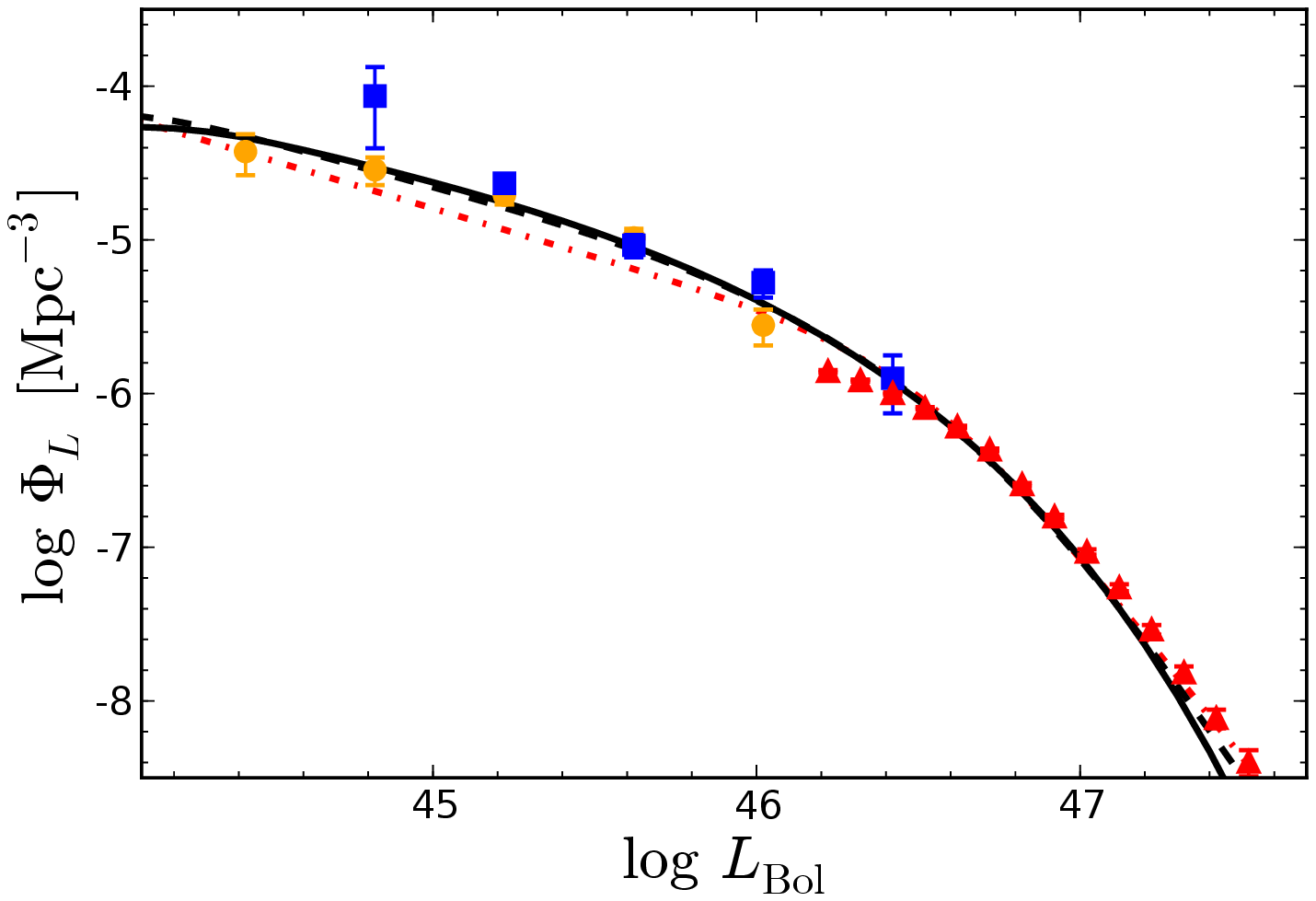}}
\caption{Comparison of the bolometric AGN LF derived by marginalizing the best-fitting bivariate distribution function $\Psi(\mbh,\lambda)$  to the one directly determined from the optical magnitudes. We show the model AGN LF based on the modified Schechter BHMF model (black solid line) and the double power law  BHMF model (black dashed line). These are compared to the binned estimates from the three surveys, VVDS (orange circles), zCOSMOS (blue squares) and SDSS (red triangles) and the best-fitting PLE model to the combined data set (red dash-dotted line).}
\label{fig:lfbol}
\end{figure}

Furthermore, the bolometric AGN LF follows directly from the bivariate distribution function $\Psi(\mbh,\lambda)$ by marginalizing over \mbh via Equation~\ref{eq:lfbol}. Thus, our model bivariate distribution function should match by design the observed bolometric LF.  In Fig.~\ref{fig:lfbol} we compare the best-fitting model bolometric LF with the observed bolometric AGN LF. Here, we use the optical AGN LF in the $M_i(z=2)$ band, presented in Appendix~\ref{sec:agnlf}, and transform it to the bolometric AGN LF for type-1 AGN, using the conversion outlined in section~\ref{sec:lbol}. We find excellent agreement between the AGN LF predicted from $\Psi(\mbh,\lambda)$ and the direct determination of the AGN LF, again confirming the consistency of our maximum likelihood fitting results.

\subsection{Comparison with previous work}  \label{sec:comp}
\begin{figure*}
\centering 
\includegraphics[width=16cm,clip]{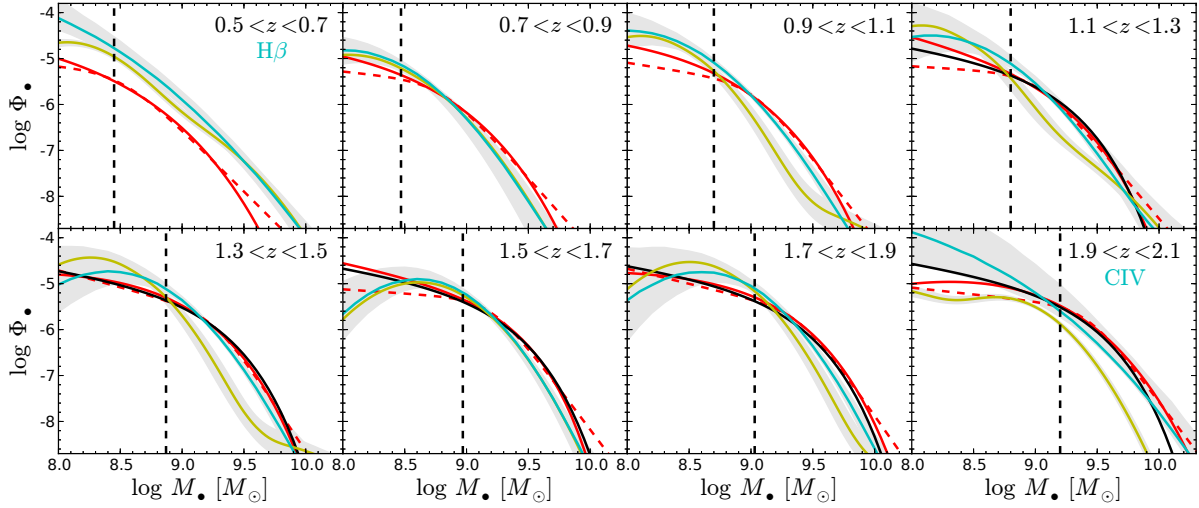}
\caption{Comparison of the active BHMF determined in this paper to previous studies based on the SDSS DR7 QSO sample. The cyan solid line and grey shaded area shows the BHMF by \citet{Kelly:2013} and the yellow solid line gives the BHMF by \citet{Shen:2012}. The red lines 
show our best-fitting results using only the SDSS data in the given redshift bin (solid line for a modified Schechter function BHMF and dashed line for double power-law BHMF). The solid black line is our best-fitting result for the combined data from all three surveys and over the full redshift range ($1.1<z<2.1$). The vertical black dashed line indicates the mass at which the completeness in the SDSS sample drops below 10~\%, as given in  \citet{Kelly:2013}. Note that the BHMF by \citet{Kelly:2013} and \citet{Shen:2012} in the lowest redshift bin is based on H$\beta$ masses and in the highest redshift bin on \ion{C}{iv} masses, while we use \ion{Mg}{ii} masses in all bins.}
\label{fig:bhmf_sdss}
\end{figure*}

\begin{figure*}
\centering 
\includegraphics[width=16cm,clip]{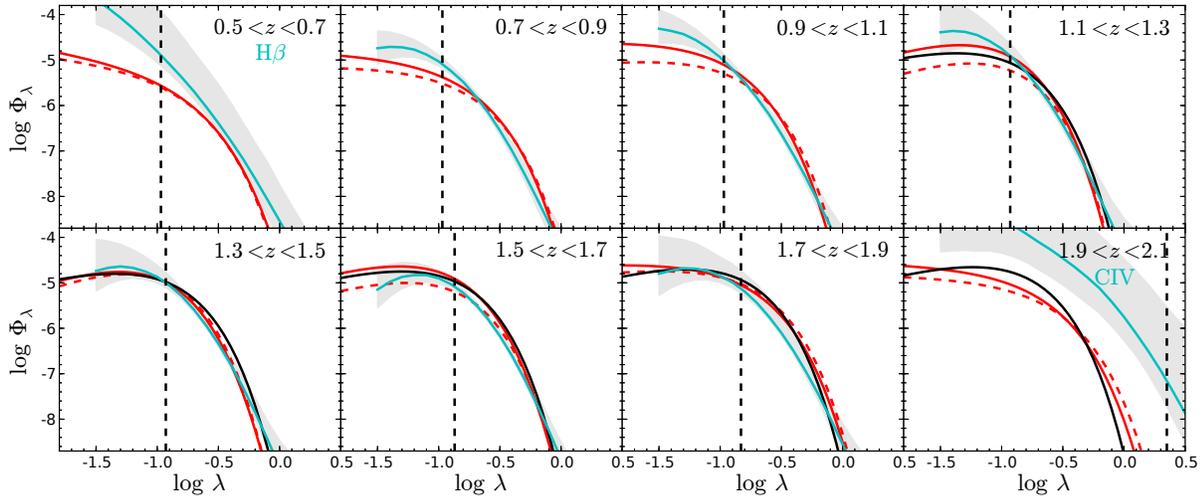}
\caption{Comparison of the ERDF determined in this paper to the previous study by  \citet{Kelly:2013}, based on the SDSS DR7 QSO sample. The lines are the same as in Fig.~\ref{fig:bhmf_sdss}. The vertical black dashed line indicates the Eddington ratio at which the completeness in the SDSS sample drops below 10~\%.}
\label{fig:erdf_sdss}
\end{figure*}

We here compare our results with previous work on the active BHMF and ERDF at $z\sim1.5$. \citet{Shen:2012} and \citet{Kelly:2013} both used the same SDSS DR7 dataset, which we are also incorporating in this study, to determine the active BHMF over the redshift range $0.3<z<5$. Both studies used a Bayesian framework to properly account for the selection function and determine the BHMF and ERDF. Their work mainly differs in the detailed model assumptions for their BHMF and ERDF. In particular, \citet{Shen:2012} focused more on the active BHMF and used a  more simple and restrictive model for the ERDF, while \citet{Kelly:2013} approached to consistently determine the joint bivariate distribution function of BH mass and Eddington ratio by employing a more flexible parametrization. Their results for the active BHMF and the ERDF are shown in Figs~\ref{fig:bhmf_sdss} and \ref{fig:erdf_sdss}.
The active BHMFs of both studies agree reasonably well, while the ERDFs show larger deviations, due to the more restrictive ansatz in the \citet{Shen:2012} study, which leads to a rather restrictive ERDF model. The work by \citet{Kelly:2013} is closer to the approach we are following here in determining the bivariate distribution function of \mbh and $\lambda$. Therefore, we will compare our results mainly to the work by \citet{Kelly:2013}. We show the BHMF from \citet{Shen:2012} for completeness and to give a direct impression of the spread of possible solutions in the Bayesian framework used in both studies.

Our work uses the same data set and the same implementation of the selection function as the two studies above. But while they used a Bayesian framework, we here employ the maximum likelihood approach presented in this paper and first outlined in \citet{Schulze:2010}. Thus, we are able to directly compare the consistency of these two methods. Our work also differs in the parametric model used to fit the data. \citet{Kelly:2013} used a superposition of multiple Gaussian functions for the bivariate distribution function, while our model parametrization is generally more restrictive, but includes less free parameters. We already demonstrated above that this parametrization is fully able to properly describe the data.

In addition to our main redshift range ($1.1<z<2.1$), we also compute the BHMF and ERDF from the SDSS data at lower redshifts, $0.5<z<1.1$, where BH masses based on \ion{Mg}{ii} are available, in three redshift bins. For this subset, we restricted the SDSS sample to $\log \mbh>7.5$ in the fitting. This allows a comparison with previous studies over a larger redshift range and will be further used in the discussion.

In Figs~\ref{fig:bhmf_sdss} and \ref{fig:erdf_sdss} we show our best-fitting results for the SDSS data only in each redshift bin by the two red lines for two different parametric models (modified Schechter function or double power-law BHMF). We generally find good agreement with the work by \citet{Kelly:2013}, while the difference with the \citet{Shen:2012} BHMF is larger. The main deviation occurs in the range where the completeness in the SDSS sample in below 10~\%, as indicated by the vertical dashed line. Thus, in this range the SDSS suffers from significant uncertainty due to the large completeness correction, as already pointed out by \citet{Kelly:2013}. Overall the agreement between the Bayesian method and the maximum likelihood approach is remarkable, demonstrating that both methods are equally able to reliably determine the BHMF and ERDF. This is further shown by comparing to our best-fitting results for the combined data, including the deep VVDS and zCOSMOS surveys, shown by the black solid line. The BHMF and ERDF by \citet{Kelly:2013} is generally consistent with the deep data over the full range they are probing, $\log \mbh>8$ and $\log \lambda>-1.5$. At $1.3<z<1.9$, their mass functions appear to turn over at the lowest masses, which is probably an artefact of their parametric model of Gaussian superpositions in the range of low completeness. Our deeper data do not show such a turn over.

\begin{figure*}
\centering
        \includegraphics[width=16cm,clip]{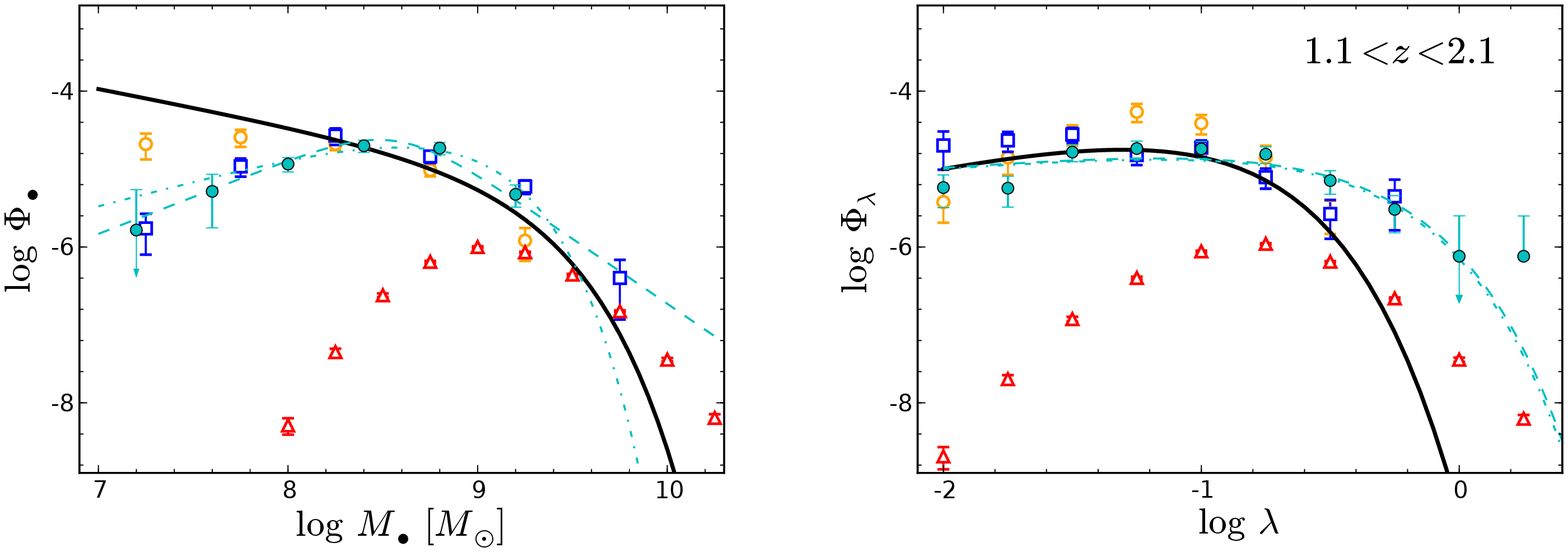}
\caption{Comparison of our results for the $1<z<2$ BHMF (left-hand panel) and ERDF (right-hand panel) with the work by \citet{Nobuta:2012}. 
The cyan circles give the luminosity-weighted $1/V_\mathrm{max}$ results from \citet{Nobuta:2012}, while the cyan dashed line and dash-dotted line give their best-fitting maximum likelihood results for their double power law BHMF and Schechter function BHMF model, respectively. Our best-fitting results are shown by the black solid line. The ERDF is normalized by integrating the BHMF until $\mbh=10^8\ M_\odot$. We also show our luminosity weighted binned estimates for the VVDS (open orange circles), zCOSMOS (open blue squares) and SDSS (open red triangles).}
\label{fig:bhmf_comp}
\end{figure*}

In the lowest redshift bin and in the highest redshift bin we actually compare to results obtained from BH masses estimated from other lines. While we always use \ion{Mg}{ii}, the BHMF and ERDF at $z\approx0.6$ from \citet{Kelly:2013} is based on H$\beta$ masses, while the ones at $z\approx2.0$ are based on \ion{C}{iv} masses. For these we see a larger disagreement, most prominent in the $z\sim2.0$ ERDF. This indicates that the currently employed virial mass calibrations do not provide consistent BHMFs and ERDFs. Thus care has to be taken when comparing distribution functions obtained from different broad lines. In this study we therefore largely focus on using only masses from a single line, namely \ion{Mg}{ii}. A more detailed investigation of this issue is beyond the scope of this paper.

Recently, \citet{Nobuta:2012} used the SXDS to determine the BHMF and ERDF at $z\sim1.4$, employing the luminosity-weighted $1/V_\mathrm{max}$ method and the maximum likelihood method from \citet{Schulze:2010}. Their sample is X-ray selected and extends significantly deeper than SDSS ($4\times10^{-15}$~erg s$^{-1}$~cm$^{-2}$ in the soft band), over an area of  $\sim1.0$~deg$^2$. Their results are shown in Fig.~\ref{fig:bhmf_comp} by the cyan circles for the luminosity-weighted $1/V_\mathrm{max}$ method and by the cyan lines for the maximum likelihood method, where the dashed line shows their double power-law BHMF model and the dash-dotted line shows the Schechter function BHMF model. We find their binned results in general in good agreement with our binned estimates from VVDS and zCOSMOS. At the low mass end, the BHMF is in better agreement with zCOSMOS than with VVDS, probably due to a similar effective flux limit of these two samples.

The maximum likelihood results from \citet{Nobuta:2012} show larger differences. The BHMF agrees well within $8.5<\log \mbh <9.5$. Similar to VVDS and zCOSMOS, SXDS is not able to reliably constrain the high-mass end of the BHMF. At the low-mass end, their work predicts a turnover in the BHMF, which we do not confirm with our data. The reason for this discrepancy is not obvious. It might be caused by cosmic variance between the small fields, differences in the model assumptions or in the target selection. The SXDS sample is soft X-ray selected, while we use basically optically selected AGN samples. 

In the right-hand panel of Fig.~\ref{fig:bhmf_comp}, we normalized the ERDF by integrating the BHMF only until $\log \mbh=8$, to alleviate the difference in normalization due to the BHMF difference at lower \mbh. At $\log \lambda <-1$, our work agrees well with  \citet{Nobuta:2012}, while they predict a much larger space density at higher $\lambda$. This is probably caused by the larger uncertainty at the high-$\lambda$ end, employing only the SXDS survey, and the fact that they have not accounted for the virial mass uncertainty in their fit, as we have done here.

Overall, our results are consistent with previous work, but provide much more robust constrains on the bivariate distribution function of \mbh and $\lambda$ at $1<z<2$ over a wide range in \mbh and $\lambda$.

\begin{figure*}
\centering
        \includegraphics[width=16cm,clip]{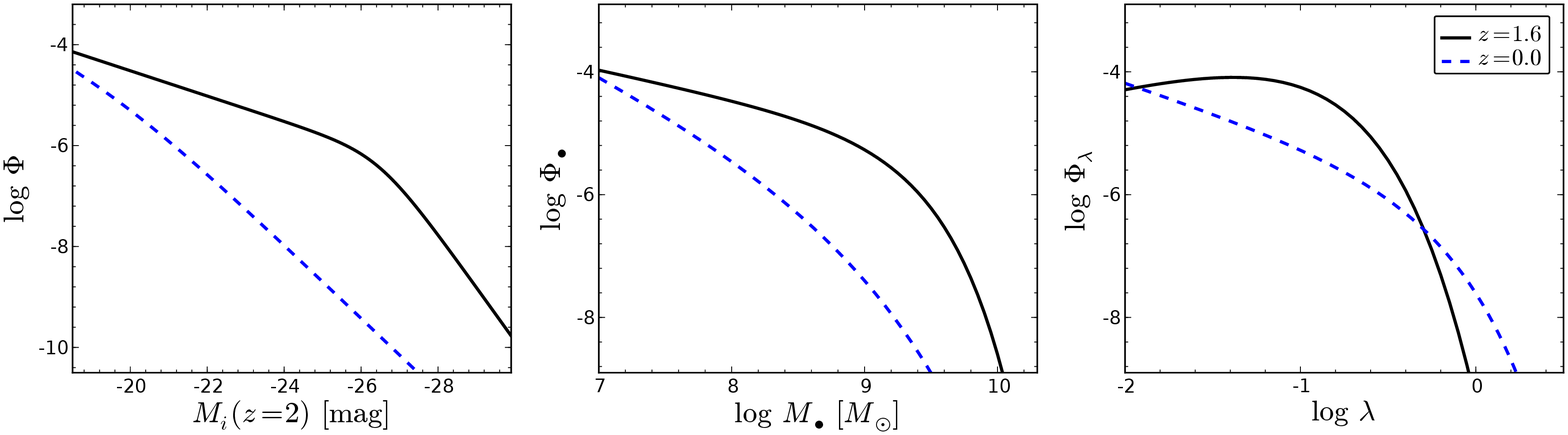}
\caption{Comparison of the AGN distribution functions at $z\sim1.6$ with $z=0$. Left-hand panel:  AGN LF in $M_i(z=2)$, Middle panel: active BHMF, Right-hand panel: ERDF. The black solid lines and blue dashed lines show the three distribution functions at $z=1.6$ and $0$, respectively.}
\label{fig:compz0}
\end{figure*}

\begin{figure*}
\centering
        \includegraphics[width=16cm,clip]{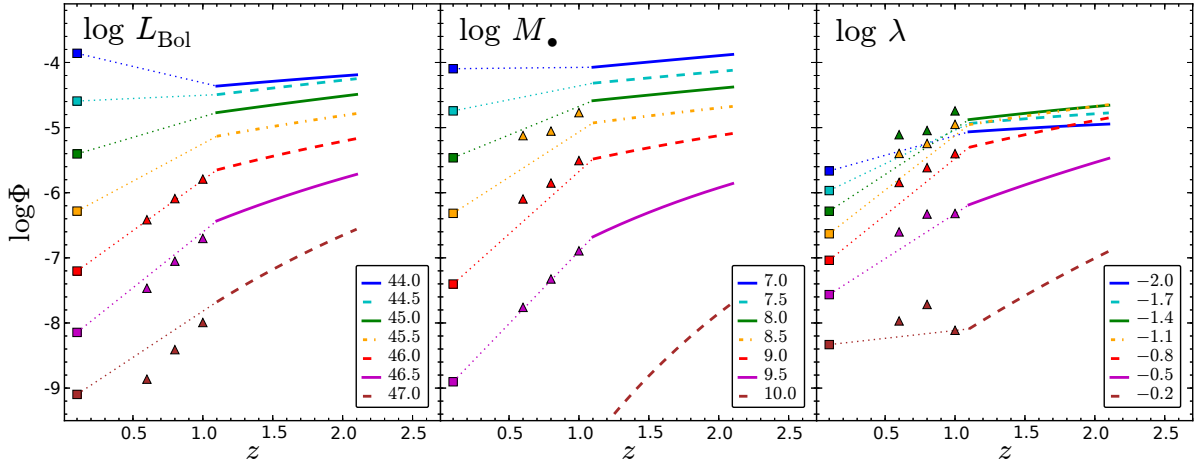}
\caption{Evolution of the type-1 AGN space density between $z=0$ and $2.1$ in bins of AGN bolometric luminosity (left-hand panel), BH mass (middle panel) and Eddington ratio (right-hand panel). The lines at $1.1<z<2.1$ give the prediction from our best-fitting model over this redshift range. 
The triangles show the best fit to the SDSS data only in the redshift range $0.5<z<1.1$ in $z$-bins of 0.2~dex. The squares give the $z=0$ results from \citet{Schulze:2009} and  \citet{Schulze:2010} and the dotted lines are simple linear extrapolations between the $z=0$ results and the our best-fitting model results at $z=1.1$. A clear trend of AGN downsizing is present in $L_\mathrm{bol}$, \mbh and $\lambda$.}
\label{fig:densityevo}
\end{figure*}

\section{Discussion}  \label{sec:discussion}

\subsection{Cosmic evolution in BH mass and accretion rate} 
The cosmic evolution of the AGN LF out to $z\sim2$ is by now well established \citep[e.g.][]{Richards:2006a,Bongiorno:2007,Croom:2009}. The space density of the most luminous QSOs strongly increases towards $z\sim2$ and peaks around this redshift. On the other hand, the space density of lower luminosity AGN shows a weaker evolution. Their space density also increases, but peaks at lower redshift. This behaviour is known as 'AGN cosmic downsizing' \citep[e.g.][]{Hasinger:2005}. In the left-hand panel of Fig.~\ref{fig:compz0} we illustrate this downsizing behaviour of the AGN LF by comparing our $z\sim1.6$ AGN LF with the AGN LF at $z=0$, as determined from the Hamburg/ESO Survey by \citet{Schulze:2009}, shown by the dashed blue line. There is not only a change in the normalization of the LF, but also a distinctive change in its shape, from being close to a single power law at $z=0$ to the presence of a prominent break at $z\sim1.6$.

With the additional information on the BHMF and ERDF, we are able to disentangle the AGN downsizing behaviour in its contribution due to BH mass downsizing and accretion rate downsizing. In the middle and left panel of Fig.~\ref{fig:compz0} we compare our $z\sim1.6$ BHMF and ERDF with the distribution functions at $z=0$, determined by \citet{Schulze:2010}. Note however, that the comparison of our results and the $z=0$ work by \citet{Schulze:2010} may be affected by systematics in the BH mass estimates, since not the same broad lines (\ion{Mg}{ii} versus. H$\beta$) have been used. Nevertheless, the general trends of evolution should be robust against these possible systematics.

The behaviour of the active BHMF is similar to the LF. We see a strong downsizing trend in the BHMF. At the high-mass end, $M_\bullet\approx 10^9 M_\odot$, there is a strong increase of the space density from $z=0$ to $\sim1.6$, while at the low-mass end, $M_\bullet\approx 10^7 M_\odot$, there is almost no change in the space density. This mass downsizing has direct implications for the active fraction (or duty cycle), as we will discuss in more detail below.

\begin{figure}
\centering 
\resizebox{\hsize}{!}{ \includegraphics[clip]{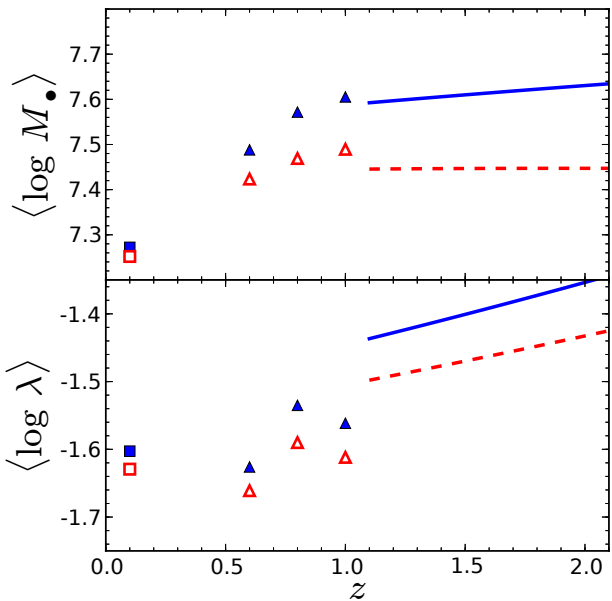}}
\caption{Evolution of the mean BH mass (upper panel) and mean Eddington ratio (lower panel). The lower integration limits for the computation of the mean are $\log \mbh=7$ and $\log \lambda=-2$. The blue solid line shows the respective mean value from our best-fitting model to the SDSS+VVDS+zCOSMOS data set, the blue triangles are for the SDSS sample only in the redshift range $0.5<z<1.1$, and the blue square is the $z=0$ results from the Hamburg/ESO Survey  \citep{Schulze:2010}. The red open symbols and the red dashed line show the mean values for the obscuration corrected AGN population as discussed in section~\ref{sec:obscured}.}
\label{fig:meanevo}
\end{figure}

In the ERDF we see a change in the shape of the distribution function, largely corresponding to a flattening of the power-law slope of the Schechter function. There is a shift of the average accretion rate towards higher values (see also Fig.~\ref{fig:meanevo}), i.e. if a SMBH is active at $z\sim1.6$, it will on average accrete at a higher rate than at $z=0$. 

To investigate these evolution trends further, in Fig.~\ref{fig:densityevo} we directly show the cosmic evolution of the type-1 AGN space density in bins of bolometric luminosity (left-hand panel), BH mass (middle panel) and Eddington ratio (right-hand panel) over the range $0<z<2.1$. Here, the lines at $1.1<z<2.1$ give the prediction from our best-fitting model to the full combined data set of VVDS, zCOSMOS and SDSS over that redshift range. 
The squares give the $z=0$ results from the Hamburg/ESO survey (based on H$\beta$ masses), while the triangles show the results from our best fit to the SDSS data only over bins in the redshift range $0.5<z<1.1$ (using \ion{Mg}{ii} for the SMBH mass estimation). The dotted lines are simple linear extrapolations between $z=0$ and $1.1$ to help guide the eye.
We normalized the ERDF by integrating the BHMF only to $\mbh=10^8 \ M_\odot$, to allow a better comparison with the SDSS only results at lower $z$.

Again, we confirm the well-known behaviour of AGN downsizing in the AGN LF. The space density is strongly  decreasing between $z=2$ and $0$ at the bright end, with much less evolution at the faint end of the AGN LF. However, our restricted redshift range does not allow us to probe the turnover present at the bright end at $z>2$. A very similar behaviour is seen in BH mass. At the high-mass end, there is strong evolution in the space density, already within $1.1<z<2.1$, but even more compared to $z=0$, with very little change in the space density at lower masses ($\mbh<10^9 \ M_\odot$). We also find evidence for evolution in Eddington ratio, though to a somewhat smaller extend. The redshift dependence in the break of the Schechter function ERDF in our best-fitting model implies a stronger decrease of the space  at the high-$\lambda$ end, visible in Fig.~\ref{fig:densityevo}.
These trends are also verified by the evolution of the mean \mbh and mean $\lambda$, as shown in Fig.~\ref{fig:meanevo}. The mean \mbh shows a strong decrease towards $z=0$, while also the mean $\lambda$ decreases from $z=2$ to $0$.

We conclude that the downsizing in the AGN LF is driven by both, the downsizing in the BHMF and evolution of the ERDF, while the BHMF evolution seems to be the dominating term. At higher redshift more massive BHs were active and these were on average accreting at a higher rate. The AGN LF evolution is therefore a combination of simultaneous evolution in BH mass and accretion rate. Disentangling both processes can resolve degeneracies on the BH growth history and accretion rate evolution.
 
\subsection{Correction for obscured AGN} \label{sec:obscured}
As noted above, our results for the active BHMF and ERDF only refer to type-1 AGN, for which we can estimate virial BH masses. However, obscured AGN represent a significant fraction of the AGN population. This obscured fraction is luminosity dependent,  with a decrease of the type-2 fraction with increasing luminosity \citep[e.g.][]{Ueda:2003,LaFranca:2005,Hao:2005,Maiolino:2007,Hasinger:2008,Merloni:2014}. In the unified scheme, this corresponds to a decrease of the opening angle of the torus with increasing luminosity. 
But also AGN feedback could be important,  where stronger winds in more luminous AGN clean the line of sight quicker \citep{Menci:2008}. 
Thus, the total active BHMF and ERDF will differ in normalization and shape from the results for type-1 AGN which we have discussed so far. However, for a better understanding of the physical processes at play and a comparison with theoretical models, the bivariate distribution function for the full AGN population would be desirable. We here apply a simple correction for the obscured fraction to our type-1 AGN results, to transform them to the expected BHMF and ERDF of the total AGN population. While such a correction still suffers from significant uncertainties, it should reveal the general trends and implications valid for the full AGN population. Note that the total active BHMF still defines an AGN as having $\lambda>0.01$, thus it does not include low accretion rate or quiescent SMBHs.

Recently, \citet{Merloni:2014}  presented an obscured fraction dependent on X-ray luminosity and consistent with no redshift evolution within $0<z<3.5$ based on an X-ray-selected AGN sample from \textit{XMM}-COSMOS. The mass and Eddington ratio dependence of the obscured fraction is consistent with a direct luminosity dependence. We use their X-ray luminosity-dependent obscuration correction, 
\begin{equation}
F_\mathrm{obs}(L_\mathrm{X}) = 0.56+\dfrac{1}{\pi}\arctan \left(  \dfrac{43.89-\log L_\mathrm{X}}{0.46}\right) \ ,
\end{equation}
to convert our bivariate distribution function $\Psi(\mbh,\lambda)$ to the distribution function for the full AGN population. To reduce the large uncertainties in this correction at low luminosities, where the type-1 fraction is low, at $\log L_X<43$ we fix the obscured fraction to the value at this luminosity, $F_\mathrm{obs}(10^{43})=0.985$ .
We convert X-ray luminosities ($2-10$~keV) into bolometric luminosities via the bolometric correction from \citet{Marconi:2004} and applied the correction factor above to $\Psi(\mbh,\lambda)$.

 \begin{figure}
\centering 
\resizebox{\hsize}{!}{ \includegraphics[clip]{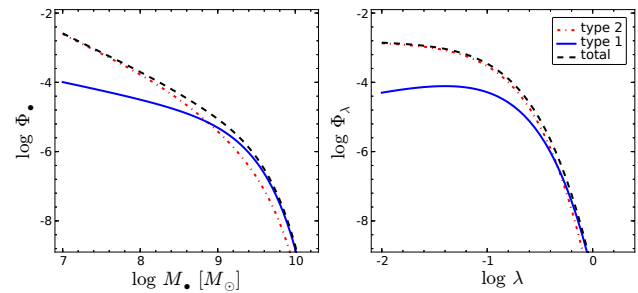}}
\caption{Obscuration correction for the active BHMF (left-hand panel) and the ERDF (right-hand panel). The blue solid line shows our best-fitting model for the active BHMF and ERDF of type-1 AGN. The black dashed line give the total BHMF/ERDF, inferred by using the luminosity-dependent obscuration correction from \citet{Merloni:2014}, while the red dash-dotted line gives the prediction for type-2 AGN.}
\label{fig:df_obscured}
\end{figure}

The derived total active BHMF and ERDF are shown in Fig.~\ref{fig:df_obscured}. This might still be an underestimate of the total AGN population, since we are missing heavily obscured, Compton thick AGN (CT AGN). The fraction of CT AGN is currently highly uncertain, but expected to be between $10$ and $40$\% \citep{Gilli:2007,Treister:2009,Akylas :2012,Alexander:2013,Vignali:2014}. This would further increase the space density of the total active BHMF and ERDF by $0.05$ and $0.15$~dex. A possible \mbh and $\lambda$ dependence of the CT AGN fraction is even more uncertain, but recently \citet{Lanzuisi:2014} found evidence that CT AGN have on average smaller \mbh and higher $\lambda$ with respect to unobscured sources, suggesting that the inclusion of this population can further change the shape of the active BHMF and ERDF. However, given the large uncertainties involved, it is at the moment not warranted to apply such corrections to our distribution functions for type-1 AGN.

At $\mbh>10^9 \  M_\odot$, the total active BHMF is dominated by type 1, with type-2 AGN becoming dominant at lower masses. This leads to a significant steepening of the low-mass slope of the active BHMF. The inferred total ERDF is also largely dominated by obscured AGN at the low-$\lambda$ end.
The accretion rate distribution function, as determined via the ERDF, links the physical quantity of the mass of an SMBH to its instantaneous observable, the AGN luminosity. It therefore contains information about the accretion process \citep{Yu:2005,Hopkins:2009,Novak:2011,Hickox:2013}. At $z=0$, the accretion rate distributions obtained from type-1 and type-2 AGN are consistent with each other \citep{Yu:2005,Schulze:2010}. Our obscuration correction above implies that at $1<z<2$ both AGN populations show differences in their accretion rate distribution. However, this indication needs to be verified by direct studies of the distribution functions for both AGN populations.

The flattened distribution of Eddington ratios found here for type-1 AGN and also inferred for the total AGN population seems to be in contrast to recent studies of the distribution of the specific accretion rate $L_\mathrm{X}/M_\ast$. \citet{Aird:2012} determined the specific accretion rate distribution function of type-2 AGN at $0.2<z<1.0$, defined as the distribution function of the ratio of AGN luminosity over stellar mass. This quantity is effectively the convolution of the ERDF with the  $M_\bullet-M_\mathrm{Bulge}$ relation. They found a power-law distribution with slope $-0.65$ over the full redshift range, while the fit to the AGN LF implies a cut-off towards the Eddington limit, which is not directly probed by their data \citep{Aird:2013}. The different  ranges in redshift, Eddington ratio (with $\lambda<0.1$) and  the different AGN population probed makes a comparison to the results presented in this work difficult.

Using an X-ray-selected COSMOS sample, including both type-1 and type-2 AGN, \citet{Bongiorno:2012} also determined the distribution of the specific accretion rate over the redshift range $0.3<z<2.5$. They also found a power-law distribution with a steeper slope, $\gamma\approx1$, but also find evidence for a sharp decrease of the space density towards the Eddington limit, consistent with our results. While a detailed comparison is again challenging, we note that for the redshift range in common, their ERDF is only well probed for $\lambda>0.1$ (see their Figure~16). Over this $\lambda$ range our ERDF, in particular the total ERDF shown in Fig.~\ref{fig:df_obscured}, is actually consistent with their data, while being inconsistent with their power law. 

On the other hand, \citet{Lusso:2012} did not find evidence for the steep power-law ERDFs reported by \citet{Aird:2012} and \citet{Bongiorno:2012}, based on a study of the \textit{XMM}-COSMOS AGN sample. Their results rather suggest a log-normal distribution of Eddington ratios, both for type-1 and for type-2 AGN.
Future, more detailed studies of the BHMF and the ERDF of both the type-1 and type-2 AGN population are required for a better understanding of the interrelation in the demographics of these two AGN populations.

 \begin{figure*}
\centering 
\includegraphics[width=11cm,clip]{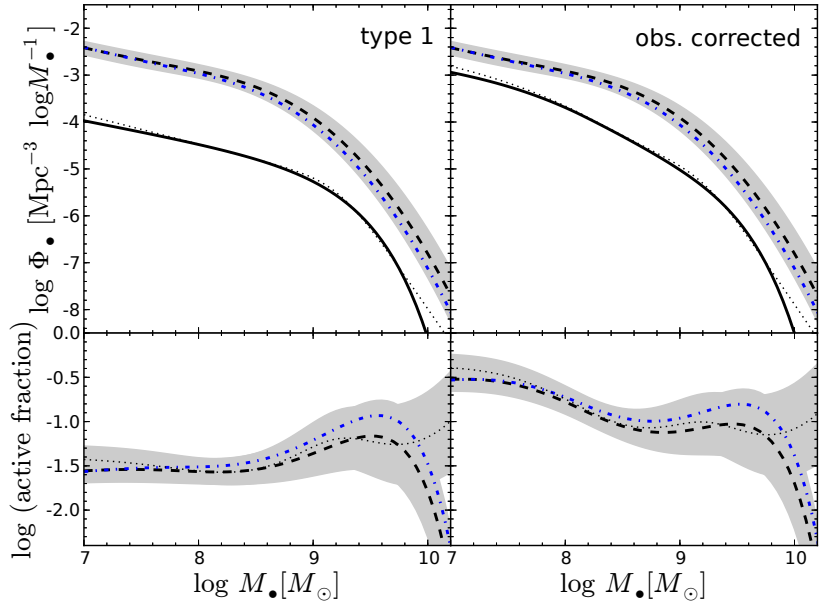}
\caption{Upper panel: Comparison of the active BH mass function at $z\sim1.5$ (black solid line for the modified Schechter BHMF model and dotted line for the double power law BHMF model) with the total black hole mass function at this redshift (black dashed line with $1\sigma$ confidence region given by the gray area). 
Lower panel: Active fraction of BHs as a function of BH mass (dashed line for the modified Schechter BHMF and dotted line for the double power law BHMF). The gray area gives the uncertainty, based on the stellar mass function uncertainty and the two active BHMF models.  
The blue dash-dotted lines gives the BHMF and active fraction for star-forming galaxies.
The active fraction for type-1 AGN at $z\sim1.5$ is consistent with being BH mass independent. The left panels present the results for type-1 AGN, while the right panels use an obscuration correction to derive the active fraction of the total AGN population.}
\label{fig:acfrac}
\end{figure*}
 
\subsection{The active fraction of black holes} \label{sec:af}
The observed BH mass downsizing has implications for the evolution in the active fraction or duty cycle of SMBHs. We here define the active fraction as the ratio of active BHs, according to our definition, to the total SMBH population. Therefore, 'active' only includes AGN above an Eddington ratio of 1\%. Our direct determination of the BHMF is limited to type-1 AGN, missing obscured AGN, while we will also adopt the obscuration correction presented in the previous section to investigate the total, obscuration-corrected active fraction.

\begin{figure*}
\centering 
 \includegraphics[width=11cm,clip]{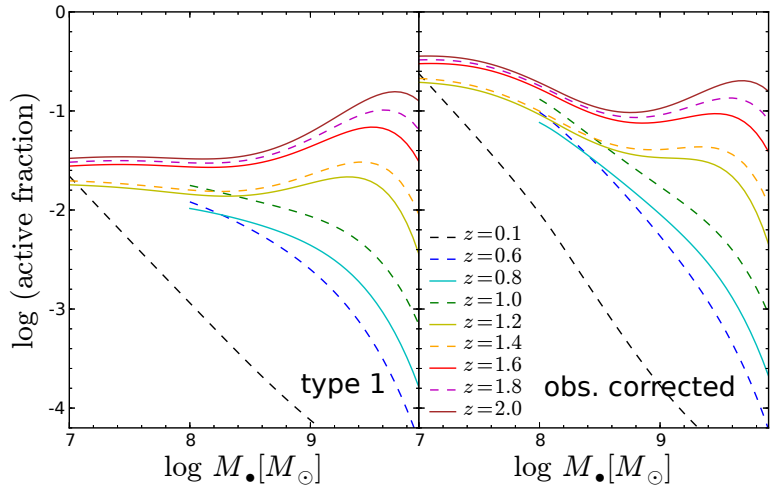}
\caption{Redshift evolution of the active fraction for the type-1 AGN population (left-hand panel) and the total AGN population (right-hand panel). The lines within $1.1<z<2.1$ show the results for our best-fitting model to the combined VVDS, zCOSMOS and SDSS dataset. The lines within $0.5<z<1.1$ are based on a maximum likelihood fit to the SDSS data only in three redshift bins. The blue dashed line gives the active fraction at $z\sim0$, based on the BHMF from \citet{Schulze:2010}. A clear AGN downsizing trend in BH mass is present in the active fraction, where we see the shutoff of AGN activity for the high \mbh population between $z=2$ and $0$.}
\label{fig:af_zevo}
\end{figure*}

At $z=0$, \citet{Schulze:2010} found a decrease of the type-1 AGN active fraction with increasing BH mass, implying that in the local Universe the most massive BHs are preferentially in a quiescent state, while lower mass BHs are still actively growing (see blue dashed line in Fig.~\ref{fig:af_zevo}). This suggests an enhanced BH growth episode of massive BHs at earlier times. With our study, we directly probe this epoch of enhanced growth of SMBHs.

The total BHMF, including quiescent BHs, is usually computed by convolving the stellar mass function (or galaxy LF), converted to a spheroid mass function, by the $M_\bullet-M_\mathrm{Bulge}$ relation. This total BHMF is reasonably well established for the local universe \citep[e.g.][]{Marconi:2004,Shankar:2009}. Their extension to higher redshift has still large uncertainties \citep{Li:2011}. First, the stellar mass function itself and the bulge-to-total ratios are less well established \citep{Bundy:2006,Fontana:2006,Ilbert:2010,Ilbert:2013}. Furthermore, the $M_\bullet-M_\mathrm{Bulge}$ relation is not well known at $z\sim1.5$. 
It has been suggested that there is redshift evolution in the $M_\bullet-M_\mathrm{Bulge}$ relation, with an increase of the $M_\bullet/M_\mathrm{Bulge}$ ratio with redshift \citep[e.g.][]{Peng:2006b,Decarli:2010,Bennert:2010}. However, the observed \textit{apparent} evolution is fully consistent with a non-evolution scenario, once sample selection effects are taken into account \citep{Schulze:2011,Schulze:2014}. We therefore assume no evolution in the $M_\bullet-M_\mathrm{Bulge}$ relation out to $z=2$.

We here use the stellar mass function from \citet{Ilbert:2013}, without applying any conversion to a spheroid mass function. Indeed, at $z>1$ it is currently not clear if SMBH mass correlates better with total stellar mass or with spheroid mass. Taking into account the observational challenges of bulge-disc decomposition at high $z$, current observations indicate that at these redshifts total stellar mass correlates at least as well with \mbh as spheroid mass  \citep{Jahnke:2009,Merloni:2010,Schramm:2013}. Furthermore, there is some evidence that even at lower redshift total stellar mass might provide an equally good correlation \citep{Marleau:2013,Lasker:2014}.

The total BHMF is given by:
\begin{equation}
 \Phi_\bullet(\mu,z)= \frac{1}{\sqrt{2 \pi} \sigma} \int  \exp \left\lbrace - \frac{(\mu-\alpha-\beta s)^2}{2 \sigma^2} \right\rbrace \, \Phi_s(s,z)\, \dd s \  ,\label{eq:qbhmf}
\end{equation}
where $\mu=\log \mbh$, $s=\log M_\ast -11$,  $\Phi_s(s,z)$ is the stellar mass function at $z$ and $\alpha$, $\beta$ and $\sigma$ are the normalization, slope and intrinsic scatter of the $\mbh-M_\ast$ relation. We use the $\mbh-M_\mathrm{bulge}$ relation from \citet{McConnell:2013}, with $\alpha$, $\beta$, $\sigma=(8.46,1.05,0.34)$  to transform the galaxy stellar mass function to the BHMF.

The resulting total BHMF at $z=1.6$ is shown by the black dashed line in the upper panels of Fig.~\ref{fig:acfrac}, with the uncertainty  associated with the underlying stellar mass function shown by the grey area. The black solid line in the upper panels of Fig.~\ref{fig:acfrac} gives the active BHMF based on the modified Schechter BHMF. The left-hand panel shows the results for  type-1 AGN and the right-hand panel gives our estimate for the total AGN population, applying the obscuration correction from \citet{Merloni:2014}. The lower panels present the active fraction for these two populations by the black dashed line. The active fraction for the type-1 AGN population is almost constant, at $\sim3$~\%, over a wide mass range. At higher masses the active fraction becomes more uncertain, due to the uncertainties in the active BHMF, the galaxy stellar mass function and the intrinsic scatter in the $\mbh-M_\ast$ relation at this redshift. There is tentative evidence for an increase of the active fraction around $\log \mbh\sim9.0$, which we will investigate in more detail below. The implied strong decrease of the active fraction towards $\log \mbh\sim10.0$ is not significant, due to the statistical uncertainties at these large masses. Already using the best-fitting double power-law BHMF would remove this turnover, as shown by the dotted black line.

The lower-right panel shows the active fraction for the total AGN population, showing a different mass dependence than the type-1 AGN active fraction. The active fraction decreases until $\log \mbh\sim8.5$ and flattens out at higher masses. Our obscuration-corrected active fraction implies that $\sim30$\% of all galaxies at $\mbh\sim10^7$ are active, i.e accrete above 1\% of the Eddington limit, while this fraction is still $\sim10-20$\% at  $\mbh>10^9$, the latter being consistent with observations of X-ray AGN in deep survey fields \citep[e.g.][]{Brusa:2009,Xue:2010, Bongiorno:2012}.

The close connection between star formation and AGN activity \citep[e.g.][]{Silverman:2009,Rosario:2012} might imply that the star-forming galaxy population might provide the main reservoir to supply  the AGN population, before AGN feedback is quenching star formation. To test the implications of this hypothesis, we compare the shape of the mass function of star-forming galaxies with the AGN population.
We use the stellar mass function of star-forming galaxies, based on the work by \citet{Ilbert:2013}, to compute the respective BHMF, shown by the dash-dotted blue line in Fig.~\ref{fig:acfrac}. We caution that  there are significant uncertainties in the assumptions made for the estimation of the BHMF for this subpopulation, while the main trends should be relative robust. 
Since at these redshift the star-forming galaxy population is dominating the total stellar mass function, the difference between the star-forming and total galaxy population is small, and hence also their differences in comparison to the AGN population.
Thus, we cannot distinguish between the star-forming and total galaxy population as the main source to supply the AGN population.

In Fig.~\ref{fig:af_zevo} we study the redshift evolution of the active fraction, both for the type-1 AGN population and for the obscuration-corrected AGN population. The individual lines show the active fraction at different redshifts. To extend the redshift coverage of our study, we augment our results by our best-fitting BHMF to the SDSS DR7 QSO sample in three redshift bins within $0.5<z<1.1$, limited in the \mbh range to $\mbh>10^8 \ M_\odot$, as presented above (note that these suffer from larger uncertainties already below $\log\mbh\approx8.5$, partly being responsible for the discontinuity seen between $z=1.2$ and $1.0$). Furthermore, we show by the blue dashed line the active fraction at $z\sim0$, derived from the local active BHMF by \citet{Schulze:2010}. We computed the $z=0$ active fraction using the \citet{McConnell:2013} $\mbh-M_\mathrm{bulge}$ relation and an estimate of the spheroid mass function, based on the local early-type and late-type galaxy stellar mass function from \citet{Bell:2003}, assuming  $B/T=0.3$ for the late type galaxy population \citep[see also][]{Schulze:2011}.\footnote{If we would instead use the total stellar mass function from \citet{Bell:2003}, the $z=0$ active fraction would shift down by $\sim0.15$~dex.}

Fig.~\ref{fig:af_zevo} again reveals a strong mass dependence in the BH growth history. While there is only moderate evolution of the active fraction at the low-mass end, we witness the shutoff of BH growth at the high-mass end between $z=2$ and $0$. Focusing only on our $1.1<z<2.1$ sample, we find that at the low redshift edge the active fraction is basically constant with BH mass, i.e. BHs are actively accreting independent of BH mass with an active fraction of $\sim2-3$\%. 
With increasing redshift, we see an increase of an upturn of the active fraction for the most massive BHs, $\mbh>10^9 \ M_\odot$. These become more active than the low-mass population. At the peak epoch of AGN activity, around $z\sim2$, about $20-30$~\% of the most massive BHs are in an active state (i.e. accrete at $\lambda>0.01$).
Adding the information from the SDSS at $z<1.1$ and the $z=0$ results, this trend continues, with the most massive BHs shutting off towards $z=0$ while there is little evolution at low \mbh, leading to an active fraction decreasing with increasing \mbh at $z<1$. The trend is present both in the type-1 AGN population and in the total AGN population. We note that the strength of the upturn at high \mbh towards $z=2$ would be reduced, but is still present, if we would use the $\mbh-M_\mathrm{bulge}$ relation from \citet{Kormendy:2013}. Contrary, it would be much more pronounced for the relation by \citet{Haering:2004}. Furthermore, assuming a larger intrinsics scatter in the $\mbh-M_\mathrm{bulge}$ relation would also flatten this upturn. The general redshift evolution trend however is robust against these uncertainties.

\subsection{Comparison to model predictions}  \label{sec:models}
Current theoretical models of galaxy evolution and BH growth are reasonably well able to reproduce the observed AGN LF and thus, the cosmic downsizing of AGN. These models range from large cosmological hydrodynamic simulations \citep[][Rosas-Guevara et al., in preparation]{Degraf:2010,Khandai:2014,Hirschmann:2014,Sijacki:2014,Bachmann:2014}, semi-analytical models \citep{Cattaneo:2005,Marulli :2008,Bonoli:2009,Fanidakis:2012,Hirschmann:2012,Menci:2014,Neistein:2014} to more phenomenological models  \citep{Wyithe:2003,Shen:2009,Conroy:2013,Veale:2014,Caplar:2014}. However, since these models differ in many details, in particular the trigger mechanisms for AGN activity, no consistent picture on the physics of BH growth can be inferred from these comparisons alone. In fact, the AGN LF alone is degenerate to physically distinct BH growth models \citep[e.g.][]{Veale:2014}.
The addition of the observed active BHMF and  ERDF provides important independent constraints  for the theoretical models of galaxy and AGN evolution. Due to varying assumptions for BH growth, these models predict or use differently evolving Eddington ratio distributions.

While a comprehensive comparison with theoretical predictions in the literature is beyond the scope of this work, we here want to illustrate the additional constraints gained from SMBH mass and Eddington ratio distribution for  theoretical models. We compare our observations to the study by \citet{Hirschmann:2014}. They analysed a subset of the cosmological, hydrodynamical simulation Magneticum Pathfinder (Dolag at al., in preparation), based on an improved version of the SPH code GADGET3 \citep{Springel:2005}, with a comoving box size of (500~Mpc)$^3$. Their model successfully reproduces the observed AGN LF at $z<3$ and explains their downsizing behaviour by the evolution of the gas density in the vicinity of the SMBH. This gas reservoir for accretion gets depleted over cosmic time, due to star formation and AGN feedback. This decrease is stronger for more massive BHs, i.e. in more massive galaxies, which therefore shut off their high accretion phase at earlier cosmic times than less massive BHs.

 \begin{figure}
\centering 
\resizebox{\hsize}{!}{ \includegraphics[clip]{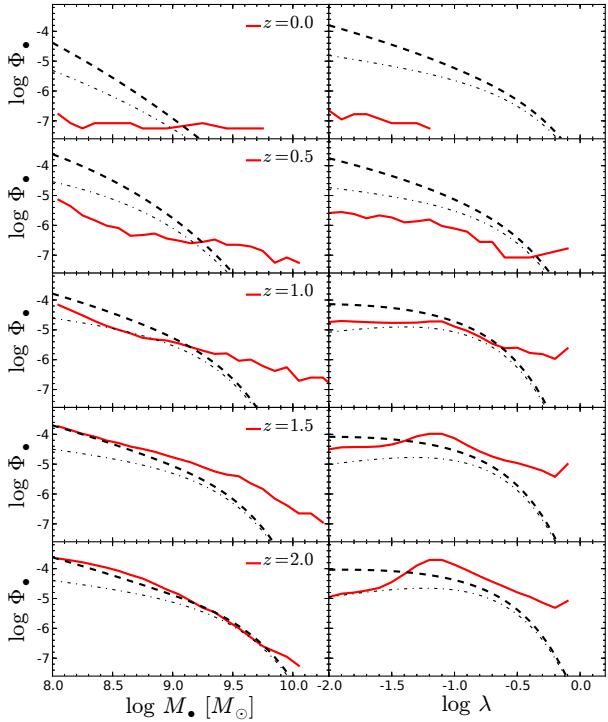}}
\caption{Comparison of our observed active BHMF and ERDF, corrected for obscuration (black dashed line) with the results of the hydrodynamical simulation from \citet{Hirschmann:2014}, shown by the red solid lines, for the redshift range $0<z<2$. We also show our observed distribution functions without obscuration correction for reference (black dash-dotted lines).}
\label{fig:df_sim}
\end{figure}

In Fig.~\ref{fig:df_sim} we show the predictions of the simulation studied in \citet{Hirschmann:2014} for the active BHMF and the ERDF by red solid lines. We here define an active SMBH consistent with our work as accreting above 1\% Eddington. Below $\log \mbh\approx 8$, their simulations under-predict the total SMBH space density (see their fig.~5) due to their relatively  low resolution in the large volume box, thus we also expect an under prediction for the active BHMF for that mass range. Therefore, we restrict our comparison to $\log \mbh>8$.
The shown ERDF is also normalized to this lower mass limit. 
We compare the simulation with our distribution function, corrected for obscuration (black dashed line), while we also show our type-1 AGN result for reference (black dash-dotted line). At $z\geq1$ and $\log \mbh<9.5$, the simulations of the active BHMF are in reasonably good agreement with the observations, in particular at $z=2$. However, they do not match our observations at lower redshift or higher masses. At $z=1$ and $1.5$ the simulations predict a higher space density at the massive end $\log \mbh>9.5$. This is probably related to the overestimation also seen in the stellar mass function at these redshifts (see their fig.~4), likely a consequence of their implementation of AGN feedback in the radio-mode, which is relatively inefficient in locking the baryons in the hot gas halo and thus, in suppressing star formation and gas accretion on to the central BH.

For $z<1$, the simulations predict a strong decrease of the AGN space density, especially at the high-mass end, in qualitative agreement with our results. However, quantitatively this shutoff of AGN activity is much stronger than we observe, leading to a significant underestimation of the AGN space density. The reason is likely again the relatively inefficient  radio-mode AGN feedback implementation. This let massive BHs grow too much at higher redshift and thus too much gas is consumed to allow active BH growth at low redshift. 

The right-hand panel in Fig.~\ref{fig:df_sim} compares our observed ERDF with the simulation result. In general, the shape of the ERDF is in reasonable agreement with our observation, especially at $z=1$. They predict a wide distribution of Eddington ratios with a decrease towards the Eddington limit. In detail, there are differences at high and low $\lambda$. The simulations show a less steep decrease of the space density towards the Eddington limit. At the same time, they tend to underestimate the space density at lower Eddington ratio, $\lambda<0.1~\%$, compared to our observations. At $z<1$ also the normalization is much lower, but this is a direct consequence of the underestimation of the active BHMF.

Overall, the agreement of the hydrodynamical simulations with the observations of the active BHMF and ERDF over certain ranges in parameters space is already encouraging. However, clear differences remain which can guide the further improvements of the theoretical models.

 \begin{figure}
\centering 
\resizebox{\hsize}{!}{ \includegraphics[clip]{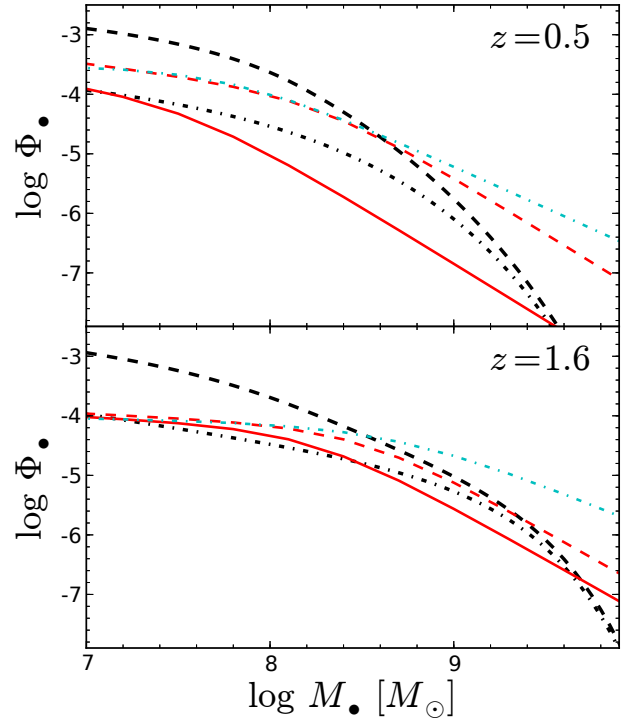}}
\caption{Comparison of our active BHMF with the semi-empirical models of BH growth from \citet{Shankar:2013} at two characteristic redshifts, $z=0.5$ and $1.6$. The black lines show our results for the obscuration corrected AGN population (dashed) and for type-1 AGN only (dash-dotted). The red and cyan lines present different models from \citet{Shankar:2013} always assuming a log-normal ERDF.  They tested models with a constant ERDF (red solid line), but also included a $z$ dependence (red dashed line) and an additional \mbh dependence (cyan dash-dotted line).
}
\label{fig:shankar}
\end{figure}

While numerical simulations potentially contain the most complete physical scenario, their interpretation can be complex. Semi-empirical or phenomenological models complement these by allowing us to easily test the effect of individual model parameters.  We here compare our results to model predictions based on the semi-empirical model from \citet{Shankar:2013}. They model BH growth via a standard continuity equation approach \citep[e.g.][]{Small:1992,Yu:2002,Marconi:2004,Merloni:2004,Shankar:2009},  using observational constraints of the AGN LF and the local total BHMF together with assumptions on the ERDF. Their ERDF model goes beyond the single Eddington ratio often adopted in these kind of studies and assumes a Gaussian distribution, with potential dependence on redshift, \mbh or the radiative efficiency $\epsilon$. This Gaussian distribution differs from the Schechter function ERDF inferred in this work, so differences in the details between these models and our results would be anticipated.
 
In Fig.~\ref{fig:shankar} we compare different models from \citet{Shankar:2013}, given by the red and cyan lines, with our active BHMF with obscuration correction (black dashed line) at two characteristic redshifts. The type-1 BHMF is again shown only for reference. 
 We compare three different models for the ERDF: (i) a Gaussian in $\log \lambda$ with dispersion of 0.3~dex and a mean at $\log \lambda_c=-0.6$ [$G$], (ii) a Gaussian with the same dispersion, but a mean decreasing with redshift as $(1+z)^{2.2}$ [$G(z)$], (iii) a $z$-dependent Gaussian  with an additional dependence on \mbh, with decreasing mean $\lambda$ with increasing mass [$G(z,\mbh)$].  
All models predict a lower space density at low masses compared to our obscuration-corrected BHMF. This is likely a consequence of the lower space density predicted by the models for the total BHMF, compared to the estimate derived above from the stellar mass function. It should be noted that the latter estimate at these redshifts is also highly uncertain at the low-mass end, so it is not clear if this indicates a shortcoming in the models.
Overall, we find the best agreement for the redshift-dependent Gaussian $G(z)$ model at both redshifts. The $G$ model provides a fair match at $z=1.6$, but underestimates the space density at low redshift, which is consistent with the conclusions in \citet{Shankar:2013}, based on a comparison to $z\sim0$. All other models overestimate the space density at the high-mass end at $z=0.5$ and partially also at $z=1.6$. This apparent disagreement at high masses is mainly a consequence of the power-law behaviour of the BHMF model in \citet{Shankar:2013}, which reflects the shape of the bolometric AGN LF, imposed in the their fitting.
At $z=1.6$, the $G(z)$ model provides clearly a better agreement with our observations than the \mbh dependent model,  $G(z,\mbh)$, while this difference is less strong at $z=0.5$. The \mbh dependent model was motivated by some observational evidence for rising active fractions of AGN with \mbh at a given $z$, a trend we do not confirm with our results. Thus the $G(z)$ model naturally performs better in comparison to our work.

In total, we find that the relative simple redshift-dependent Gaussian model presented in \citet{Shankar:2013} already provides a reasonable match to our observations.  We note that this agreement with their models which assume a log-normal ERDF does not directly support such a distribution function. The comparison is mostly valid around the break and above of the BHMF. This regime is mainly regulated by the highest $\lambda$ sources, where a log-normal model and a Schechter function model are basically consistent. The disagreement we see at lower masses could be a consequence of the mismatch in the ERDF. 

This comparison rather highlights the potential of combining our observational constraints with such models for constraining the BH growth history. The standard continuity equation approach uses only the AGN LF as observational constraint to study the BH growth history. While this provides strong constraints on integrated quantities like the BH density (i.e. the Soltan argument), the AGN LF is degenerate in constraining the details of the accretion history \citep[see e.g.][]{Veale:2014}. Therefore, \citet{Shankar:2013} already included additional observational constraints on active fractions (i.e. the active BHMF) to derive more detailed models of BH growth. Directly including the observational determination of the active BHMF and the ERDF presented in this work  into these kind of empirical models has the potential to refine these studies, in particular in regard to the detailed accretion and triggering history.

\section{Summary and Conclusions}  \label{sec:conclusions}
We here present a census of the broad line active SMBH population at redshifts $1.1<z<2.1$. We employ three different type-1 AGN surveys, together covering a wide range in AGN luminosity. These are the VVDS epoch-2 AGN sample, the zCOSMOS 20k AGN sample and the well-defined subset of the SDSS DR7 QSO sample. All three samples have well defined selection functions, allowing the determination of the AGN distribution functions. Besides the AGN LF we here focus on the more fundamental distribution function, namely the bivariate distribution function of BH mass and accretion rate. The marginalization of this bivariate distribution function to one dimension gives the active BHMF, the ERDF or alternatively the AGN LF.

We obtain virial BH mass estimates and Eddington ratios for the majority of objects either from the literature or from our own spectral fitting. Due to the bright flux limit of the SDSS, this survey is limited to more luminous AGN and to high BH masses ($M_\bullet \gtrsim 10^8\,M_\odot$) and Eddington ratios ($\lambda \gtrsim0.03$). The deeper surveys from the VVDS and zCOSMOS extend this range down by $\sim2$~dex in AGN luminosity, $\sim1$~dex in BH mass and $\sim0.5$~dex in Eddington ratio. They fill in the space in the mass-luminosity plane not occupied by the SDSS due to its flux limit. On the other hand, due to the large survey area SDSS provides an excellent  coverage of the bright end of the LF and thus of the high mass, high Eddington ratio regime, which is poorly constrained by VVDS and zCOSMOS due to their limited volume. Therefore, it is essential to use a multi layer approach and combine the information from bright, large area surveys with deep, small area surveys, like we have performed it in this work.

We employ a maximum likelihood fitting technique, based on our previous work \citep{Schulze:2010}, but generalized to be more flexible and applicable to our specific surveys. The approach consists of fitting a parametric model for the bivariate distribution function of BH mass and Eddington ratio and its redshift evolution to the observations. It accounts for incompleteness due to the sample selection and also for the uncertainty in the virial BH mass estimates that broadens the observed mass and Eddington ratio distributions. This way we determine the active BHMF and ERDF. We stress that our definition of an \textit{active} BH refers to type-1 (broad line) AGN accreting above an Eddington ratio of $\log \lambda=-2$. 
Our best-fitting model accounts for a BH mass dependence in the ERDF and allows for a linear redshift dependence in the normalization, the break mass in the BHMF and in the break and low $\lambda$ slope of the ERDF. Our combined bivariate distribution function of \mbh and $\lambda$, $\Psi(\mbh,\lambda)$, is consistent with the best-fitting result for the individual surveys and with previous work, but  significantly improves the reliability and the dynamical range of each individual study. The thereby-predicted AGN LF is also in excellent agreement with their direct determination.

We investigate the cosmic evolution of the AGN population as a function of AGN luminosity, BH mass and Eddington ratio between $z=0$ and $2$.  To extend our results into the redshift range $0.5<z<1.1$, we additionally use the SDSS QSO sample to consistently determine the bivariate distribution function within this redshift range, down to $\mbh=10^8 M_\odot$.
The bivariate AGN distribution function $\Psi(\mbh,\lambda)$ allows us to disentangle the well known downsizing behaviour we see in the AGN LF into its contribution due to BH mass downsizing and Eddington ratio evolution. The downsizing is well represented in the active BHMF, with the most massive SMBHs experiencing the strongest decrease of their space density, while the space density of low-mass SMBHs is approximately constant throughout $0<z<2$. The ERDF shows a flattening at the low-$\lambda$ end, with an almost constant distribution at $\log \lambda<-1$, which is in contrast to the shape at $z=0$. We find evidence for an increase of the average Eddington ratio with redshift.

Further evidence for AGN BH mass downsizing comes from the cosmic evolution of the \mbh dependence of the active fraction or duty cycle. Overall, we find an active fraction of type-1 AGN almost constant with BH mass, implying a phase of active accretion on to BHs throughout the galaxy population. Furthermore, we see evidence for an upturn of AGN activity at the highest masses, $\mbh>10^9 M_\odot$ between $z=1$ and $2$. At $z\sim2$, during the peak period of AGN activity, the most massive BHs reach a peak in their activity.
At lower redshifts, more and more massive BHs shut off their mass accretion and become inactive, while lower mass BHs continue to accrete actively. At the same time, the average accretion rate shifts to lower values, as accretion at low rates becomes more frequent. 

While our results here strictly only apply to the type-1 AGN population, we make an attempt to account for obscured AGN. Including a correction for obscured sources changes the AGN space density, while our main conclusions remain unchanged. 

Finally, we compare our results to model predictions. First, to the large size, hydrodynamical simulation by \citet{Hirschmann:2014}. We find a reasonably good agreement at $z>1$ and $\log \mbh<9.5$, but also identify distinct differences between our observations and the simulation results. Robust observational results of the ERDF are expected to provide important constraints on the implementation of the accretion process and their evolution for numerical simulations and semi-analytic models. Secondly, we compared with the semi-empirical models by \citet{Shankar:2013}, finding a reasonable agreement with their relative simple model using a redshift dependent Gaussian ERDF to trace BH growth through cosmic time.

Our work strongly supports the picture of BH mass downsizing or 'anti-hierarchical' BH growth, suggested by previous studies \citep{Merloni:2004,Heckman:2004,Shankar:2009,Vestergaard:2009,Shen:2012,Nobuta:2012}. The downsizing in the AGN LF is  driven by mass downsizing and by a change of the accretion rate distribution function with time, suggesting a complex process driving the cosmic evolution of BH growth.

\section*{Acknowledgements}
We thank Knud Jahnke for fruitful discussions on this work and Micol Bolzonella for providing the zCOSMOS target sampling rates.
AS  acknowledges support by the China Postdoctoral Scientific Foundation, grant 2012M510256. This work was supported by World Premier International Research Center Initiative (WPI Initiative), MEXT, Japan. ABÕs work is supported by the INAF-Fellowship Program and through PRIN-INAF 2011 'BH growth and AGN feedback through the cosmic time'. IG acknowledges support from FONDECYT through grant 11110501. MH acknowledges financial support from the European Research Council under the European CommunityÕs Seventh Framework Programme (FP7/2007-2013)/ERC grant agreement no. 202781 and from the European Research Council via an Advanced Grant under grant agreement no. 321323 NEOGAL.

Based on observations made with ESO Telescopes at the La Silla or Paranal Observatories under programme ID 175.A-0839.
We gratefully acknowledge the contribution of the entire VVDS and COSMOS collaborations. 

Funding for the SDSS and SDSS-II was provided by the Alfred P. Sloan Foundation, the Participating Institutions, the National Science Foundation, the US Department of Energy, the National Aeronautics and Space Administration, the Japanese Monbukagakusho, the Max Planck Society, and the Higher Education Funding Council for England. The SDSS website is http://www.sdss.org/.


\appendix
\section[]{Selection functions}  \label{sec:selfunc}
For the study of the statistical properties of the AGN samples and the determination of the distribution functions (AGN LF, BHMF, ERDF), we need to statistically account for objects that are not observed due to our survey layout. This information is included in the selection function. Besides the flux limit of the survey, there are several levels where we can lose objects. We here largely adopt  the terminology used within the VVDS and zCOSMOS survey.

\textit{Target sampling rate (TSR):} based on the photometric input catalog only a fraction of the objects above the survey limit have been observed spectroscopically. We account for this fraction by correcting the effective area of the respective survey $j$ by this TSR $f^j_\mathrm{TSR} =N_\mathrm{spec}/N_\mathrm{phot}$.

\textit{Spectroscopic success rate (SSR):} for the objects that have been observed spectroscopically, the spectrum has to allow the secure identification as a broad line AGN and the measurement of a redshift. The SSR gives the probability of successful AGN identification which depends on the apparent magnitude, redshift and SED of the object. We include the flux limit of the respective survey into the SSR. The SSR of the respective survey is a function of $z$ and magnitude, $f^j_\mathrm{SSR}(m,z)=N_z/N_\mathrm{spec}$.

\textit{Mass measurement rate (MMR):} for the determination of the BHMF, we additionally require a reliable BH mass estimate for the AGN in the sample. The MMR is the probability that for an object such a mass estimate is possible. It can be defined by the fraction of objects in the sample that have a \mbh estimate. In general, it will depend on redshift and the S/N (or magnitude) of an object, i.e on the location of the broad emission line in the spectral coverage and on the quality of the spectrum. As discussed below, we here only account for a redshift dependence of the MMR, $f^j_\mathrm{MMR}(z)=N_{\mbh}/N_z$.

\textit{Redshift degeneracy weight ($W_{z\mathrm{d}}$):}  for the VVDS sample, some objects have a degenerate redshift and thus only a certain probability $f_{z\mathrm{d}}$ to be intrinsically within the redshift range we study. For all sources with a secure redshift, $f_{z\mathrm{d}}=1$. We account for this redshift degeneracy weight via an individual weight factor applied to each object in the VVDS.

The combined selection function times area for each survey then follows as $\Omega^j_\mathrm{e}(m,z)=\Omega^j \times f^j_\mathrm{TSR} \times f^j_\mathrm{SSR}(m,z) \times f^j_\mathrm{MMR}(z)$.

\subsection{VVDS} \label{sec:selfunc_vvds}
The TSR is given by the mean sampling rate of a specific VVDS field, taken from \citet{Garilli:2008}, namely  0.23, 0.24, 0.24, 0.22 and 0.22 for CDFS, VVDS-0226-04, VVDS-1003+01, VVDS-1400+05 and VVDS-2217+00, respectively. They also provide the effective area of the fields with 0.13, 0.48, 0.6, 0.9  and 3.0 deg$^2$, respectively. The corrected total area is given by the product of area and TSR over all fields $\Omega = \sum_{i=1}^N f^i_\mathrm{TSR} \Omega^i$.

The SSR has been estimated in \citet{Gavignaud:2006} by simulations of VIMOS pointings and the identification of broad line AGN in the simulated spectra.

For the MMR, we computed the ratio of AGN with successful mass measurements over all detected AGN over several bins in redshift or magnitude. We could not find any evidence in the sample for a $z$ or magnitude dependence, thus we assume that any unaccounted mild dependence is negligible for our study. We therefore use a constant weight factor for the MMR, $f_\mathrm{MMR}=0.94$.

To statistically account for the redshift degenerate objects, we include them into the sample but weight them by the probability that the broad line is in fact \ion{Mg}{ii}. For some of the objects with an initially degenerate redshift, a secure redshift could be assigned from additional spectroscopy. The broad line was resolved into \ion{Mg}{ii} in 17 cases, \ion{C}{iii} in 7 cases and \ion{C}{iv} for 4 objects. We use these empirical numbers to compute the weight factor for the individual redshift degenerate objects via $f_{z\mathrm{d}}=\sum N_\mathrm{MgII} / N_\mathrm{line} $, where the sum is over all possible redshift solutions for the object, as given in \citet{Gavignaud:2008}.

\subsection{zCOSMOS}
The TSR is determined by dividing the number of objects spectroscopically observed and the number of objects in the respective photometric input catalogue. For the zCOSMOS field the TSR is 0.966 for the compulsory targets and 0.550 for the random targets. The total TSR for our sample is then given by $f_\mathrm{TSR}=N_\mathrm{target}/N_\mathrm{total}=(N_c+N_r)/(N_c/\mathrm{TSR}_c+N_r/\mathrm{TSR}_r)$, where $N_c$ and $N_r$ are the number of compulsory and random targets in our sample. This yields $f_\mathrm{TSR}=0.67$ for our zCOSMOS sample. The effective area of the survey is 1.648 deg$^2$.

zCOSMOS used the same instrument (VIMOS) and the same integration time as the VVDS-wide survey. The main difference is the different, higher resolution grism used (MR versus. LRRED), which will not significantly affect the detection probability of broad line AGN. Therefore, we can adopt the SSR for the VVDS-wide, as given in \citet{Gavignaud:2006}, also for the zCOSMOS sample. We tested the consistency of the SSR, based on the X-ray-detected AGN sample from \textit{XMM}-COSMOS. The location in the $z-I_\mathrm{AB}$ diagram of X-ray-detected type-1 AGN that have been targeted by zCOSMOS and are detected versus not detected in the spectroscopy is fully consistent with our SSR, confirming our used SSR, while the number statistics are too low to use them for a direct empirical determination of the zCOSMOS SSR.

We estimate $f_\mathrm{MMR}$ for our zCOSMOS sample as the ratio of targets with mass measurements over all targets in several $z$ bins. This mass success rate  shows a redshift dependence in our sample, with a lower success rate at the lower and upper redshift edges, while we achieve high completeness in between. We find a mean of 0.94, varying between 1.0 and 0.79 over our redshift range.

\subsection{SDSS}
For the SDSS, the TSR is already included in the effective area, thus we can assume $f_\mathrm{TSR}=1$. In contrast to the VVDS and zCOSMOS samples the SDSS QSO sample involves a colour pre-selection. This pre-selection significantly reduces the selection efficiency at redshifts $2<z<3.5$ where QSO colours move into the stellar locus. For the redshift range studied in this paper, the QSO selection efficiency  is not strongly affected by the colour selection, i.e. the selection function is unity within the flux limit, $15.0<i<19.1$ \citep{Richards:2006a}. We compute the MMR over several $z$ bins. The mean is 0.97, with a constant value over most of the redshift range and a slight decrease towards the upper edge.

\section{Results from the luminosity-weighted $1/V_\mathrm{max}$ method}  \label{sec:vmax}

\begin{figure*}
\centering
        \includegraphics[width=16cm,clip]{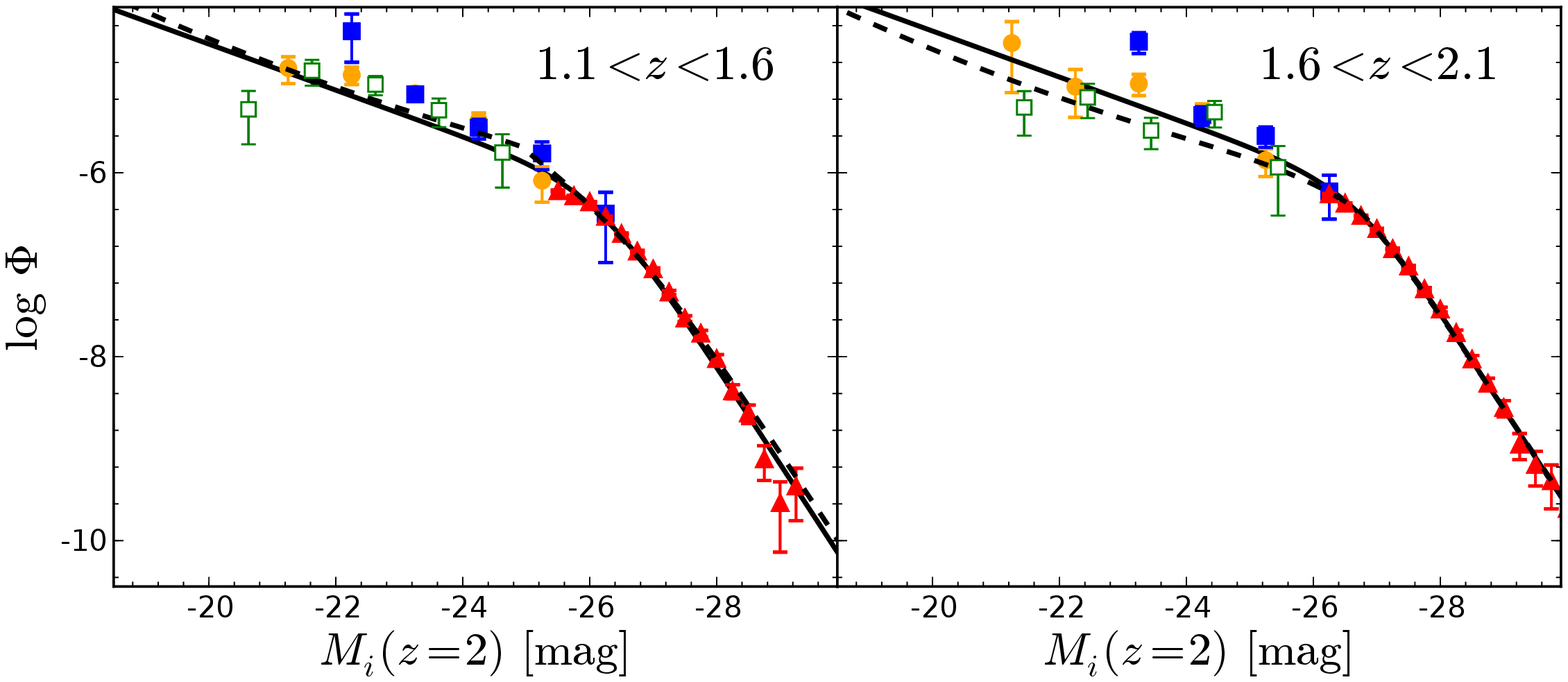}
\caption{AGN LF in $M_i(z=2)$ from the SDSS sample (red triangles), the VVDS sample (orange circles) and the zCOSMOS sample (blue squares). The AGN LF is shown in two redshift bins, $1.1<z<1.6$ (left-hand panel) and $1.6<z<2.1$ (right-hand panel). The black solid  line and black dashed line show our best-fitting PLE model and LDDE model to the combined sample. The open green squares show the AGN LF from \citet{Bongiorno:2007}, based on VVDS epoch~1.}
\label{fig:lf}
\end{figure*}

\begin{table*}
\caption{Best fit model parameters for the combined AGN LF from SDSS, VVDS and zCOSMOS over $1.1<z<2.1$. We also compare our maximum likelihood fit models to the binned AGN LFs of the three samples in three redshift bins via a $\chi^2$ test.}
\label{tab:qlf}
\centering
\begin{tabular}{lcccccccccc}
\hline \hline \noalign{\smallskip}
Evolution model & $\log (\Phi^\ast)$  & $M^\ast_i(z=0)$ & $\alpha$ & $\beta$ & $k_1$ / $p_1$ & $k_2$ / $p_2$ & $\gamma$ & $z_{c,0}$ & $M_{c,0}$ & $\chi^2/\nu$\\ 
\noalign{\smallskip} \hline \noalign{\smallskip}
PLE & $-$6.14 & $-$22.64 & $-$1.63 & $-$3.66 & 1.469 & $-0.317$ &  $-$  & $-$ & $-$ & 129/70 \\ \noalign{\smallskip} 
LDDE & $-$8.61 & $-$26.57 & $-$2.20 & $-$3.64 & 5.77 & $-$1.39 & 0.19 & 2.0 & $-$27.19& 192/67 \\
\noalign{\smallskip} \hline
\end{tabular}
\end{table*}

We here discuss our results, adopting the directly the $1/V_\mathrm{max}$ method. We re-emphasize that this approach introduces unaccounted sample incompleteness at the low-mass end of the BHMF and a potential overestimation of the space density at the high-mass end, due to not accounting for the effect of virial mass uncertainties.

In general, the $1/V_\mathrm{max}$ method would be able to determine the intrinsic BHMF and ERDF when the proper volume weights are applied \citep{Schulze:2010}. The accessible volume of an AGN in the sample is given by
\begin{equation} 
	V_\mathrm{max} 
	= \int_{z_\mathrm{min}}^{z_\mathrm{max}} \Omega(L,z) \frac{\dd V}{\dd z} \dd z  	\label{eq:vmax} \ ,
\end{equation}
where $\Omega(L,z)$ is the \textit{luminosity} selection function of the respective survey, defined in Appendix~\ref{sec:selfunc}. Using Equation~\ref{eq:vmax} gives the proper volume weights for the determination of the LF, but in general not for the determination of the BHMF or ERDF. For these, we would have to use $\Omega(\mbh,z)$, i.e. compute mass-weighted volumes, or $\Omega(\lambda,z)$, compute $\lambda$-weighted volumes. With $\Omega(L,z)=\Omega(\mbh,\lambda,z)$, these volume weights are given by weighting $\Omega(L,z)$ with the ERDF (for $\Omega(\mbh,z)$) or the BHMF (for $\Omega(\lambda,z)$). Because these distribution functions are a priori unknown beforehand, this approach is not practical, but at best serves as a consistency check \citep[see][]{Schulze:2010}. Alternatively, given a sufficiently large sample size, the bivariate distribution function can be determined from the $1/V_\mathrm{max}$ method by binning in both, \mbh and $\lambda$ \citep[e.g.][]{Yu:2005,Kauffmann:2009}.

On the other hand, the determination of the binned BHMF and ERDF using  the luminosity-weighted $1/V_\mathrm{max}$ method can serve as a useful tool. It provides a non-parametric and model-independent estimate of the space density, even if only as a lower limit, and therefore can guide the refined but parametric analysis presented in Section~\ref{sec:df}.

The binned distribution function of the property $X$ is given by \citep{Schmidt:1968}:
\begin{equation}
\Phi(X) = \frac{1}{\Delta X}\sum_{i} \frac{1}{V_{\mathrm{max, i}}}  \ .
\end{equation}

\subsection{The AGN LF}   \label{sec:agnlf}
We first compute the AGN LF for the three samples, which can be compared to previous results on the SDSS QSO sample \citep{Richards:2006a,Shen:2012} and the VVDS AGN sample \citep{Bongiorno:2007}. We derive the AGN LF for absolute magnitudes in the SDSS $i$-band defined at $z=2$, i.e. $X=M_i (z=2)$ \citep{Richards:2006a,Ross:2013}. As discussed in Section~\ref{sec:lbol}, we employ the SDSS $i$-band 
$K$-correction from \citet{Richards:2006a}, where we transform it to the CFHT/CFH12K $I$-band for VVDS and the $HST$/ACS $F814W$ filter  for zCOSMOS.

The binned AGN LFs for the three samples are shown in Fig.~\ref{fig:lf} in two redshift bins. We find excellent agreement with the original SDSS DR7 AGN LF by \citet{Shen:2012}. We also find good agreement with the previous VVDS epoch~1 AGN LF by \citet{Bongiorno:2007}, when accounting for the difference in the applied $K$-correction. The zCOSMOS AGN LF is in good agreement with the VVDS result, spanning a similar luminosity range.

Our data improves the statistical accuracy both on the faint end and on the bright end of the AGN LF, compared to the work of \citet{Bongiorno:2007}. Thus, we here also present a determination of the combined AGN LF for the three samples. We fit a parametric AGN LF model to the full unbinned data over $1.1<z<2.1$ by performing  a maximum likelihood fit \citep{Marshall:1983}. As discussed in section~\ref{sec:ml}, we combine the three surveys, following the strategy outlined by \citet{Avni:1980}.

Following common practice we fit the AGN LF as a double power law:
\begin{equation}
\Phi_L(M,z) = \frac{\Phi^\ast_M}{10^{0.4(\alpha+1)(M-M^\ast(z))} + 10^{0.4(\beta+1)(M-M^\ast(z))}}  \ . \label{eq:qlf}
\end{equation}
The normalization of the AGN LF is determined by the ratio of the number of observed AGN to the number of AGN expected from the best-fitting model, $\Phi^\ast_M=N_\mathrm{obs}/N_\mathrm{mod}$.
Over the restricted redshift range we are probing ($1.1<z<2.1$), the redshift evolution in the AGN LF can be well approximated with a simple pure luminosity evolution (PLE) model \citep{Croom:2004,Ross:2013}. We follow \citet{Boyle:2000} by modelling the redshift evolution in the break of the LF with a second-order polynomial:
\begin{equation}
M_i^\ast(z) = M_i^\ast(z=0) -2.5(k_1 z+k_2 z^2) \ .
\end{equation}
The best-fitting PLE model is shown as solid black line in Fig.~\ref{fig:lf} and the best-fitting parameters are given in Table~\ref{tab:qlf}.
Additionally, we also adopt a luminosity dependent density evolution (LDDE) model, following the parametrization in \citet{Bongiorno:2007} \citep[see also][]{Hasinger:2005}. The best-fitting LDDE model is shown as dashed black line in Fig.~\ref{fig:lf} and we give the best-fitting parameters in Table~\ref{tab:qlf}. Both models provide a reasonably good fit to our data. Since we here only probe a limited redshift range, further discussions  on the redshift evolution of the AGN LF are beyond the scope of this work.

\begin{figure}
\centering
       \resizebox{\hsize}{!}{ \includegraphics[width=8.5cm,clip]{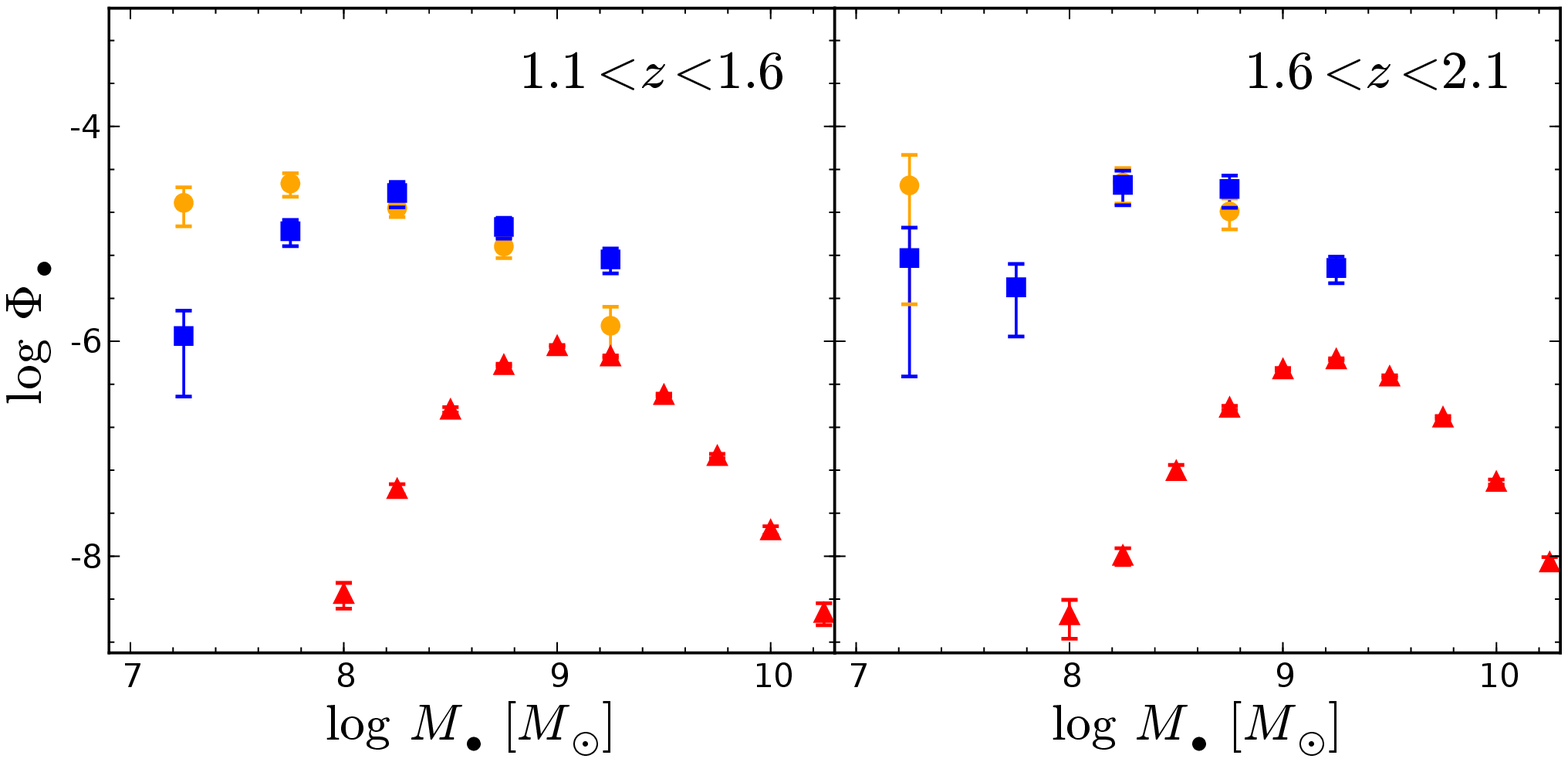}}
\caption{Binned, luminosity-weighted active BHMF in two redshift bins, $1.1<z<1.6$ and $1.6<z<2.1$, for the SDSS sample (red triangles), VVDS sample (orange circles) and zCOSMOS sample (blue squares). While all three distribution function estimates are affected by inherent incompleteness, this is most severe for the SDSS sample, due to the brighter flux limit.}
\label{fig:binned_bhmf}
\end{figure}

\begin{figure}
\centering
         \resizebox{\hsize}{!}{\includegraphics[width=8.5cm,clip]{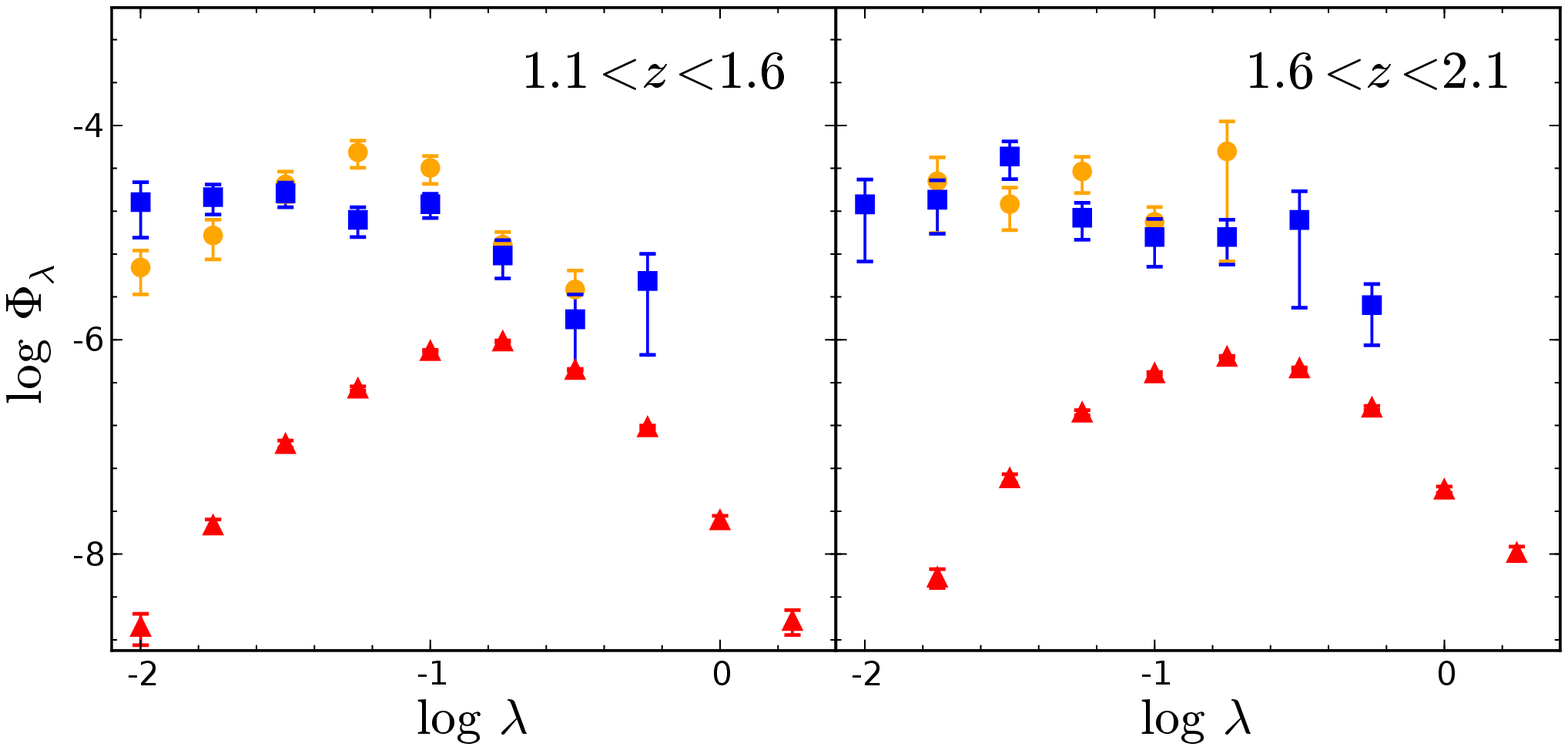}}
\caption{Binned, luminosity-weighted ERDF in two redshift bins. Symbols are the same as for the BHMF in Fig.~\ref{fig:binned_bhmf}.}
\label{fig:binned_erdf}
\end{figure}

\subsection{The binned BHMF and ERDF}    \label{sec:binnedDF}
Next, we determine the binned, luminosity-weighted BHMF ($X=\log \mbh$) and ERDF ($X=\log \lambda$) for the three samples. The results are shown in Figs~\ref{fig:binned_bhmf} and \ref{fig:binned_erdf}. The binned BHMF and ERDF for the SDSS sample shows an \textit{apparent} turnover towards low masses and low Eddington ratio. This is an artefact of the incompleteness introduced to the binned estimate due to the use of \textit{luminosity}-weighted volumes instead of \textit{mass}-weighted or \textit{Eddington-ratio}-weighted volumes.
This becomes evident when adding the VVDS and zCOSMOS samples. Due to the deeper flux limit, they are less affected by this kind of  incompleteness and thus have a higher space density at low masses and Eddington ratios. Thus, incompleteness corrections are smaller than for the SDSS, while still being required. In particular, the BHMFs from the VVDS and zCOSMOS also show an apparent turnover, which occurs at lower masses than for the SDSS sample. Based on the luminosity-weighted binned estimates, it is less clear if this turnover is intrinsic or due to incompleteness in the luminosity-weighted BHMF. The binned zCOSMOS BHMF has a more prominent turnover at $\log \mbh<8$ than the VVDS BHMF, caused by the deep fields included in the VVDS with a flux limit of $I_\mathrm{AB}=24$. Restricting the VVDS sample to the wide fields only gives a  luminosity-weighted BHMF fully consistent with the zCOSMOS result.

A further limitation of the binned approach is the fact that it does not correct for the uncertainty in the virial BH mass estimates \citep{Kelly:2009,Shen:2012}. This will generally broaden the observed \mbh and $\lambda$ distributions and in particular lead to an overestimate of the space density at the steeply decreasing parts of the distribution functions, i.e. at the high-mass end in the BHMF and at high $\lambda$ in  the ERDF. In our maximum likelihood approach, this effect has been taken into account.

\label{lastpage}


\begin{thebibliography}{}

\bibitem[Aird et al.(2010)]{Aird:2010} Aird, J., Nandra, K., 
Laird, E.~S., et al.\ 2010, \mnras, 401, 2531

\bibitem[Aird et al.(2012)]{Aird:2012} Aird, J., Coil, A.~L., 
Moustakas, J., et al.\ 2012, \apj, 746, 90 

\bibitem[Aird et al.(2013)]{Aird:2013} Aird, J., Coil, A.~L., 
Moustakas, J., et al.\ 2013, \apj, 775, 41 

\bibitem[Akylas et  al.(2012)]{Akylas :2012} Akylas, A., Georgakakis, A., Georgantopoulos, I., Brightman, M., \& Nandra, K.\ 2012, \aap, 546, AA98

\bibitem[Alexander et al.(2013)]{Alexander:2013} Alexander, D.~M., Stern, D., Del Moro, A., et al.\ 2013, \apj, 773, 125 

\bibitem[{{Aller} \& {Richstone}(2002)}]{Aller:2002}
{Aller}, M.~C. \& {Richstone}, D. 2002, \aj, 124, 3035

\bibitem[Angl{\'e}s-Alc{\'a}zar et al.(2013)]{Angles:2013} 
Angl{\'e}s-Alc{\'a}zar, D., {\"O}zel, F., 
\& Dav{\'e}, R.\ 2013, \apj, 770, 5 

\bibitem[Assef et al.(2011)]{Assef:2011} Assef, R.~J., Kochanek, 
C.~S., Ashby, M.~L.~N., et al.\ 2011, \apj, 728, 56 

\bibitem[Avni \& Bahcall(1980)]{Avni:1980} Avni, Y., \& Bahcall, J.~N.\ 1980, \apj, 235, 694

\bibitem[{{Babi{\'c}} {et~al.}(2007){Babi{\'c}}, {Miller}, {Jarvis}, {Turner},
  {Alexander}, \& {Croom}}]{Babic:2007}
{Babi{\'c}}, A., {Miller}, L., {Jarvis}, M.~J., {et~al.} 2007, \aap, 474, 755

\bibitem[Bachmann et al.(2014)]{Bachmann:2014} Bachmann, L.~K., Dolag, K., Hirschmann, M., Almudena Prieto, M., \& Remus, R.-S.\ 2014, arXiv:1409.3221 

\bibitem[{{Bell} {et~al.}(2003){Bell}, {McIntosh}, {Katz}, \&
  {Weinberg}}]{Bell:2003}
{Bell}, E.~F., {McIntosh}, D.~H., {Katz}, N., \& {Weinberg}, M.~D. 2003, \apjs,
  149, 289

\bibitem[{{Bennert} {et~al.}(2010){Bennert}, {Treu}, {Woo}, {Malkan}, {Le
  Bris}, {Auger}, {Gallagher}, \& {Blandford}}]{Bennert:2010}
{Bennert}, V.~N., {Treu}, T., {Woo}, J., {et~al.} 2010, \apj, 708, 1507

  
  \bibitem[Bentz et al.(2009)]{Bentz:2009} Bentz, M.~C., Peterson, 
B.~M., Netzer, H., Pogge, R.~W., \& Vestergaard, M.\ 2009, \apj, 697, 160 
  
 \bibitem[Bentz et al.(2013)]{Bentz:2013} Bentz, M.~C., Denney, 
K.~D., Grier, C.~J., et al.\ 2013, \apj, 767, 149 

\bibitem[{{Blandford} \& {McKee}(1982)}]{Blandford:1982}
{Blandford}, R.~D. \& {McKee}, C.~F. 1982, \apj, 255, 419

\bibitem[{{Bongiorno} {et~al.}(2007){Bongiorno}, {Zamorani}, {Gavignaud},
  {Marano}, {Paltani}, {Mathez}, {M{\o}ller}, {Picat}, {Cirasuolo},
  {Lamareille}, {Bottini}, {Garilli}, {Le Brun}, {Le F{\`e}vre}, {Maccagni},
  {Scaramella}, {Scodeggio}, {Tresse}, {Vettolani}, {Zanichelli}, {Adami},
  {Arnouts}, {Bardelli}, {Bolzonella}, {Cappi}, {Charlot}, {Ciliegi},
  {Contini}, {Foucaud}, {Franzetti}, {Guzzo}, {Ilbert}, {Iovino}, {McCracken},
  {Marinoni}, {Mazure}, {Meneux}, {Merighi}, {Pell{\`o}}, {Pollo}, {Pozzetti},
  {Radovich}, {Zucca}, {Hatziminaoglou}, {Polletta}, {Bondi}, {Brinchmann},
  {Cucciati}, {de La Torre}, {Gregorini}, {Mellier}, {Merluzzi}, {Temporin},
  {Vergani}, \& {Walcher}}]{Bongiorno:2007}
{Bongiorno}, A., {Zamorani}, G., {Gavignaud}, I., {et~al.} 2007, \aap, 472, 443

\bibitem[Bongiorno et al.(2012)]{Bongiorno:2012} Bongiorno, A., Merloni, A., Brusa, M., et al.\ 2012, \mnras, 427, 3103 

\bibitem[Bonoli et al.(2009)]{Bonoli:2009} Bonoli, S., Marulli, F., Springel, V., et al.\ 2009, \mnras, 396, 423 


\bibitem[{{Booth} \& {Schaye}(2011)}]{Booth:2011}
{Booth}, C.~M. \& {Schaye}, J. 2011, \mnras, 413, 1158

\bibitem[Boyle \& Terlevich(1998)]{Boyle:1998} Boyle, B.~J., \& Terlevich, R.~J.\ 1998, \mnras, 293, L49

\bibitem[{{Boyle} {et~al.}(2000){Boyle}, {Shanks}, {Croom}, {Smith}, {Miller},
  {Loaring}, \& {Heymans}}]{Boyle:2000}
{Boyle}, B.~J., {Shanks}, T., {Croom}, S.~M., {et~al.} 2000, \mnras, 317, 1014

\bibitem[Brusa et al.(2009)]{Brusa:2009} Brusa, M., Fiore, F., Santini, P., et al.\ 2009, \aap, 507, 1277 

\bibitem[Brusa et al.(2010)]{Brusa:2010} Brusa, M., Civano, F., Comastri, A., et al.\ 2010, \apj, 716, 348

\bibitem[Bundy et al.(2006)]{Bundy:2006} Bundy, K., Ellis, R.~S., 
Conselice, C.~J., et al.\ 2006, \apj, 651, 120

\bibitem[Caplar et al.(2014)]{Caplar:2014} Caplar, N., Lilly, S., \& Trakhtenbrot, B.\ 2014, arXiv:1411.3719 

\bibitem[Cattaneo et al.(2005)]{Cattaneo:2005} Cattaneo, A., Blaizot, 
J., Devriendt, J., \& Guiderdoni, B.\ 2005, \mnras, 364, 407 


\bibitem[Conroy \& White(2013)]{Conroy:2013} Conroy, C., \& White, M.\ 2013, \apj, 762, 70

\bibitem[Cowie et al.(1996)]{Cowie:1996} Cowie, L.~L., Songaila, 
A., Hu, E.~M., \& Cohen, J.~G.\ 1996, \aj, 112, 839 

\bibitem[{{Croom} {et~al.}(2004){Croom}, {Smith}, {Boyle}, {Shanks}, {Miller},
  {Outram}, \& {Loaring}}]{Croom:2004}
{Croom}, S.~M., {Smith}, R.~J., {Boyle}, B.~J., {et~al.} 2004, \mnras, 349,
  1397

\bibitem[{{Croom} {et~al.}(2009){Croom}, {Richards}, {Shanks}, {Boyle},
  {Strauss}, {Myers}, {Nichol}, {Pimbblet}, {Ross}, {Schneider}, {Sharp}, \&
  {Wake}}]{Croom:2009}
{Croom}, S.~M., {Richards}, G.~T., {Shanks}, T., {et~al.} 2009, \mnras, 399,
  1755
  \bibitem[{{Croton}(2006)}]{Croton:2006}
{Croton}, D.~J. 2006, \mnras, 369, 1808



\bibitem[{{Decarli} {et~al.}(2010){Decarli}, {Falomo}, {Treves}, {Labita},
  {Kotilainen}, \& {Scarpa}}]{Decarli:2010}
{Decarli}, R., {Falomo}, R., {Treves}, A., {et~al.} 2010, \mnras, 402, 2453

\bibitem[Denney et al.(2013)]{Denney:2013} Denney, K.~D., Pogge, 
R.~W., Assef, R.~J., et al.\ 2013, \apj, 775, 60 

\bibitem[Degraf et al.(2010)]{Degraf:2010} Degraf, C., Di Matteo, 
T., \& Springel, V.\ 2010, \mnras, 402, 1927 

\bibitem[Di Matteo et al.(2008)]{DiMatteo:2008} Di Matteo, T., 
Colberg, J., Springel, V., Hernquist, L., 
\& Sijacki, D.\ 2008, \apj, 676, 33

\bibitem[Draper 
\& Ballantyne(2012)]{Draper:2012} Draper, A.~R., \& Ballantyne, D.~R.\ 2012, \apj, 751, 72

\bibitem[{{Dubois} {et~al.}(2011){Dubois}, {Devriendt}, {Slyz}, \&
  {Teyssier}}]{Dubois:2011}Dubois, Y., Devriendt, J., Slyz, A., \& Teyssier, R.\ 2012, \mnras, 420, 2662
  
\bibitem[Elbaz et  al.(2007)]{Elbaz:2007} Elbaz, D., Daddi, E., Le Borgne, D., et al.\ 2007, \aap, 468, 33
  
 \bibitem[Fanidakis et al.(2012)]{Fanidakis:2012} Fanidakis, N., Baugh, 
C.~M., Benson, A.~J., et al.\ 2012, \mnras, 419, 2797 

\bibitem[{{Ferrarese} \& {Merritt}(2000)}]{Ferrarese:2000}
{Ferrarese}, L. \& {Merritt}, D. 2000, \apjl, 539, L9

\bibitem[Fine et al.(2008)]{Fine:2008} Fine, S., Croom, S.~M., 
Hopkins, P.~F., et al.\ 2008, \mnras, 390, 1413

\bibitem[Fiore et al.(2012)]{Fiore:2012} Fiore, F., Puccetti, S., Grazian, A., et al.\ 2012, \aap, 537, A16

\bibitem[Fontana et 
al.(2006)]{Fontana:2006} Fontana, A., Salimbeni, S., Grazian, A., et al.\ 2006, \aap, 459, 745 

\bibitem[Gallo et al.(2010)]{Gallo:2010} Gallo, E., Treu, T., 
Marshall, P.~J., et al.\ 2010, \apj, 714, 25 

\bibitem[Garilli et al.(2008)]{Garilli:2008} Garilli, B., Le F{\`e}vre, O., Guzzo, L., et al.\ 2008, \aap, 486, 683

\bibitem[Gavignaud et al.(2006)]{Gavignaud:2006} Gavignaud, I., Bongiorno, A., Paltani, S., et al.\ 2006, \aap, 457, 79

\bibitem[{{Gavignaud} {et~al.}(2008){Gavignaud}, {Wisotzki}, {Bongiorno},
  {Paltani}, {Zamorani}, {M{\o}ller}, {Le Brun}, {Husemann}, {Lamareille},
  {Schramm}, {Le F{\`e}vre}, {Bottini}, {Garilli}, {Maccagni}, {Scaramella},
  {Scodeggio}, {Tresse}, {Vettolani}, {Zanichelli}, {Adami}, {Arnaboldi},
  {Arnouts}, {Bardelli}, {Bolzonella}, {Cappi}, {Charlot}, {Ciliegi},
  {Contini}, {Foucaud}, {Franzetti}, {Guzzo}, {Ilbert}, {Iovino}, {McCracken},
  {Marano}, {Marinoni}, {Mazure}, {Meneux}, {Merighi}, {Pell{\`o}}, {Pollo},
  {Pozzetti}, {Radovich}, {Zucca}, {Bondi}, {Busarello}, {Cucciati}, {de La
  Torre}, {Gregorini}, {Mellier}, {Merluzzi}, {Ripepi}, {Rizzo}, \&
  {Vergani}}]{Gavignaud:2008}
{Gavignaud}, I., {Wisotzki}, L., {Bongiorno}, A., {et~al.} 2008, \aap, 492, 637

\bibitem[{{Gebhardt} {et~al.}(2000){Gebhardt}, {Bender}, {Bower}, {Dressler},
  {Faber}, {Filippenko}, {Green}, {Grillmair}, {Ho}, {Kormendy}, {Lauer},
  {Magorrian}, {Pinkney}, {Richstone}, \& {Tremaine}}]{Gebhardt:2000}
{Gebhardt}, K., {Bender}, R., {Bower}, G., {et~al.} 2000, \apjl, 539, L13

\bibitem[Gilli et al.(2007)]{Gilli:2007} Gilli, R., Comastri, A., \& Hasinger, G.\ 2007, \aap, 463, 79

\bibitem[Goulding et al.(2010)]{Goulding:2010} Goulding, A.~D., 
Alexander, D.~M., Lehmer, B.~D., \& Mullaney, J.~R.\ 2010, \mnras, 406, 597

\bibitem[{{Greene} \& {Ho}(2005)}]{Greene:2005}
{Greene}, J.~E. \& {Ho}, L.~C. 2005, \apj, 630, 122

\bibitem[{{Greene} \& {Ho}(2007)}]{Greene:2007}
{Greene}, J.~E. \& {Ho}, L.~C. 2007, \apj, 667, 131

\bibitem[{{Hao} {et~al.}(2005){Hao}, {Strauss}, {Fan}, {Tremonti}, {Schlegel},
  {Heckman}, {Kauffmann}, {Blanton}, {Gunn}, {Hall}, {Ivezi{\'c}}, {Knapp},
  {Krolik}, {Lupton}, {Richards}, {Schneider}, {Strateva}, {Zakamska},
  {Brinkmann}, \& {Szokoly}}]{Hao:2005}
{Hao}, L., {Strauss}, M.~A., {Fan}, X., {et~al.} 2005, \aj, 129, 1795

\bibitem[{{H{\"a}ring} \& {Rix}(2004)}]{Haering:2004}
{H{\"a}ring}, N. \& {Rix}, H.-W. 2004, \apjl, 604, L89

\bibitem[Hasinger et al.(2007)]{Hasinger:2007} Hasinger, G., 
Cappelluti, N., Brunner, H., et al.\ 2007, \apjs, 172, 29 

\bibitem[{{Hasinger}(2008)}]{Hasinger:2008}
{Hasinger}, G. 2008, \aap, 490, 905

\bibitem[{{Hasinger} {et~al.}(2005){Hasinger}, {Miyaji}, \&
  {Schmidt}}]{Hasinger:2005}
{Hasinger}, G., {Miyaji}, T., \& {Schmidt}, M. 2005, \aap, 441, 417

\bibitem[{{Heckman} {et~al.}(2004){Heckman}, {Kauffmann}, {Brinchmann},
  {Charlot}, {Tremonti}, \& {White}}]{Heckman:2004}
{Heckman}, T.~M., {Kauffmann}, G., {Brinchmann}, J., {et~al.} 2004, \apj, 613,
  109
  
  \bibitem[Hickox et al.(2009)]{Hickox:2009} Hickox, R.~C., Jones, 
C., Forman, W.~R., et al.\ 2009, \apj, 696, 891

\bibitem[Hickox et al.(2014)]{Hickox:2013} Hickox, R.~C., Mullaney, 
J.~R., Alexander, D.~M., et al.\ 2014, \apj, 782, 9
  
 \bibitem[Hirschmann et al.(2012)]{Hirschmann:2012} Hirschmann, M., 
Somerville, R.~S., Naab, T., \& Burkert, A.\ 2012, \mnras, 426, 237  

\bibitem[Hirschmann et al.(2014)]{Hirschmann:2014} Hirschmann, M., 
Dolag, K., Saro, A., et al.\ 2014, \mnras, 442, 2304 

\bibitem[Ho(2008)]{Ho:2008} Ho, L.~C.\ 2008, \araa, 46, 475

\bibitem[Hopkins et al.(2007)]{Hopkins:2007} Hopkins, P.~F., 
Richards, G.~T., \& Hernquist, L.\ 2007, \apj, 654, 731 

\bibitem[{{Hopkins} \& {Hernquist}(2009)}]{Hopkins:2009}
{Hopkins}, P.~F. \& {Hernquist}, L. 2009, \apj, 698, 1550

\bibitem[Ilbert et al.(2010)]{Ilbert:2010} Ilbert, O., Salvato, M., 
Le Floc'h, E., et al.\ 2010, \apj, 709, 644

\bibitem[Ilbert et al.(2013)]{Ilbert:2013} Ilbert, O., McCracken, H.~J., Le F{\`e}vre, O., et al.\ 2013, \aap, 556, A55

\bibitem[{{Jahnke} {et~al.}(2009){Jahnke}, {Bongiorno}, {Brusa},
  {Capak}, {Cappelluti}, {Cisternas}, {Civano}, {Colbert}, {Comastri}, {Elvis},
  {Hasinger}, {Ilbert}, {Impey}, {Inskip}, {Koekemoer}, {Lilly}, {Maier},
  {Merloni}, {Riechers}, {Salvato}, {Schinnerer}, {Scoville}, {Silverman},
  {Taniguchi}, {Trump}, \& {Yan}}]{Jahnke:2009}
{Jahnke}, K., {Bongiorno}, A., {Brusa}, M., {et~al.} 2009{\natexlab{a}}, \apjl,
  706, L215

\bibitem[Jin et al.(2012)]{Jin:2012} Jin, C., Ward, M., 
\& Done, C.\ 2012, \mnras, 425, 907 

\bibitem[{{Kaspi} {et~al.}(2000){Kaspi}, {Smith}, {Netzer}, {Maoz}, {Jannuzi},
  \& {Giveon}}]{Kaspi:2000}
{Kaspi}, S., {Smith}, P.~S., {Netzer}, H., {et~al.} 2000, \apj, 533, 631

\bibitem[{{Kauffmann} \& {Heckman}(2009)}]{Kauffmann:2009}
{Kauffmann}, G. \& {Heckman}, T.~M. 2009, \mnras, 397, 135

\bibitem[{{Kelly} {et~al.}(2009){Kelly}, {Vestergaard}, \& {Fan}}]{Kelly:2009}
{Kelly}, B.~C., {Vestergaard}, M., \& {Fan}, X. 2009, \apj, 692, 1388

\bibitem[Kelly et al.(2010)]{Kelly:2010} Kelly, B.~C., 
Vestergaard, M., Fan, X., et al.\ 2010, \apj, 719, 1315


\bibitem[Kelly 
\& Shen(2013)]{Kelly:2013} Kelly, B.~C., \& Shen, Y.\ 2013, \apj, 764, 45

\bibitem[Khandai et al.(2014)]{Khandai:2014} Khandai, N., Di Matteo, 
T., Croft, R., et al.\ 2014, arXiv:1402.0888

\bibitem[{{Kollmeier} {et~al.}(2006){Kollmeier}, {Onken}, {Kochanek}, {Gould},
  {Weinberg}, {Dietrich}, {Cool}, {Dey}, {Eisenstein}, {Jannuzi}, {Le Floc'h},
  \& {Stern}}]{Kollmeier:2006}
{Kollmeier}, J.~A., {Onken}, C.~A., {Kochanek}, C.~S., {et~al.} 2006, \apj,
  648, 128

\bibitem[Kormendy \& Ho(2013)]{Kormendy:2013} Kormendy, J., \& Ho, L.~C.\ 2013, \araa, 51, 511 

\bibitem[Kormendy 
\& Richstone(1995)]{Kormendy:1995} Kormendy, J., \& Richstone, D.\ 1995, \araa, 33, 581

\bibitem[Koekemoer et al.(2007)]{Koekemoer:2007} Koekemoer, A.~M., 
Aussel, H., Calzetti, D., et al.\ 2007, \apjs, 172, 196

\bibitem[La Franca et al.(2005)]{LaFranca:2005} La Franca, F., Fiore,  F., Comastri, A., et al.\ 2005, \apj, 635, 864 

\bibitem[Lanzuisi et al.(2014)]{Lanzuisi:2014} Lanzuisi, G., Ranalli, P., Georgantopoulos, I., et al.\ 2014, arXiv:1409.1867 

\bibitem[L{\"a}sker et al.(2014)]{Lasker:2014} L{\"a}sker, R., 
Ferrarese, L., van de Ven, G., \& Shankar, F.\ 2014, \apj, 780, 70 

\bibitem[{{Lauer} {et~al.}(2007){Lauer}, {Tremaine}, {Richstone}, \&
  {Faber}}]{Lauer:2007}
{Lauer}, T.~R., {Tremaine}, S., {Richstone}, D., \& {Faber}, S.~M. 2007, \apj,
  670, 249
  
 \bibitem[Le F{\`e}vre et al.(2004)]{LeFevre:2004} Le F{\`e}vre, O., Mellier, Y., McCracken, H.~J., et al.\ 2004, \aap, 417, 839
  
  \bibitem[Le Fevre et 
al.(2005)]{LeFevre:2005} Le F{\`e}vre, O., Vettolani, G., Garilli, B., et al.\ 2005, \aap, 439, 845
  
  \bibitem[Le Fevre et al.(2013)]{LeFevre:2013} Le F{\`e}vre, O., Cassata, P., Cucciati, O., et al.\ 2013, \aap, 559, A14

\bibitem[Li et al.(2011)]{Li:2011} Li, Y.-R., Ho, L.~C., 
\& Wang, J.-M.\ 2011, \apj, 742, 33

\bibitem[Lilly et al.(2007)]{Lilly:2007} Lilly, S.~J., Le 
F{\`e}vre, O., Renzini, A., et al.\ 2007, \apjs, 172, 70

\bibitem[Lilly et al.(2009)]{Lilly:2009} Lilly, S.~J., Le Brun, 
V., Maier, C., et al.\ 2009, \apjs, 184, 218

\bibitem[Lusso et al.(2012)]{Lusso:2012} Lusso, E., Comastri, A., 
Simmons, B.~D., et al.\ 2012, \mnras, 425, 623 

\bibitem[Lutz et al.(2008)]{Lutz:2008} Lutz, D., Sturm, E., Tacconi, L.~J., et al.\ 2008, \apj, 684, 853

\bibitem[{{Magorrian} {et~al.}(1998){Magorrian}, {Tremaine}, {Richstone},
  {Bender}, {Bower}, {Dressler}, {Faber}, {Gebhardt}, {Green}, {Grillmair},
  {Kormendy}, \& {Lauer}}]{Magorrian:1998}
{Magorrian}, J., {Tremaine}, S., {Richstone}, D., {et~al.} 1998, \aj, 115, 2285

\bibitem[{{Maiolino} {et~al.}(2007){Maiolino}, {Shemmer}, {Imanishi}, {Netzer},
  {Oliva}, {Lutz}, \& {Sturm}}]{Maiolino:2007}
{Maiolino}, R., {Shemmer}, O., {Imanishi}, M., {et~al.} 2007, \aap, 468, 979

\bibitem[{{Marconi} \& {Hunt}(2003)}]{Marconi:2003}
{Marconi}, A. \& {Hunt}, L.~K. 2003, \apjl, 589, L21

\bibitem[{{Marconi} {et~al.}(2004){Marconi}, {Risaliti}, {Gilli}, {Hunt},
  {Maiolino}, \& {Salvati}}]{Marconi:2004}
{Marconi}, A., {Risaliti}, G., {Gilli}, R., {et~al.} 2004, \mnras, 351, 169

\bibitem[Marleau et al.(2013)]{Marleau:2013} Marleau, F.~R., Clancy, 
D., \& Bianconi, M.\ 2013, \mnras, 435, 3085

\bibitem[{{Marshall} {et~al.}(1983){Marshall}, {Tananbaum}, {Avni}, \&
  {Zamorani}}]{Marshall:1983}
{Marshall}, H.~L., {Tananbaum}, H., {Avni}, Y., \& {Zamorani}, G. 1983, \apj,
  269, 35
  
\bibitem[Marulli et al.(2008)]{Marulli :2008} Marulli, F., Bonoli, 
S., Branchini, E., Moscardini, L., \& Springel, V.\ 2008, \mnras, 385, 1846

\bibitem[McConnell \& Ma(2013)]{McConnell:2013} McConnell, N.~J., \& Ma, C.-P.\ 2013, \apj, 764, 184

\bibitem[McCracken et al.(2003)]{McCracken:2003} McCracken, H.~J., Radovich, M., Bertin, E., et al.\ 2003, \aap, 410, 17 

\bibitem[McGill et al.(2008)]{McGill:2008} McGill, K.~L., Woo, 
J.-H., Treu, T., \& Malkan, M.~A.\ 2008, \apj, 673, 703 

\bibitem[{{McLure} \& {Dunlop}(2004)}]{McLure:2004}
{McLure}, R.~J. \& {Dunlop}, J.~S. 2004, \mnras, 352, 1390

\bibitem[Menci et al.(2008)]{Menci:2008} Menci, N., Fiore, F., Puccetti, S., \& Cavaliere, A.\ 2008, \apj, 686, 219


\bibitem[Menci et al.(2014)]{Menci:2014} Menci, N., Gatti, M., Fiore, F., \& Lamastra, A.\ 2014, \aap, 569, A37 


\bibitem[{{Merloni}(2004)}]{Merloni:2004}
{Merloni}, A. 2004, \mnras, 353, 1035

\bibitem[{{Merloni} \& {Heinz}(2008)}]{Merloni:2008}
{Merloni}, A. \& {Heinz}, S. 2008, \mnras, 388, 1011

\bibitem[{{Merloni} {et~al.}(2010){Merloni}, {Bongiorno}, {Bolzonella},
  {Brusa}, {Civano}, {Comastri}, {Elvis}, {Fiore}, {Gilli}, {Hao}, {Jahnke},
  {Koekemoer}, {Lusso}, {Mainieri}, {Mignoli}, {Miyaji}, {Renzini}, {Salvato},
  {Silverman}, {Trump}, {Vignali}, {Zamorani}, {Capak}, {Lilly}, {Sanders},
  {Taniguchi}, {Bardelli}, {Carollo}, {Caputi}, {Contini}, {Coppa}, {Cucciati},
  {de la Torre}, {de Ravel}, {Franzetti}, {Garilli}, {Hasinger}, {Impey},
  {Iovino}, {Iwasawa}, {Kampczyk}, {Kneib}, {Knobel}, {Kova{\v c}},
  {Lamareille}, {Le Borgne}, {Le Brun}, {Le F{\`e}vre}, {Maier}, {Pello},
  {Peng}, {Perez Montero}, {Ricciardelli}, {Scodeggio}, {Tanaka}, {Tasca},
  {Tresse}, {Vergani}, \& {Zucca}}]{Merloni:2010}
{Merloni}, A., {Bongiorno}, A., {Bolzonella}, M., {et~al.} 2010, \apj, 708, 137

\bibitem[Merloni et al.(2014)]{Merloni:2014} Merloni, A., Bongiorno, 
A., Brusa, M., et al.\ 2014, \mnras, 437, 3550 

\bibitem[Miller et al.(2012)]{Miller:2012} Miller, B., Gallo, E., 
Treu, T., \& Woo, J.-H.\ 2012, \apj, 747, 57

\bibitem[Neistein \& Netzer(2014)]{Neistein:2014} Neistein, E., \& Netzer, H.\ 2014, \mnras, 437, 3373 

\bibitem[Netzer et al.(2007)]{Netzer:2007} Netzer, H., Lutz, D., 
Schweitzer, M., et al.\ 2007, \apj, 666, 806

\bibitem[{{Netzer} \& {Trakhtenbrot}(2007)}]{Netzer:2007b}
{Netzer}, H. \& {Trakhtenbrot}, B. 2007, \apj, 654, 754

\bibitem[Nicastro(2000)]{Nicastro:2000} Nicastro, F.\ 2000, \apjl, 530, L65 

\bibitem[Nobuta et al.(2012)]{Nobuta:2012} Nobuta, K., Akiyama, M., 
Ueda, Y., et al.\ 2012, \apj, 761, 143

\bibitem[Novak et al.(2011)]{Novak:2011} Novak, G.~S., Ostriker, 
J.~P., \& Ciotti, L.\ 2011, \apj, 737, 26 

\bibitem[{{Onken} {et~al.}(2004){Onken}, {Ferrarese}, {Merritt}, {Peterson},
  {Pogge}, {Vestergaard}, \& {Wandel}}]{Onken:2004}
{Onken}, C.~A., {Ferrarese}, L., {Merritt}, D., {et~al.} 2004, \apj, 615, 645

\bibitem[{{Peng} {et~al.}(2006){Peng}, {Impey}, {Rix}, {Kochanek}, {Keeton},
  {Falco}, {Leh{\'a}r}, \& {McLeod}}]{Peng:2006b}
{Peng}, C.~Y., {Impey}, C.~D., {Rix}, H., {et~al.} 2006, \apj, 649, 616

\bibitem[Peng et al.(2010)]{Peng:2010} Peng, Y.-j., Lilly, S.~J., 
Kova{\v c}, K., et al.\ 2010, \apj, 721, 193 

\bibitem[{{Peterson} {et~al.}(2004){Peterson}, {Ferrarese}, {Gilbert}, {Kaspi},
  {Malkan}, {Maoz}, {Merritt}, {Netzer}, {Onken}, {Pogge}, {Vestergaard}, \&
  {Wandel}}]{Peterson:2004}
{Peterson}, B.~M., {Ferrarese}, L., {Gilbert}, K.~M., {et~al.} 2004, \apj, 613,
  682
  
\bibitem[Park et al.(2012)]{Park:2012} Park, D., Woo, J.-H., 
Treu, T., et al.\ 2012, \apj, 747, 30
  
\bibitem[Park et al.(2013)]{Park:2013} Park, D., Woo, J.-H., 
Denney, K.~D., \& Shin, J.\ 2013, \apj, 770, 87 

\bibitem[P{\'e}rez-Gonz{\'a}lez et al.(2008)]{Perez:2008} 
P{\'e}rez-Gonz{\'a}lez, P.~G., Rieke, G.~H., Villar, V., et al.\ 2008, 
\apj, 675, 234 

\bibitem[Richards et al.(2002)]{Richards:2002} Richards, G.~T., Fan, 
X., Newberg, H.~J., et al.\ 2002, \aj, 123, 2945

\bibitem[{{Richards} {et~al.}(2006{\natexlab{b}}){Richards}, {Lacy},
  {Storrie-Lombardi}, {Hall}, {Gallagher}, {Hines}, {Fan}, {Papovich}, {Vanden
  Berk}, {Trammell}, {Schneider}, {Vestergaard}, {York}, {Jester}, {Anderson},
  {Budav{\'a}ri}, \& {Szalay}}]{Richards:2006b}
{Richards}, G.~T., {Lacy}, M., {Storrie-Lombardi}, L.~J., {et~al.}
  2006{\natexlab{a}}, \apjs, 166, 470

\bibitem[{{Richards} {et~al.}(2006{\natexlab{a}}){Richards}, {Strauss}, {Fan},
  {Hall}, {Jester}, {Schneider}, {Vanden Berk}, {Stoughton}, {Anderson},
  {Brunner}, {Gray}, {Gunn}, {Ivezi{\'c}}, {Kirkland}, {Knapp}, {Loveday},
  {Meiksin}, {Pope}, {Szalay}, {Thakar}, {Yanny}, {York}, {Barentine},
  {Brewington}, {Brinkmann}, {Fukugita}, {Harvanek}, {Kent}, {Kleinman},
  {Krzesi{\'n}ski}, {Long}, {Lupton}, {Nash}, {Neilsen}, {Nitta}, {Schlegel},
  \& {Snedden}}]{Richards:2006a}
{Richards}, G.~T., {Strauss}, M.~A., {Fan}, X., {et~al.} 2006{\natexlab{b}},
  \aj, 131, 2766
  
  \bibitem[Rosario et  al.(2012)]{Rosario:2012} Rosario, D.~J., Santini, P., Lutz, D., et al.\ 2012, \aap, 545, A45
  
  \bibitem[Rosario et al.(2013)]{Rosario:2013} Rosario, D.~J., Trakhtenbrot, B., Lutz, D., et al.\ 2013, \aap, 560, A72
  
 \bibitem[Ross et al.(2013)]{Ross:2013} Ross, N.~P., McGreer, I.~D., White, M., et al.\ 2013, \apj, 773, 14 
  
  \bibitem[Runnoe et al.(2012)]{Runnoe:2012} Runnoe, J.~C., 
Brotherton, M.~S., \& Shang, Z.\ 2012, \mnras, 422, 478 
  
\bibitem[{{Salucci} {et~al.}(1999){Salucci}, {Szuszkiewicz}, {Monaco}, \&
  {Danese}}]{Salucci:1999}
{Salucci}, P., {Szuszkiewicz}, E., {Monaco}, P., \& {Danese}, L. 1999, \mnras,
  307, 637  
  
\bibitem[{{Schechter}(1976)}]{Schechter:1976}
{Schechter}, P. 1976, \apj, 203, 297

\bibitem[{{Schmidt}(1968)}]{Schmidt:1968}
{Schmidt}, M. 1968, \apj, 151, 393

\bibitem[Schneider et al.(2010)]{Schneider:2010} Schneider, D.~P., 
Richards, G.~T., Hall, P.~B., et al.\ 2010, \aj, 139, 2360

\bibitem[Schramm 
\& Silverman(2013)]{Schramm:2013} Schramm, M., \& Silverman, J.~D.\ 2013, \apj, 767, 13

\bibitem[Schulze et 
al.(2009)]{Schulze:2009} Schulze, A., Wisotzki, L., \& Husemann, B.\ 2009, \aap, 507, 781

\bibitem[{{Schulze} \& {Wisotzki}(2010)}]{Schulze:2010}
{Schulze}, A. \& {Wisotzki}, L. 2010, \aap, 516, A87+

\bibitem[{{Schulze} \& {Wisotzki}(2011)}]{Schulze:2011}
{Schulze}, A. \& {Wisotzki}, L. 2011, \aap, 535, A87

\bibitem[{{Schulze}(2011)}]{Schulze:2011b} Schulze, A.\ 2011, PhD thesis, urn:nbn:de:kobv:517-opus-54464

\bibitem[Schulze \& Wisotzki(2014)]{Schulze:2014} Schulze, A., \& Wisotzki, L.\ 2014, \mnras, 438, 3422 

\bibitem[Sijacki et al.(2014)]{Sijacki:2014} Sijacki, D., 
Vogelsberger, M., Genel, S., et al.\ 2014, arXiv:1408.6842 


\bibitem[Silverman et al.(2008)]{Silverman:2008} Silverman, J.~D., 
Green, P.~J., Barkhouse, W.~A., et al.\ 2008, \apj, 679, 118 

\bibitem[Silverman et al.(2009)]{Silverman:2009} Silverman, J.~D., 
Lamareille, F., Maier, C., et al.\ 2009, \apj, 696, 396

\bibitem[{{Shankar} {et~al.}(2009){Shankar}, {Weinberg}, \&
  {Miralda-Escud{\'e}}}]{Shankar:2009}
{Shankar}, F., {Weinberg}, D.~H., \& {Miralda-Escud{\'e}}, J. 2009, \apj, 690, 20

\bibitem[Shankar et al.(2013)]{Shankar:2013} Shankar, F., Weinberg, 
D.~H., \& Miralda-Escud{\'e}, J.\ 2013, \mnras, 428, 421
  
  
  \bibitem[{{Shen} {et~al.}(2008){Shen}, {Greene}, {Strauss},
  {Richards}, \& {Schneider}}]{Shen:2008}
{Shen}, Y., {Greene}, J.~E., {Strauss}, M.~A., {Richards}, G.~T., \&
  {Schneider}, D.~P. 2008, \apj, 680, 169
  
  \bibitem[Shen(2009)]{Shen:2009} Shen, Y.\ 2009, \apj, 704, 89
  
  \bibitem[Shen et al.(2011)]{Shen:2011} Shen, Y., Richards, G.~T., 
Strauss, M.~A., et al.\ 2011, \apjs, 194, 45 
  
 \bibitem[Shen \& Kelly(2012)]{Shen:2012} Shen, Y., \& Kelly, B.~C.\ 2012, \apj, 746, 169
 
 \bibitem[Shen(2013)]{Shen:2013} Shen, Y.\ 2013, Bulletin of the 
Astronomical Society of India, 41, 61

\bibitem[Small \& Blandford(1992)]{Small:1992} Small, T.~A., \& Blandford, R.~D.\ 1992, \mnras, 259, 725
  
\bibitem[{{Soltan}(1982)}]{Soltan:1982}
{Soltan}, A. 1982, \mnras, 200, 115

\bibitem[Soria et al.(2006)]{Soria:2006} Soria, R., Fabbiano, G., 
Graham, A.~W., et al.\ 2006, \apj, 640, 126 

\bibitem[Springel(2005)]{Springel:2005} Springel, V.\ 2005, \mnras, 364, 1105 

\bibitem[Steinhardt \& Elvis(2010)]{Steinhardt:2010} Steinhardt, C.~L., \& Elvis, M.\ 2010, \mnras, 406, L1

\bibitem[Trakhtenbrot 
\& Netzer(2012)]{Trakhtenbrot:2012} Trakhtenbrot, B., \& Netzer, H.\ 2012, \mnras, 427, 3081

\bibitem[Treister et al.(2009)]{Treister:2009} Treister, E., Urry, C.~M., \& Virani, S.\ 2009, \apj, 696, 110 

\bibitem[{{Tremaine} {et~al.}(2002){Tremaine}, {Gebhardt}, {Bender}, {Bower},
  {Dressler}, {Faber}, {Filippenko}, {Green}, {Grillmair}, {Ho}, {Kormendy},
  {Lauer}, {Magorrian}, {Pinkney}, \& {Richstone}}]{Tremaine:2002}
{Tremaine}, S., {Gebhardt}, K., {Bender}, R., {et~al.} 2002, \apj, 574, 740

\bibitem[{{Trump} {et~al.}(2009){Trump}, {Impey}, {Kelly}, {Elvis}, {Merloni},
  {Bongiorno}, {Gabor}, {Hao}, {McCarthy}, {Huchra}, {Brusa}, {Cappelluti},
  {Koekemoer}, {Nagao}, {Salvato}, \& {Scoville}}]{Trump:2009}
{Trump}, J.~R., {Impey}, C.~D., {Kelly}, B.~C., {et~al.} 2009, \apj, 700, 49

\bibitem[Trump et al.(2011)]{Trump:2011} Trump, J.~R., Impey, 
C.~D., Kelly, B.~C., et al.\ 2011, \apj, 733, 60 

\bibitem[{{Ueda} {et~al.}(2003){Ueda}, {Akiyama}, {Ohta}, \&
  {Miyaji}}]{Ueda:2003}
{Ueda}, Y., {Akiyama}, M., {Ohta}, K., \& {Miyaji}, T. 2003, \apj, 598, 886

\bibitem[Ueda et al.(2008)]{Ueda:2008} Ueda, Y., Watson, M.~G., Stewart, I.~M., et al.\ 2008, \apjs, 179, 124 

\bibitem[Ueda et al.(2014)]{Ueda :2014} Ueda, Y., Akiyama, M., 
Hasinger, G., Miyaji, T., \& Watson, M.~G.\ 2014, \apj, 786, 104 

\bibitem[{{Vanden Berk} {et~al.}(2001){Vanden Berk}, {Richards}, {Bauer},
  {Strauss}, {Schneider}, {Heckman}, {York}, {Hall}, {Fan}, {Knapp},
  {Anderson}, {Annis}, {Bahcall}, {Bernardi}, {Briggs}, {Brinkmann}, {Brunner},
  {Burles}, {Carey}, {Castander}, {Connolly}, {Crocker}, {Csabai}, {Doi},
  {Finkbeiner}, {Friedman}, {Frieman}, {Fukugita}, {Gunn}, {Hennessy},
  {Ivezi{\'c}}, {Kent}, {Kunszt}, {Lamb}, {Leger}, {Long}, {Loveday}, {Lupton},
  {Meiksin}, {Merelli}, {Munn}, {Newberg}, {Newcomb}, {Nichol}, {Owen}, {Pier},
  {Pope}, {Rockosi}, {Schlegel}, {Siegmund}, {Smee}, {Snir}, {Stoughton},
  {Stubbs}, {SubbaRao}, {Szalay}, {Szokoly}, {Tremonti}, {Uomoto}, {Waddell},
  {Yanny}, \& {Zheng}}]{VandenBerk:2001}
{Vanden Berk}, D.~E., {Richards}, G.~T., {Bauer}, A., {et~al.} 2001, \aj, 122,
  549

\bibitem[Vasudevan 
\& Fabian(2007)]{Vasudevan:2007} Vasudevan, R.~V., \& Fabian, A.~C.\ 2007, \mnras, 381, 1235 

\bibitem[Veale et al.(2014)]{Veale:2014} Veale, M., White, M., 
\& Conroy, C.\ 2014, \mnras, 445, 1144 

\bibitem[Vestergaard \& Wilkes(2001)]{Vestergaard:2001} Vestergaard, M., \& Wilkes, B.~J.\ 2001, \apjs, 134, 1 

\bibitem[{{Vestergaard}(2002)}]{Vestergaard:2002}
{Vestergaard}, M. 2002, \apj, 571, 733

\bibitem[{{Vestergaard} \& {Peterson}(2006)}]{Vestergaard:2006}
{Vestergaard}, M. \& {Peterson}, B.~M. 2006, \apj, 641, 689

\bibitem[{{Vestergaard} {et~al.}(2008){Vestergaard}, {Fan}, {Tremonti},
  {Osmer}, \& {Richards}}]{Vestergaard:2008}
{Vestergaard}, M., {Fan}, X., {Tremonti}, C.~A., {Osmer}, P.~S., \& {Richards},
  G.~T. 2008, \apjl, 674, L1
  
\bibitem[{{Vestergaard} \& {Osmer}(2009)}]{Vestergaard:2009}
{Vestergaard}, M. \& {Osmer}, P.~S. 2009, \apj, 699, 800

\bibitem[Vignali et al.(2014)]{Vignali:2014} Vignali, C., Mignoli, M., Gilli, R., et al.\ 2014, \aap, 571, AA34

\bibitem[Wang et al.(2006)]{Wang:2006} Wang, J.-M., Chen, Y.-M., 
\& Zhang, F.\ 2006, \apjl, 647, L17 

\bibitem[Wang et al.(2009)]{Wang:2009} Wang, J.-G., Dong, X.-B., 
Wang, T.-G., et al.\ 2009, \apj, 707, 1334 

\bibitem[{{Wisotzki} {et~al.}(2000){Wisotzki}, {Christlieb}, {Bade},
  {Beckmann}, {K{\"o}hler}, {Vanelle}, \& {Reimers}}]{Wisotzki:2000}
{Wisotzki}, L., {Christlieb}, N., {Bade}, N., {et~al.} 2000, \aap, 358, 77

\bibitem[Wolf et 
al.(2003)]{Wolf:2003} Wolf, C., Wisotzki, L., Borch, A., et al.\ 2003, \aap, 408, 499

\bibitem[{{Woo} {et~al.}(2010){Woo}, {Treu}, {Barth}, {Wright}, {Walsh},
  {Bentz}, {Martini}, {Bennert}, {Canalizo}, {Filippenko}, {Gates}, {Greene},
  {Li}, {Malkan}, {Stern}, \& {Minezaki}}]{Woo:2010}
{Woo}, J., {Treu}, T., {Barth}, A.~J., {et~al.} 2010, \apj, 716, 269

\bibitem[Woo et al.(2013)]{Woo:2013} Woo, J.-H., Schulze, A., 
Park, D., et al.\ 2013, \apj, 772, 49


\bibitem[Wyithe \& Loeb(2003)]{Wyithe:2003} Wyithe, J.~S.~B., \& Loeb, A.\ 2003, \apj, 595, 614

\bibitem[Xue et al.(2010)]{Xue:2010} Xue, Y.~Q., Brandt, W.~N., Luo, B., et al.\ 2010, \apj, 720, 368 

\bibitem[{{Yu} \& {Lu}(2004)}]{Yu:2004}
{Yu}, Q. \& {Lu}, Y. 2004, \apj, 602, 603

\bibitem[{{Yu} \& {Lu}(2008)}]{Yu:2008}{Yu}, Q. \& {Lu}, Y. 2008, \apj, 689, 732

\bibitem[{{Yu} {et~al.}(2005){Yu}, {Lu}, \& {Kauffmann}}]{Yu:2005}
{Yu}, Q., {Lu}, Y., \& {Kauffmann}, G. 2005, \apj, 634, 901

\bibitem[{{Yu} \& {Tremaine}(2002)}]{Yu:2002} {Yu}, Q. \& {Tremaine}, S. 2002, \mnras, 335, 965

\bibitem[Yuan(2007)]{Yuan:2007} Yuan, F.\ 2007, The Central 
Engine of Active Galactic Nuclei, 373, 95

\end{thebibliography}
\end{document}